\documentclass[useAMS,usenatbib,usegraphicx]{mn2e}
\title[Box Size]{The Effect of Large-Scale Power on Simulated Spectra
of the \lyaf}

\author[D. Tytler \etal]{David Tytler\thanks{E-mail: tytler@ucsd.edu},
  Pascal Paschos,
  David Kirkman, Michael L. Norman, and
  Tridivesh Jena \newauthor\\
  Center for Astrophysics and Space Sciences,
  University of California San Diego,
  La Jolla, CA, 92093-0424 \newauthor
}

\usepackage{color}
\usepackage{graphicx}
 
\newcommand{\Lya}{\mbox{Ly$\alpha$}}
\newcommand{\lya}{\mbox{Ly$\alpha$}}

\newcommand{\kms}{\mbox{km s$^{-1}$}}
\newcommand{\cmm}{\mbox{cm$^{-2}$}}

\newcommand{\msun}{$M_{\odot}~$}

\newcommand{\ob}{\mbox{$\Omega_b$}}

\newcommand{\om}{\mbox{$\Omega_m$}}
\newcommand{\ol}{\mbox{$\Omega_{\Lambda } $}}
\newcommand{\sig}{\mbox{$\sigma_8$}}
\newcommand{\gammahe}{\mbox{$X_{228}$}}
\newcommand{\gammah}{\mbox{$\gamma_{912}$}}

\newcommand{\dc }{\mbox{$\delta _{CDM}$}}
\newcommand{\stwol }{\mbox{$s_l^2$}}
\newcommand{\varl }{\mbox{$\sigma _l^2$}}

\newcommand{\zem}{\mbox{$z_{\rm em}$}}

\newcommand{\nhi}{\mbox{N$_{\rm H I}$}}
\newcommand{\lnhi}{\mbox{log \nhi}}

\newcommand{\bsig}{\mbox{$b_{\sigma}$}}

\newcommand{\etal}{{\it et al.}}
\newcommand{\lyaf} {\lya\ forest}

\newif\ifdraftmodep
\draftmodepfalse

\newif\ifapjp
\apjpfalse

\newcommand{\aap}{A\&A}
\newcommand{\apjs}{ApJS}

\begin{document}

\date{\today}

\maketitle

\begin{abstract}
We explore the effects of size of the box that we use for
simulations of the intergalactic medium (IGM) at redshift two.
We examine six simulations from the hydrodynamic code ENZO
using the same cosmological and astrophysical
input parameters and cell size, but different box size.
We study the CDM distribution and
many statistics of the \lyaf\ absorption from the IGM.
Larger boxes have fewer pixels with significant absorption
(flux $< 0.96$), more pixels in longer stretches with little or no absorption,
and they have wider \lya\ lines.
The larger boxes differ only because they include power from long
wavelength modes that do not fit inside the periodic conditions of the
smaller boxes.
The long modes change the density, velocity and temperature fields
and these increase in the gas temperature.
Small simulations are too cold compared to larger ones.
When we deliberately increase the heat we put into the IGM, we can
approximate the \lyaf\ in a simulation of twice the size.
When we double the box size, the difference of most statistics
from their value in our largest 76.8~Mpc
box is reduced by approximately a factor of two.
Most of the statistics converge towards their value in the simulation
with the largest box size, though line widths are not yet converged
and the most common value of the CDM density
shows no sign of converging, because the
larger boxes include places with ever higher densities. These regions
are not in the IGM, but they may produce the strongest of \lya\ lines.

When we double the box size from 38.4~Mpc to 76.8~Mpc,
the mean \lya\ absorption decreases 0.5\%,
the frequency with which we encounter different common CDM densities
changes by 2\%,
typical \lya\ line widths, the frequency of flux values and
the power spectrum of the flux all change by 4--7\%, and the
column density distribution changes by up to 15\%.
When we compare to the errors in data, we find that our
76.8~Mpc box is larger than we need for the mean flux, barely large
enough for the
column density distribution and the power spectrum of the flux, and too
small for the line widths that increase by 1~\kms\  when we increase the
box from 38.4~Mpc to 76.8~Mpc, which is approximately the error in data.
We can most readily see the effects of the long wavelength modes in
measurement on the smallest scales in the \lyaf , the line widths,
because they are easier to measure than the long wavelength power.
Our optically thin simulations
have a factor of several too few lines with H~I column densities
$> 10^{17}$~\cmm . Reducing the cell size from 75 to 18.75~kpc is not
a solution.
Our simulated spectra have 20\% less power than data on small scales and
50\% less on large scales, and  their \lya\ lines are 2.6~\kms\ too wide.
We do not see how our simulations might match all data at $z=2$.
Reducing the cell size to 18.75~kpc lowers the \lya\ line widths by 1.8 \kms , but
radiative transfer effects can increase them by as much as 1.3 \kms~
at z=2.5.
We might reduce line widths using
a softer ionizing spectrum to reduce heating, or we could use \sig ~$>0.9$
that has the additional benefit of increasing the large scale power.
It is hard to see how simulations using popular
cosmological and astrophysical parameters can match the \lyaf\ data at $z=2$.

\end{abstract}


\section{Introduction}

We are exploring the physical conditions in the IGM and the history
stored in those conditions. We retrieve the physical conditions by
finding numerical simulations that give simulated spectra that have
H~I \lyaf\ absorption that are statistically similar to real spectra.
Data on the IGM can give relative errors of the
order of a few percent for statistical properties of the forest.
In \citet{jena05a} (J05) we showed that at redshift 1.95 our simulations
using typical cosmological and
astrophysical parameters gave a good match to both the mean flux transmitted
in the \lyaf\ and
the $b$-value distribution that we use to describes the \lya\ line widths.
We noted that the power spectrum of the flux for these
simulations was broadly similar to that of data. However, we spent little
time with the power, and we did not attempt
to run any simulations that gave exactly the mean flux and $b$-values of data.

Here we explore one aspect of the accuracy of our simulations, the
dependence on size of the simulation box.
We discuss various statistics, including
the mean flux, the flux probability distribution function (pdf), the
typical $b$-value,
the pdf of the $b$-values, the power and autocorrelation of the flux and the
density and power of the CDM. We restrict our attention to one cosmological model at
one epoch, and we say little or nothing about other relevant factors such as
redshift evolution, the tilt of the initial power spectrum of fluctuations and
a measure of the amplitude of the density fluctuations today, \sig . We also
do not discuss other important aspects of the simulations,
including the accuracy of the initial conditions, the
redshift where the simulations begin \citep{heitmann06a,lukic07a}
the accuracy of the potential and hydrodynamical evolution \citep{regan07a},
the ionization and heating and the lack of radiative transfer.
In J50 we showed how cosmological and astrophysical
parameters, and the box and cell size change the mean flux and $b$-values.
Here we cover many more statistics of the \lyaf\ in a more quantitative manner.

It is now well known
\citep{kauffmann92,pen97a,barkana04a,sirko07a}, that we need box sizes of many hundreds of Mpc
to measure the power of the matter accurately.
For CDM (gravity) alone we can now run simulations
that are large enough to capture most of the variations from large scale
modes \citep{neto07a}.
However, we can not yet run large enough hydrodynamic simulations with the
$\simeq 20$~kpc cell size required \citep{meiksin04a} in the IGM, although we
could instead run an ensemble of simulations, each with a different mean
density \citep{mandelbaum03}.
\citet{meiksin04a}, for example,  study the convergence properties of the flux
power spectrum and autocorrelation function and recommend a box size of
25~$h^{-1}$ Mpc but only for high redshifts (${z > 3}$) and even then they do
not find convergence to better than 10\%.
Similarly, \citep{bagla05a} study the effects of box size
on halo mass functions and indicate that a minimum box size of
several 100 ${h^{-1}}$ Mpc is needed.

A secondary goal of this work is to make it easier to obtain
validated and reproducible results on the \lyaf , in accord with the
sentiments of the
``Cosmic Code Comparison Project" \citep{heitmann07a}.
Hence we deliberately include many tables
and figures to aid comparisons with other simulations.

In \S 2 below, we briefly describe the simulation code and parameters we have
adopted.
In \S 3 we describe the statistics of the cold
dark matter distribution.
\S 4 describes the statistics of the flux in the
\lyaf\ including the mean flux, flux distribution, and line
$b$-values and column densities.
In \S 5 we give the power of the flux spectra and the autocorrelation.
In \S 6 we give the velocity field, baryon temperature and density.
In \S 7 we show how  putting more heat into a simulation
makes its \lyaf\ appear like a simulation of twice the box length.
In \S 8 we discuss an ambiguity present in the way flux power is calculated.
In \S 9 we discuss how cell size, or resolution changes the \lyaf\ statistics.
In \S 10 we show how the different statistics converge on the values in large boxes
and we compare to data.
In \S 11 we review the physical causes of the changes we saw with box size.
The appendices contain technical details:
A how we make spectra,
B how we evolve them,
C how we make extended sight lines, and
D the lack of realistic variations in the density field.

Overall, the comparison with data shows some large differences that make it
hard to see how simulations will 
be able to exactly match the current \lyaf\ data at $z=2$ using the popular
cosmological and astrophysical parameters.

\section{ENZO IGM Simulations}
\label{secenzo}

The numerical simulations \citep{bodenheimer07a} that we describe in this paper
use the Eulerian hydrodynamic cosmological code ENZO
\citep{bryan95a,bryan97a,norman99,oshea04a,oshea05a,regan07a,norman07a}.
The simulations contain both CDM and baryons in the
form of gas, but no stars.
The simulations were all run with the same cosmological parameters: a
flat geometry $\Omega_{total}$ = 1, comprising a vacuum energy density of
$\Omega_{\Lambda}=0.73$,
$\Omega_{m}=0.27$ (CDM plus baryons), a baryon density of
$\Omega_{b}=0.044$,
a Hubble constant of $H_0$ = 71 km s$^{-1}$ Mpc$^{-1}$ and
an initial power spectrum scalar slope of $n_s = 1.0$ with a current amplitude
of \sig $=0.9$.

The ENZO code follows the evolution of the gas using non-equilibrium chemistry
and cooling for hydrogen and helium ions \citep{abel97a,anninos97a}.
After reionization at $z=6$, photoionization
is provided using the \citet{haardt01a}
volume average UV background (UVB) from an evolving population of
galaxies and QSOs. This gives
$1.330 \times 10^{-12}$ photoionizations per second per H~I atom at $z=2$ and
$1.041 \times 10^{-12}$ photoionizations per second at $z=3$.
The simulations are optically thin so that all cells experience the
same UV intensity at a given time.
We do not treat the transfer of radiation inside the volume, and we include no
feedback from individual stars or QSOs except for that implied by the
uniform UVB.

As in J05, we use two parameters to describe the intensity of the UVB.
The parameter \gammah\ is the
rate of ionization per H~I atom in units of the Haardt \& Madau model discussed
above, while \gammahe\ measures the heat input per He~II ionization, again
in units of the rate for the Haardt \& Madau spectrum.

We initiate the simulations using an \citet{eisenstein99} power spectrum
for the dark matter perturbations, that we insert at $z=99$.
The simulated volumes are all cubes with strictly periodic boundary conditions.
Hence the power is
input at a finite number of discrete wavenumbers. When we increase the box
size, we insert the new modes that now fit inside the box, but we do not change the
amplitudes of the smaller modes.

The amplitude of the power that we insert varies smoothly with wavenumber.
We insert the amplitude expected for the universe as a whole, with no
random variations associated with the finite box sizes.
Since all our simulations use the same cosmological parameters, they all
have exactly the same initial power for all modes that fit inside their box.
The power in the simulations is not adjusted to include the variations
in mean density that we see in the universe on the scale of the boxes.
Hence, all the simulations are
more similar to the mean of the universe than would be any observational
measurement.  We discuss this more in Appendix D. 
In this limited sense, the boxes contain information on
scales much larger than their sizes.

We initiated all simulations using the same random number seed to generate
the phases of all the modes. The phases are assigned to modes according to the
mode direction and size measured in units of the box size (and not Mpc).
Hence, in box units, the simulations have the same distribution of matter
on the largest scales, as we show below.

We ran the simulations to $z=2$, and all the results that we give refer to
$z=2$, except for specific cases discussed in Appendix \ref{secevol}.

\subsection{Series of Simulations}
We will discuss three series of simulations with parameters listed in
Table \ref{sim_table}.
The main A series have identical input parameters except for the
box size. The larger boxes contain more total volume and mass and they contain
power on scales that does not fit in the smaller boxes.

The A and KP series simulations have identical cosmological
parameters and exactly the same comoving cell size of 53.25$h^{-1}$
or 75~kpc comoving.
Each simulation has one CDM particle for each cell initially, and each
dark matter particle, in each simulation, has a mass of
$M_{CDM} \simeq 9.5 \times 10^{6}~h^{-1} M_{\odot}$.
All the simulations are fully constrained by the input
parameters since we do not re-scale any of
the simulations outputs, such as the densities, H~I, or fluxes.

\begin{table}
\caption{\label{sim_table} Parameters input to specify the simulations.
Box and Cell size are comoving distances.
Simulation A4 is in both the A series that explores box size and the
B series that explores the cell size. The KP series are variants on
the A series with different UVB intensity (\gammah )
and heating per He~II ionization (\gammahe ).
}
\begin{tabular}{lrrcll}
\hline
Simulation & $N$     & Box   & Cell & \gammah & \gammahe \cr
or box     & (cells) & size  & size &         &          \cr
name       &         & (Mpc) & (kpc)&         &          \cr
\hline
A    & 1024  & 76.8  & 75   & 1.0    & 1.8    \cr
A2   &  512  & 38.4  & 75   & 1.0    & 1.8    \cr
A3   &  256  & 19.2  & 75   & 1.0    & 1.8    \cr
A4   &  128  & 9.6   & 75   & 1.0    & 1.8    \cr
A6   &   64  & 4.8   & 75   & 1.0    & 1.8    \cr
A7   &   32  & 2.4   & 75   & 1.0    & 1.8    \cr
A2kp &  512  & 38.4  & 75   & 0.9217 & 2.165  \cr
A3kp &  256  & 19.2  & 75   & 0.836  & 2.579  \cr
A4kp &  128  & 9.6   & 75   & 0.738  & 3.045  \cr
B2   &  512  & 9.6   & 18.75 & 1.0    & 1.8    \cr
B    &  256  & 9.6   & 37.5 & 1.0    & 1.8    \cr
A4   &  128  & 9.6   & 75   & 1.0    & 1.8    \cr
B3   &   64  & 9.6   & 150  & 1.0    & 1.8    \cr
\end{tabular}
\end{table}

Each A and KP series simulation is a cube with side length $N \times 75$~kpc.
The A series simulations differ in size by factors of two,
from the largest simulation A with $N^3 = 1024^3$ cells to the smallest A7 with
$N^3 = 32^3$. The A simulation has comoving box side length of
54.528h$^{-1}$ or 76.8 Mpc , while the A7 has sides of 1.704h$^{-1}$ or 2.4 Mpc.
The box for simulation A is 32 times larger in each dimension than A7, giving
it a volume $2^{15} = 32,768$ times larger.
We also ran a simulation
with $16^3$ cells but found it significantly different and of no value to us.

We discussed simulations A, A2, A3 and A4 in J05 for other purposes.
We do not use the label A5 because this was a version of A4 described by J05.
In J05 we noted a problem with A2. We have now re-run this simulation
and we found that the problem was incorrect joining of the sub-grids of
A2 presented in J05. The results presented for A2 in J05 were incorrect
for the power spectrum, but correct for the flux and $b$-values.

In the next few sections we discuss the A series simulations alone.
We discuss the three simulations, A2kp, A3kp and A4kp, the KP series, in \S \ref{fixedout}.
They explore the effect of changing the heat input per He~II ionization.
We discuss the 4 simulations in the B series, which include A4, in \S \ref{cellres}
when we discuss changing the resolution of the simulations by changing the cell size.

\section{The CDM Density Distribution}
\label{sec.cdmden}

We discuss how the CDM density distribution
varies with box size in the A series of simulations at $z=2$.
We do not discuss the baryons until \S 4.

\begin{figure}
\includegraphics[width=80mm]{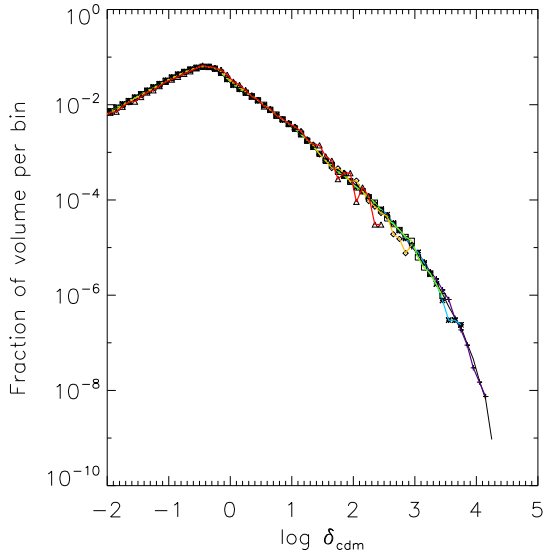} 
\caption{\label{pdfdelta} The frequency distribution of density of CDM,
\dc\ per cell.
We plot the fraction of the cells (or volume) in a given simulation that have
CDM density, in units of the cosmological mean density, in bins of width
log \dc $= 0.1$.
The A simulation is shown with a solid (black) line and the other simulations
are shown using a symbol in each bin, connected by straight line segments, as follows:
A2 (violet pluses), A3 (blue stars), A4 (green squares), A6
(orange diamonds) and A7 (red triangles), where the larger simulations extend farther
to the right.
}
\end{figure}

In Figure \ref{pdfdelta} we show the frequency distribution of the
CDM in the cells. We define the normalized-density of CDM as
\begin{equation}
\dc \equiv \rho _{CDM}/ \bar{ \rho }_{CDM}
\label{eqn.delta}
\end{equation}
where the denominator is the mean density of CDM in the universe at $z=2$
which is also the mean density of the CDM in all our simulations. The shape of the
curve is expected from semi-analytical work \citep{lacey94} on the formation
of dark matter halos.

The distributions of the densities for all simulations continue
towards much lower densities than we show.
The distributions become steeper with decreasing density, but otherwise we
see no conspicuous features.
Since the simulations initially have an average of one CDM particle per cell, most cells
with \dc $< 1$
contain no particles. Their densities can be non-zero because density is defined
by distributing mass with immediately neighboring cells whenever a
particle is not exactly centered in its cell. Hence a particle at the corner
of a cube would contribute \dc = 0.125 to each of the eight cells.

In Table \ref{tabsim1}
we list the percentage of the cells with zero density.
This is 13.78\% in simulation A, increasing systematically to 14.72\% in A7.
These cells have no immediate neighbors containing a CDM particle and we
assigned them a nominal density of
$10^{-22}$ particles per cell, which we can ignore.
We also list the minimum and maximum density in any cell.

The lowest density portions of the distribution are not physically realistic
because we have too few particles per
cell to accurately simulate densities far  below the mean.
We are more interested in the gravitational potential than in the density in
a cell, and the potential is much smoother than the
density distribution. The dark matter is smoothed twice, once when CDM is
assigned to cells using the piecewise linear cloud-in-cell algorithm
\citep{hockney88} and again when the potential is calculated.
Where we have low dark matter densities, we may have low-level fluctuations
in the potential due to particle discreteness.
At worst, the CDM particles become mildly collisional.

The simulations in larger boxes contain higher densities, and hence
they extend further to the right in Fig. \ref{pdfdelta}.
However, on the scale of this plot, all the simulations contain approximately
the same frequencies at each density, with no variation with box size,
which suggests that we will see little effect of box size on the statistics of
the flux beyond those coming from the larger volumes and higher maximum densities
in larger boxes. We do, however,  see one minor difference between the boxes.
Several simulations have frequencies lower by about a factor of two
for the densities within a factor of a few
of the largest value found in that simulation. This is conspicuous for
A7 (red), A6 (orange) and A3 (blue), but not for A4 (green)
and A2 (violet) which seem to follow A (black).

It is striking that we see almost no change with box size
in the probability distribution function (pdf) of the CDM density per cell for the
overdensities responsible for the \lyaf\ absorption, approximately
0.5 $\leq$ \dc$~ \leq$ 16. 
We obtain the typical baryon density corresponding to a given \lnhi\
from \citet[Eqn. 10]{schaye01a} who developed an analytic model that gives an
excellent fit to simulations.
Using the cosmology for the A series (\om = 0.27, \ob = 0.044, H=71~km/s/Mpc)
and the temperature-density relation from simulation A2
($T = 12910~{\rm  K} ( \rho _b / \bar{ \rho }_b )^{0.6}$, J05 Table 9) we find at $z=2$
\begin{equation}
N_{HI} \simeq 7.8 \times 10^{12} (\rho _b/ \bar{ \rho } _b)^{1.34}
\end{equation}
Lines in the \lyaf\ with \lnhi = 12.5 -- 14.5 ~\cmm\
then come typically from baryon overdensities of
$ \rho _b/ \bar{ \rho }_b  = $ 0.5 -- 15.7.
A larger range of densities is involved in making significant absorption
in the \lyaf\ \citep{schaye03a}.
For this discussion we assume that the baryon and CDM density fluctuations
are similar in amplitude. \citet{gnedin03a} show that the baryon fluctuations are
similar to the CDM fluctuations on scales larger than the filtering scale, with
$\delta _b / \delta _{CDM} = exp(-k^2/k_F^2)$, where the filtering scale $k_F$
depends on the integral over time of the Jeans length and is
 approximately 0.055 s/km (11 Mpc$^{-1}$) at $z=2$ (their Fig. 2).

\begin{table}
\caption{\label{tabsim1} Statistics of the distribution of
the density of CDM in the A-series simulations.
Zeros are the percentage of cells in simulation with zero density.
NonL gives the percentage of cells in simulation with \dc $> 3$.
}
\begin{tabular}{lrcccc}
\hline
Name & $N$ & Zeros & Min  & NonL & Max \cr
     & (cells)& (\%)      & \dc\ & (\%)     & \dc\ \cr
\hline
A    & 1024 & 13.78 & 3.68$\cdot$10$^{-11}$ & 4.67 & 1.64$\cdot$10$^{4}$ \cr
A2   &  512 &  14.07 & 2.32$\cdot$10$^{-10}$ & 4.68 & 1.41$\cdot$10$^{4}$ \cr
A3   &  256 &  14.37 & 4.86$\cdot$10$^{-9}$  & 4.71 & 5.71$\cdot$10$^{3}$ \cr
A4   &  128 &   14.44 & 3.92$\cdot$10$^{-8}$  & 4.79 & 3.12$\cdot$10$^{3}$ \cr
A6   &   64 &   14.67 & 3.50$\cdot$10$^{-8}$  & 4.85 & 8.25$\cdot$10$^{2}$ \cr
A7   &   32 &  14.72 & 2.09$\cdot$10$^{-7}$  & 4.91 & 3.01$\cdot$10$^{2}$ \cr

\end{tabular}
\end{table}

The statistics on the distribution of the density of CDM in
Table \ref{tabsim1} show that both the maximum and minimum
density of CDM in any cell increases systematically with
the box size. The minimum is not relevant to us; since we work with the density
and not the log(density), these values are all essentially zero, but the maxima are
important for the flux spectra.

\begin{figure}
\includegraphics[width=84mm]{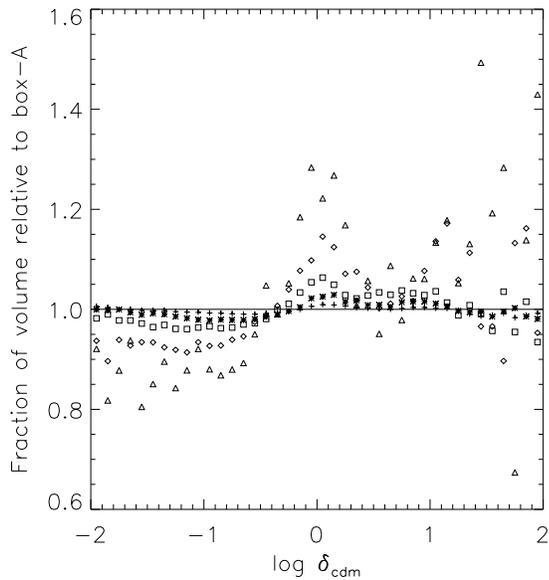}
\caption{\label{pdfdeltaa} As Fig. \ref{pdfdelta} but showing the
frequency of the densities of CDM in units of the frequency in simulation A.
The points from the larger boxes are higher at log \dc $=1$.
}
\end{figure}

In Figure \ref{pdfdeltaa}
we zoom in on the densities that are more
important for the flux in the \lyaf , and we expand the sensitivity by
dividing by the frequencies found
in simulation A. We now see systematic trends with box size.
The larger boxes have higher frequencies of small
densities, approximately log(\dc) $ < -0.3$, and lower
frequencies of higher densities.
All boxes have approximately the same frequency for
log(\dc) $ \simeq -0.3$, near the most common density.
At densities below the most common, the largest difference from the
A simulation is seen at lower densities in the smaller boxes.
Above the most common densities,  the largest differences from A are seen
at near the mean density.
The larger the box, the less the deviation from A.
We anticipate that these differences will manifest as changes
in the baryon density and hence the \lyaf .

In Table \ref{tabsim2}
we list two further statistics showing the changes in the CDM density per cell
with box size,
 the RMS and the mean of the absolute difference (MAD), each relative to the value in
  to box A and
averaged over $-2 < \log (\dc) < 2$.
Both statistics decrease by about a factor of two for each doubling in box
size, reaching an RMS of only 0.7\% for simulation A2.
We  will see a similar rate of convergence for other statistics of the
\lyaf\ \citep{mcdonald01b}.

\begin{table}
\caption{\label{tabsim2} Statistics of the Distribution of
the Density of CDM Relative to Simulation A
}
\begin{tabular}{lrr}
\hline
Name & MAD  & RMS \cr
     & (\%)     & (\%)    \cr
\hline
A  &  0.00   &   0.00  \cr
A2 &  0.57   &   0.70  \cr
A3 &  1.23   &   1.44  \cr
A4 &  3.01   &   3.32  \cr
A6 &  7.11   &   8.23  \cr
A7 &  14.30   &  17.79 \cr

\end{tabular}
\end{table}

\subsection{Distribution of the variance of the density of CDM
amongst sight lines}
\label{vardc}

We will be examining the power of the CDM in the next section because this
helps us understand how the simulations in general and particularly
the power of the flux change with box size.
Here we look first at the variance of the CDM
because this is related to the sum of the power over all modes.

\begin{figure*} 
\includegraphics[width=168mm]{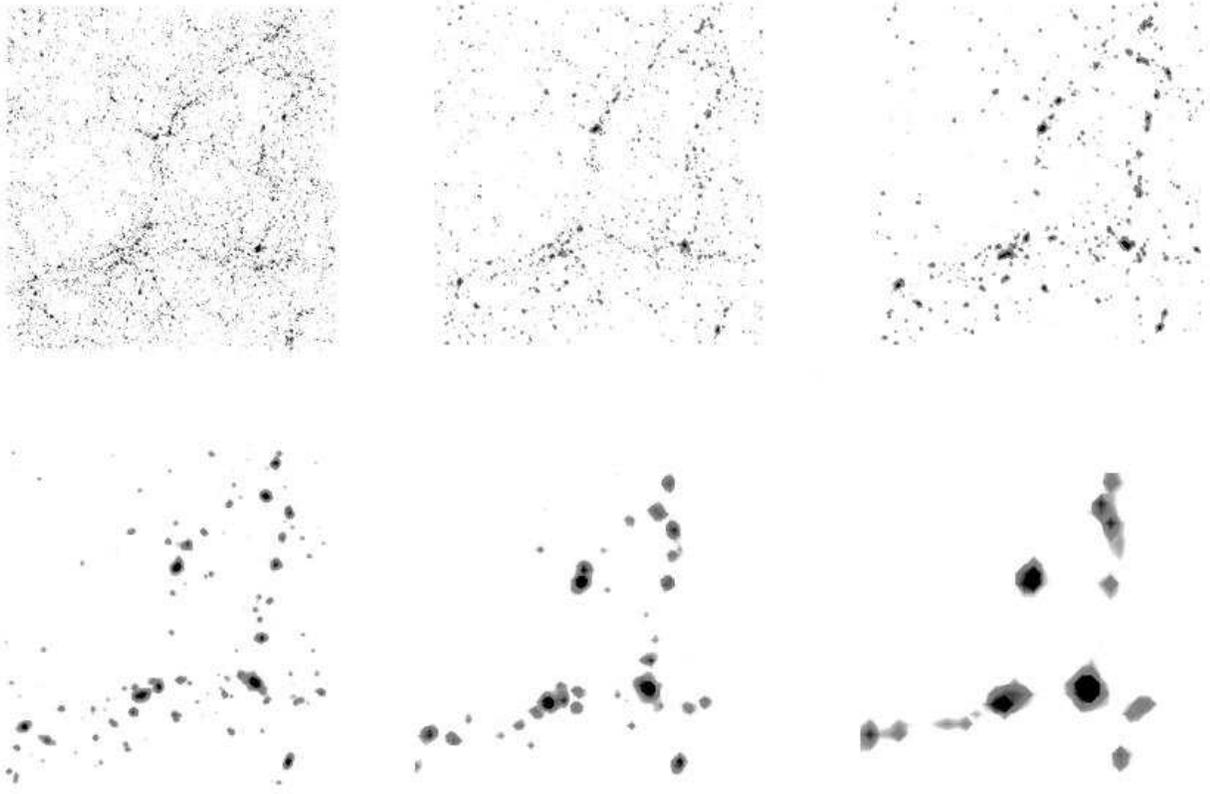}  
\caption{
The contribution to the total variance of the density of CDM
from each sight line. The sight lines
are parallel to the $z$ axis that goes into the page.
Simulation are, from upper left to lower right, A, A2, A3, A4, A6 and A7.
Pixels that have variance below the mean are shown as white.
Darker pixels have larger log \stwol .
}
\label{fig.varimage}
\end{figure*}

In Figure \ref{fig.varimage}
we show the $xy$ faces of the series A simulations.
For each position in the $xy$ plane, we show
\begin{equation}
\stwol \equiv (1/N) \sum_z (\dc -1)^2,
\end{equation}
the mean value of $(\delta _{CDM} -1)^2$ along the $N$ cells
parallel to the $z$ axis which goes into the page.
We consider these rows as sight lines, which we label
with the subscript ``$l$" to indicate a choice of both $x$ and $y$.
We show this quantity because $N\stwol $ is the contribution of that
sight line to the variance of \dc\ in the whole simulation box, since
\begin{eqnarray}
N^3 Var(\delta _{CDM}) \equiv \sum_{x,y,z}(\delta _{CDM} - \bar{\delta }_{CDM})^2
\nonumber \\
= \sum_{x,y,z} (\delta _{CDM}  -1)^2 = \sum_{x,y}N \stwol ,
\label{vardel}
\end{eqnarray}
where $N^3$ is the number of cells in the simulation box,
and $ \bar{\delta }_{CDM} = 1$ if and only if
it is the mean of the \dc\ values of all cells in the box,
following the definition of $\delta _{CDM} $ in Eqn. \ref{eqn.delta}.

The quantity \stwol\ tells us how much that sight line
contributes to the mean power of all sight lines.
The quantity \stwol\ is not the variance along each sight line, \varl ,
since that is the mean of
$(\delta _{CDM} - \bar{\delta }_{CDM,z})^2$,
where the $\bar{\delta }_{CDM,z} $ is the mean
along each sight line. These means differ from sight line to sight line, and
can be very different from unity.

Figure \ref{fig.varimage} looks similar to the projection of the density,
since the \stwol\ is largest where we encounter a cell with a high density.
If we shrink the squares from the smaller simulations
to give constant Mpc per mm on the page, then
the density and size of structures looks approximately the same in all
simulations, although they are not the same, for example because the smaller
boxes are also smaller in the $z$ direction.

\begin{figure*}
\includegraphics[width=3in,height=3in]{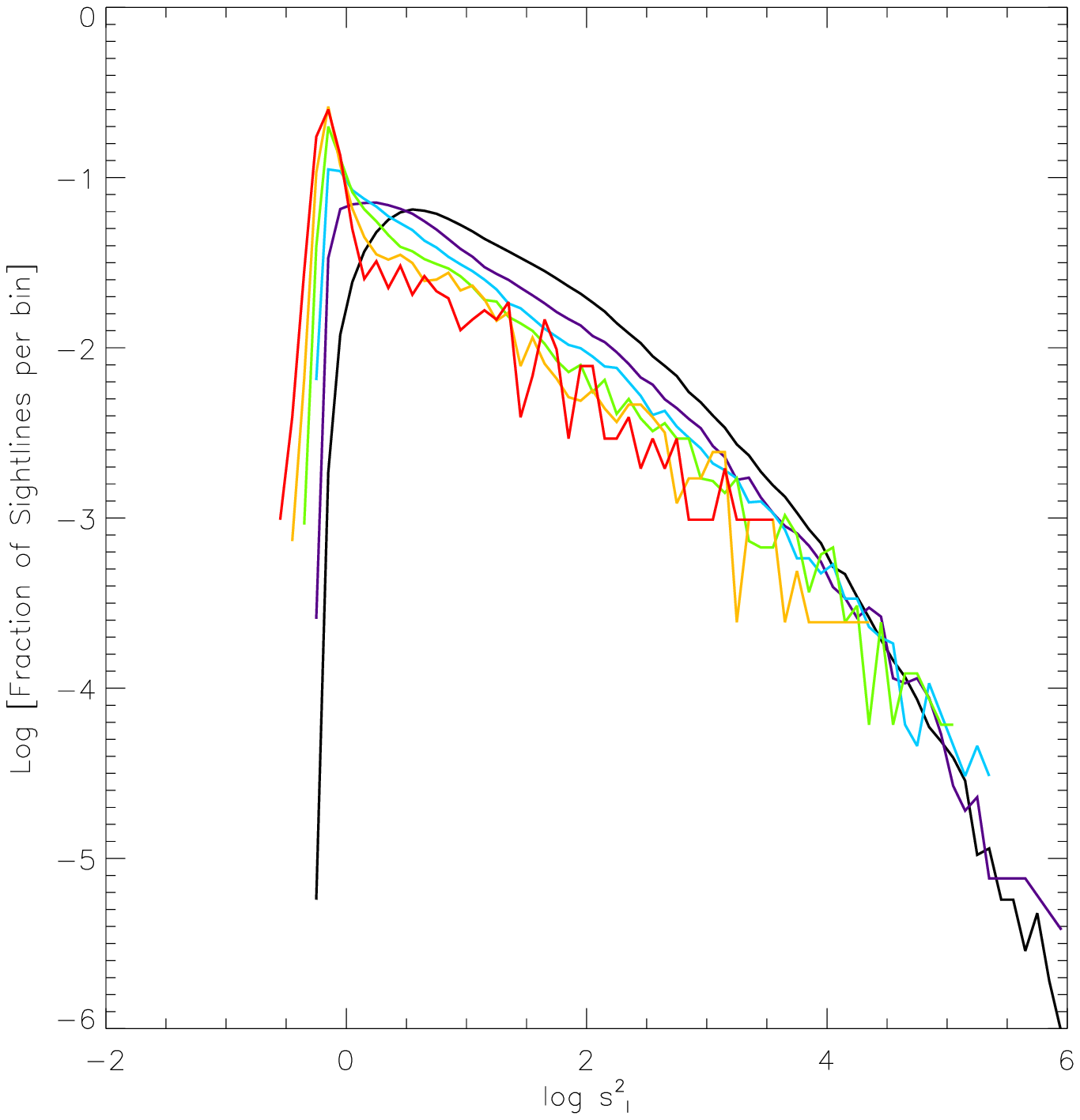}
\includegraphics[width=3in,height=3in]{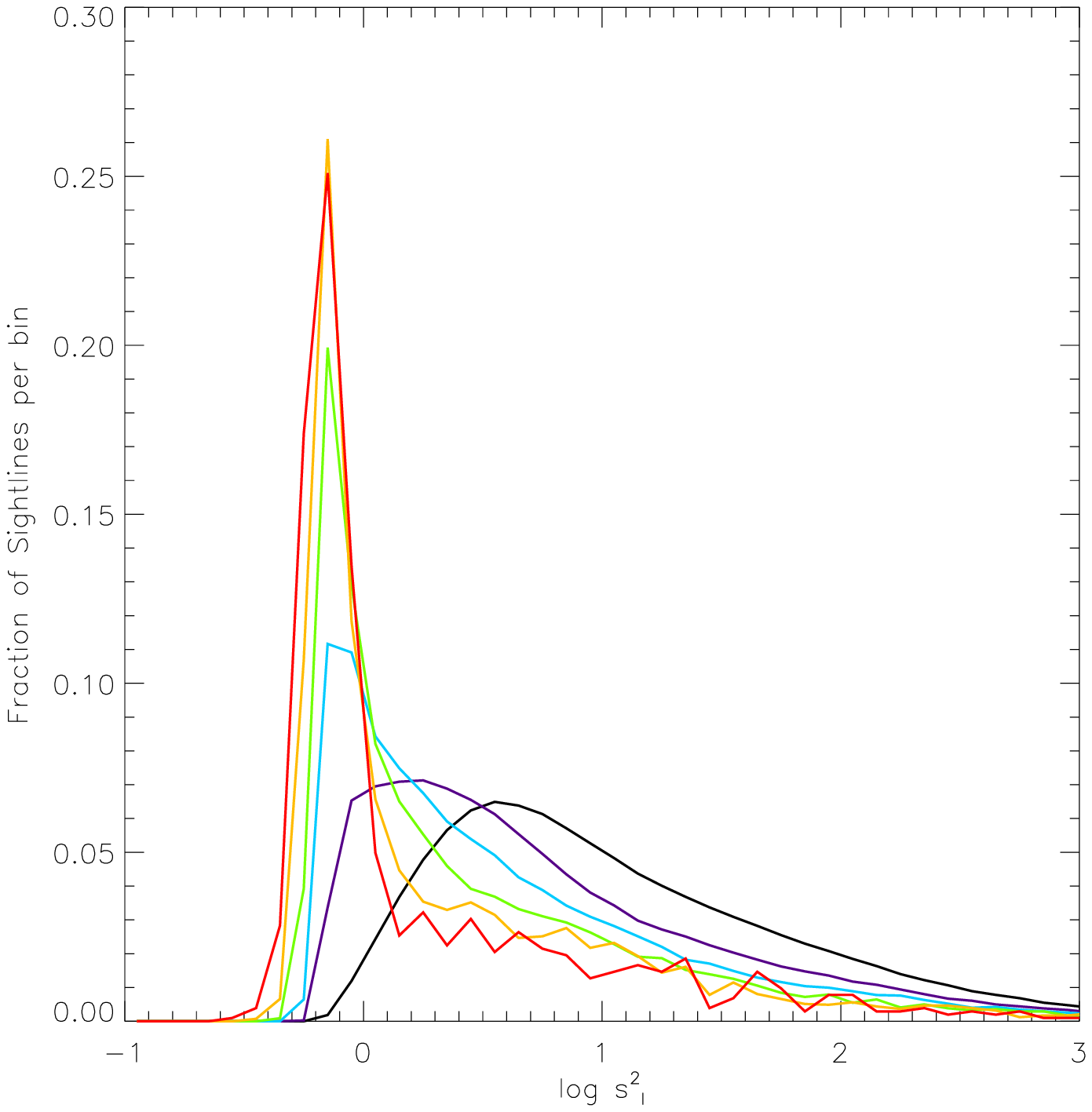}
\caption{
Distribution of the \stwol .
The \stwol\ is measured along sight lines parallel to the $z$ axis
and of length equal to the size of each box.
The vertical axis shows the fraction of sight lines (pixels in the $xy$ plane)
with variance in bins of log $ s_l^2 = 0.1$.
The vertical scale is log (fraction) on the left panel and linear fraction
on the right.
The larger boxes have larger modes (right panel), and extend to larger \stwol\
values (left panel).
Both A and A2 have approximately the same maximum \stwol .
}
\label{fig.cdm_pdf_updated}
\end{figure*}

In Figure \ref{fig.cdm_pdf_updated}
we show the distribution of the \stwol\ and in Table \ref{tabvar}
we give some statistics.
We have one \stwol\ for each sight line parallel to the $z$ axis of
each simulation box.
Larger boxes have a lower frequency of sight lines with small \stwol ,
their most common (mode) \stwol\ is larger, they have a higher frequency of
larger \stwol , and larger maximum \stwol .
This is because the larger boxes have more sight lines, each of which is
longer, and there are higher densities in the larger boxes.

\begin{table}
\caption{\label{tabvar} Statistics of the distribution \stwol\ amongst sight lines.
}
\begin{tabular}{lllrl}
\hline
Name & Min & Mode & Mean & Max \cr
\hline
A    &  0.61 & 3.55 &  175.78 & 1.14$\cdot$10$^{6}$ \cr
A2   &  0.58 & 1.78 &  146.12 & 8.39$\cdot$10$^{5}$ \cr
A3   &  0.50 & 0.71 &  122.84 & 2.49$\cdot$10$^{5}$ \cr
A4   &  0.43 & 0.71 &  100.66 & 1.11$\cdot$10$^{5}$ \cr
A6   &  0.36 & 0.71 &   53.38 & 2.24$\cdot$10$^{4}$ \cr
A7   &  0.29 & 0.71 &   26.61 & 3.42$\cdot$10$^{3}$ \cr

\end{tabular}
\end{table}

The small boxes lack the high density peaks of the larger boxes because they
lack volume, and they lack long modes. They do not contain enough
particles to produce the highest densities.
To make a peak with $10^6$ particles in a cell, we must collect particles
from $10^6$ cells, more than are contained in the A7 simulation.

\citet{bagla05a} have explored how the frequency of high density
CDM collapsed structures changes with effective box size.
They use simulations with $N=256$ CDM particles in $300 h^{-1}$Mpc
boxes with a softening length of $0.47 h^{-1}$Mpc.
They find that the number of high density peaks decreases when they
truncate the  initial power spectra at lengths less than the full box size.
They see a factor of three fewer collapsed structures with mass
$10^{15}$\msun\ when they truncate the power at 1/4 of the box size.
However, when they truncate at 1/2 the box size they see only an 80\%
reduction in the number, showing convergence to the result expected for a
much larger box. This convergence happens for boxes 3 -- 6 times larger than A.

\subsection{Power of Normalized CDM Density}

We compute the Fourier transform $D(k)$ of \dc $- 1$,
the normalized-density of CDM minus one,
\begin{eqnarray}\label{eqnft}
D(k) = {1 \over (\Delta u)^{1/2}}\int_{u} (\dc (u) -1) e^{- j k u} du
\cr
\simeq (\Delta u)^{1/2} \sum_{i=1}^{N_{p}} (\dc (u_{i})-1)
e^{- j k u_{i}}
\end{eqnarray}
where $k$ is the wavenumber, $i$ is the pixel index,
$u$ is velocity
and $\Delta u = c \Delta z/(1+z)$ is the velocity width of a pixel
with redshift width $\Delta z$.
Subtracting one has no effect on $D(k)$ except for
the  mode with zero frequency.
We take the transform of the density along each sight line parallel to and
extending the full length of the $z$ axis.
We did not explore the $x$ and $y$ directions.
We use a discrete Fast Fourier Transformation algorithm, and we use
\begin{equation}\label{eqnpower}
P(k) = < D(k) D^{*}(k) >  \rm~~~~( \kms ),
\end{equation}
as our estimate for the one-dimensional power,
where the brackets refer to averaging over all sight lines parallel to the
$z$ axis.

Since the density distributions in the box is always strictly periodic,
the power spectrum (of the signal along sight lines parallel to the $z$ axis)
is non-zero for the discrete set of modes
$k = 2\pi s /L_v$,
where s=1,2,3... $N$, and $L_v= LH(z)/(1+z)$ is
the length of each spectrum in velocity units, corresponding to $L$ comoving
Mpc.  At redshift $z=2$, $H=201.069$ \kms\/Mpc$^{-1}$, and simulation A
has $L_v = 5147.37$ \kms .

\begin{figure}
\includegraphics[width=84mm]{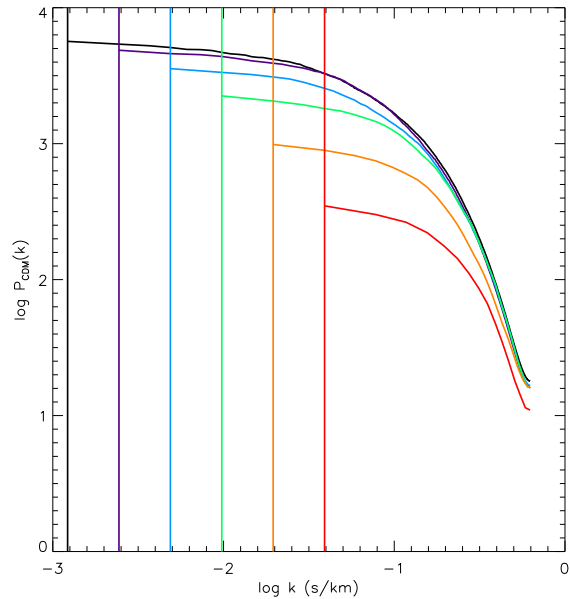}
\caption{The 1D power spectra of  \dc -1,
taken along all 1D line of sight lines parallel to the $z$ axis of each A
series box.
The vertical lines show the largest modes for each box, those with
$k_{box} = 2 \pi / L_v$. The power for the larger box are larger and
extend farther to the left (large distances).
}
\label{fig.pdmspc}
\end{figure}

In Figure \ref{fig.pdmspc}
we show the power spectrum of \dc $-1$ for the A series boxes.
We begin each spectrum at the left at its fundamental mode, the smallest wavenumber
$k_{box} = 2\pi/L_v$ for that box.
For simulation A $k_{box} = 2\pi/L_v = 1.22066 \times 10^{-3}$ s/km.
We end the plots on the right at the Nyquist frequency,
$k_{cell} = 2\pi / (2L_{cell}) = 0.682$ s/km, or log $k_{cell} = -0.20$ s/km,
where $L_{cell}$ is the velocity width of a simulation cell, 5.027 \kms\ at $z=2$.

The power is larger at all $k$ in the larger boxes, which we expect because
the variance like quantity \stwol\
in Fig. \ref{fig.cdm_pdf_updated} is also larger.
The increase in power with box size is most pronounced on the largest scales.
The power is larger in larger boxes even on the smallest scales. We now
present a figure that shows that the increase in the power with box size is
intrinsic to the density distribution, and not an artifact of the length of
the sight lines or the number of sight lines through the boxes.

\begin{figure}
   \includegraphics[width=84mm]{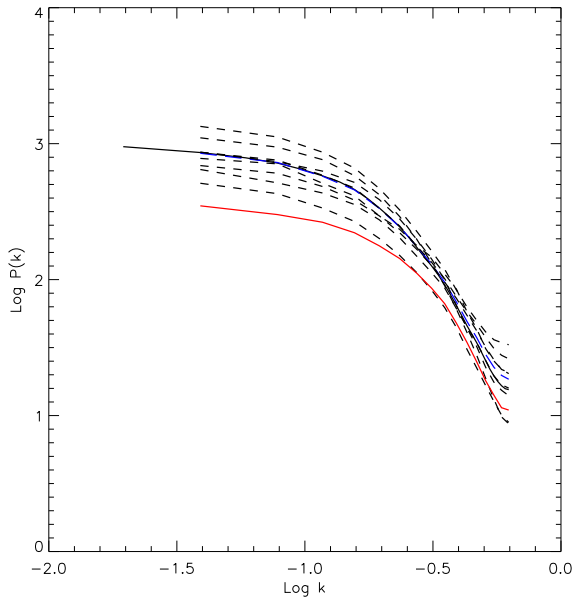}
   \caption{\label{subpowerrandom} The 1D power spectra of \dc -1 for sight
   lines of length 64 cells
   drawn from the A6 ($128^3$) simulation. The solid (black) line extending farthest
to the left is the power for all sight lines through A6. The eight short
dashed
lines show the power for sub-samples, each with $64^2$ sight lines of
length 64 cells. The lowest lines on the left (short solid red) is the power for the
A7 box, which also comprises $64^2$ sight lines each of length 64 cells.
   }
\end{figure}

In Figure \ref{subpowerrandom}
we show power spectra obtained from simulation A6 using a reduced number of
shorter sight lines. We divide the A6 cube into the eight sub-cubes, each of
size A7, that together exactly fill A6.
We made spectra from all the sight lines restricted to each sub-cube, and we
took the power spectrum of each. We then distribute these power spectra
randomly into the eight means,
each of which contains some sight lines for each of the eight sub-cubes,
and the same total number of sight lines as does the mean power spectrum of A7.
We see that the power in the sub-cubes is
distributed about that in A6, and not A7. This shows that the extra power
in A6 is intrinsic to the
density distribution in A6, and not from the length or number of sight lines.
The dispersion of the power in the 8 spectra gives one indication of the random
error in the power of the A7 simulation.

\subsection{Power from Density Peaks}

Parseval's Theorem states that sum of the power at all modes is
proportional to the sum of the
square of the signal. Using the signal from Eqn. \ref{vardel}, we have
\begin{equation}
\label{parseval1}
{1 \over N^3}\sum_{x,y,z} (\delta _{CDM} -1)^2 = {k_{box} \over 2\pi } \sum_{k}P(k)
= {1 \over 2\pi}\int _k P(k)dk
\end{equation}
where $N^3$ is the number of cells in a box, and P(k) has velocity units
 that cancel the inverse velocity units on the $k_{box}$. This can also
be written as
\begin{equation}
\label{parseval2}
{1 \over N^2} \sum_{x,y} \stwol = {1 \over L_v } \sum_{k}P(k),
\end{equation}
which shows that the mean value of \stwol\ averaged over all cells in the
box, is equal to the sum of all modes in the power (which is also averaged
over all sight lines), divided by the length of each spectrum.
This is the normalization used by
\citet{mcdonald06a} (Eqn. 12).
It is the ``System 2" normalization from \citet{bracewell} (p. 7).

The \stwol\ values show how the sum of the power at all modes is distributed
amongst the sight lines. We saw in
Fig. \ref{fig.cdm_pdf_updated} and Table \ref{tabvar} that a few sight lines
have   \stwol\ values vastly larger than the mean. This means that these few
sight lines also dominate the total power of each simulation.

\begin{figure*}
\includegraphics[width=7in,height=3in]{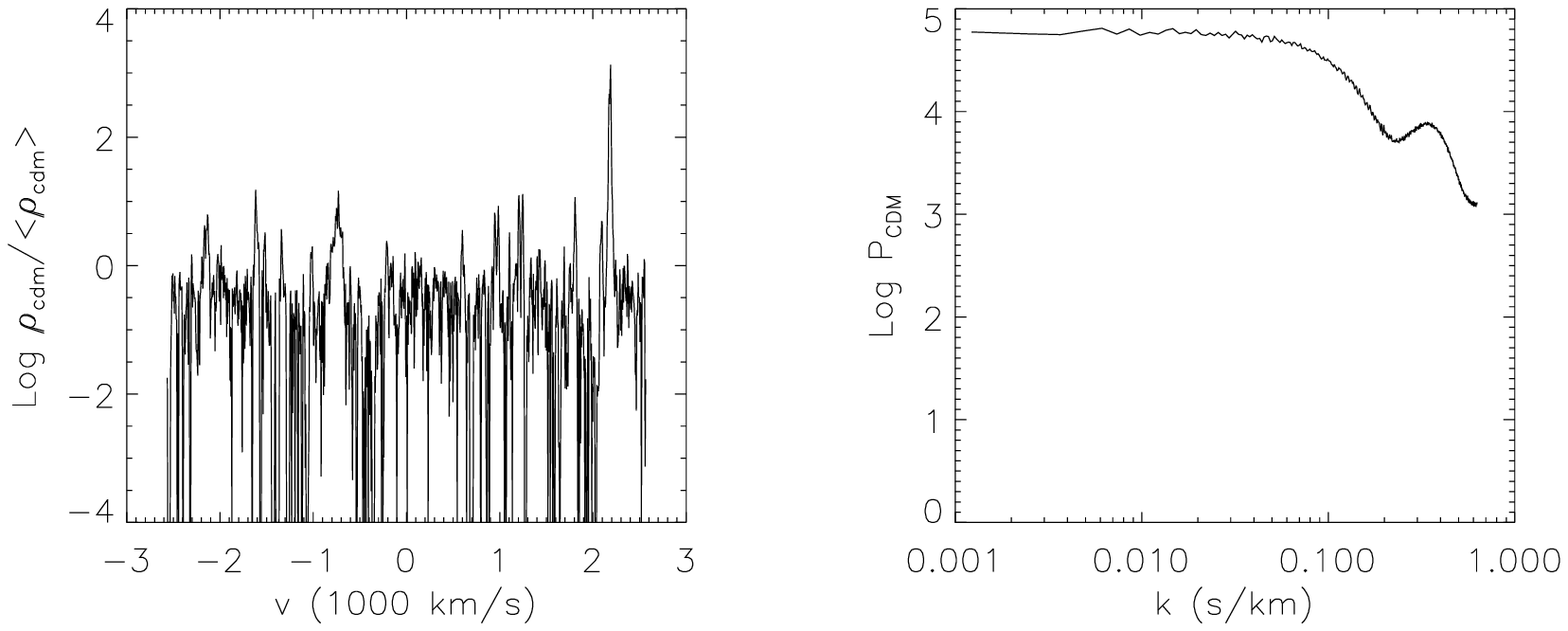}
\includegraphics[width=7in,height=3in]{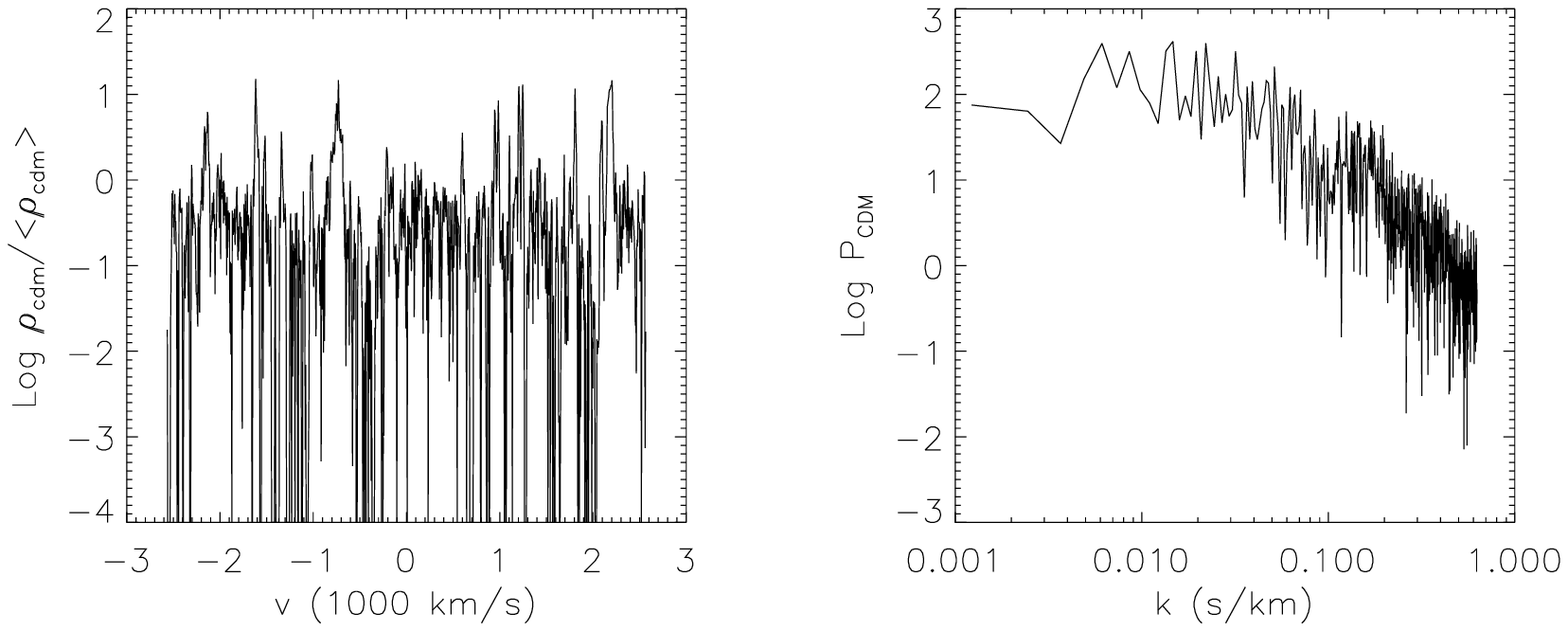}
\caption{
The normalized-density of CDM \dc\ along a single sight line (208) in simulation A as a
function of velocity in the full length through the box (top left).
On the top right we show the 1D power spectrum of this normalized-density.
In the lower left panel we have removed the density peak by smoothing the CDM
between 2200 and 2250~\kms\  with a box car average of 20 pixels or
100 km/s.  The corresponding power is on the lower right.
We show the log of the normalized-density to show both the typical
variations and the peak. We take the power of the linear, not the
log normalized-density. The vertical scale on the two lower panels is different
from on the corresponding upper panels.
}
\label{fig.los208}
\end{figure*}

In Figure \ref{fig.los208}
we show a single 1024 pixel line of sight from simulation A with a density
peak of \dc $\simeq 3 \times 10^{3}$. This peak increases the power about 1000
times over a wide range of wavenumbers.
The density spike is narrow in velocity space, and hence wide in
$k$ space (e.g. Fig. 6.2 of \citet{bracewell}).
In the lower right panel we see that smoothing the
density spike removes most of the excess power.

We can estimate the amount of power added quantitatively.
A typical sight line from simulation A has \stwol\ of order 3
(Table \ref{tabvar}).
One pixel with a normalized-density of 1700 increases the variance to
$(3 \times 1023 + 1700^2)/1024 = 2800$,
approximately the increase we see in the integrated power.
Hence, a single sight line with a normalized-density \dc $= 10^6$
will contain as much power as the whole cube of $10^9$ sight lines each with
typical variance. Simulation A contains
77 cells with \dc $ >  10^5$ and
12 with \dc $>  10^{5.5}$, sufficient that these few cells,
and the sight lines that pass through them, will dominate the
power spectrum of the (un-smoothed) CDM.

\begin{figure}
\includegraphics[width=84mm]{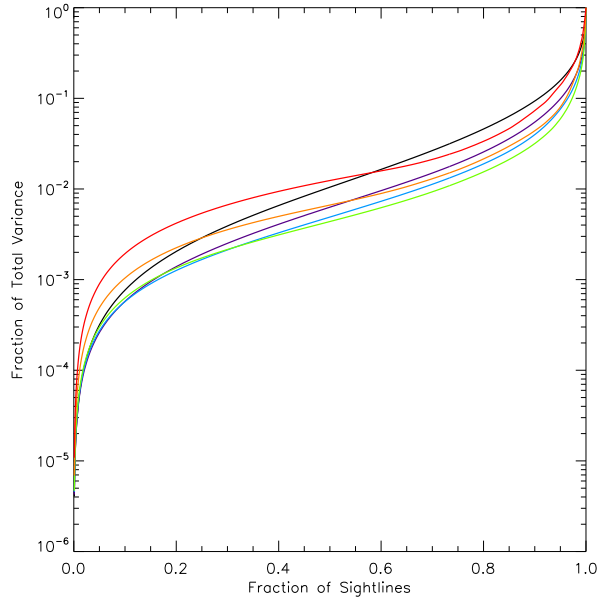}
\caption{
The cumulative sum of the contribution of the sight lines through a simulation
to the total power of all sight lines.
The vertical axis is the cumulative sum of \stwol .
The horizontal axis shows the rank of the sight line ordered in increasing
\stwol , which is its contribution to the total variance.
We express the rank as a fraction of the total number of sight lines,
to simplify comparison of the simulations.
The vertical axis is the sum of the \stwol\ in all sight lines
with \stwol\ smaller than indicated on the horizontal axis,
divided by the total \stwol . From top to bottom at a fraction of 0.8
the boxes are A (black), A7 (red), A2 (violet), A6 (orange), A3 (blue) and
A4 (green).
}
\label{sumvar}
\end{figure}

In Figure \ref{sumvar}
we show how the sum of the \stwol\ of the sight lines
increase with the number of sight lines included. We start on the left with
the sight lines with the smallest \stwol , ending on the right with those
with the largest.
We see that, for all simulations, the few sight lines with the largest \stwol\
completely dominate the total. Depending on the simulation, 90\% of the total
\stwol\ comes from only 2 to 10\% of the sight lines. The total \stwol , and
hence the total power, is an unstable quantity, which can change
significantly as the few highest density peaks changes in number and density.

We have shown that the power of the CDM density
is larger in larger boxes primarily because
larger boxes contain rare regions with higher density. We also see higher
power because we use longer sight lines in the larger boxes. As
\citet{bagla05a} showed, there is some effect from having longer modes in
the larger boxes. However, Figs. \ref{pdfdelta}, \ref{pdfdeltaa}
showed only small changes in the frequencies of different densities per cell,
which suggest that the extra long modes have a small effect on the quantities
that we are evaluating.

When we examine the flux transmitted through the IGM we are  most
interested in densities near the mean. Although the larger boxes
have higher maximum densities, they have slightly lower portions
of their volume above a moderate density.
In Table \ref{tabsim1} the column NonL gives the fraction of cells with CDM density
exceeding 3 times the mean. This fraction decreases systematically with box size,
from 4.91\% for A7 to 4.67\% for A.

\section{Statistics of the Flux in the \lya\ Forest }

We make flux spectra using code described in \citet{zhang97} and J05.
We make each spectrum along a row of cells parallel to the $z$ axis of the
boxes, just as we did for the \stwol\ values that describe the variance of
\dc\ in \S \ref{vardc}. We use a number of pixels equal to the box side
length $N$.

\begin{figure}
\includegraphics[width=84mm]{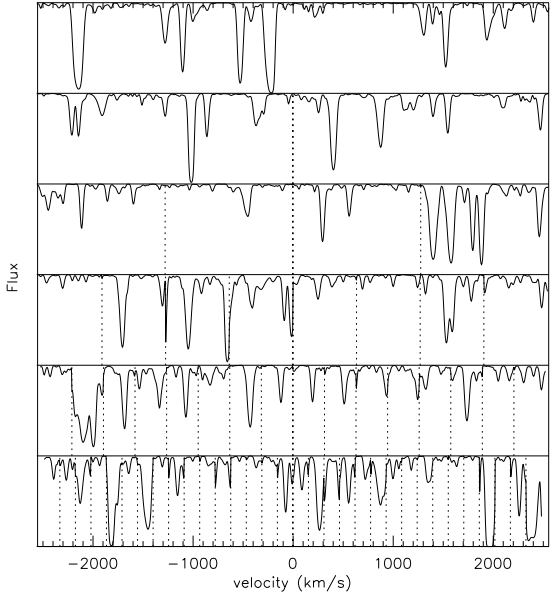}
\caption{
Spectra of the flux from sight lines drawn from the A series simulations, with
the largest, A at the top, A2 next through A7 the smallest
at the bottom. All are shown on the same
velocity scale to aid comparison
of line widths. To fill the length of simulation A, we show 2 disjoint,
unrelated spectra for simulation A2, separated by the vertical line in
the middle. For simulation A7 we show 32 randomly chosen spectra from
this $32^3$ box.When the end of a spectrum occurs in an absorption line
we see a vertical discontinuity in the plot.
Averaging along the spectra shown, the mean fluxes are $0.887,
0.907,0.900,0.891,0.870$ and 0.821 from the largest to the smallest box.
}
\label{fluxsp}
\end{figure}

In Figure \ref{fluxsp}
we show spectra of the flux along some random,
unrelated sight lines through the A series simulations. We show equal total
velocity length for all simulations, and hence for A7 we show 32 disjoint
spectra, each separated by a vertical dotted line. We should ignore the
vertical discontinuities where spectra end inside an absorption line. These
tend to make lines look narrower, especially in the smaller boxes where
there are more discontinuities.
We see two major trends with box size.

The larger boxes contain large velocity intervals with very little
absorption. These stretch over many hundreds of km/s, longer than the size
of the smaller boxes. The larger the box, the longer the regions with
little absorption. These low absorption regions are clearly showing
correlation in the density
on large scales. We would not expect to see them as often if we truncated
the power to include only short modes.

The smaller boxes seem to have more absorption in total. The values that we
list in the figure caption show this is correct for the A6 and A7, but
the larger simulations have nearly identical mean flux. The smaller
simulations also have a higher proportion of pixels with flux within
a few percent of the continuum, and they have fewer lines of
depth 5 -- 50\%. The number of the deepest lines seems approximately constant.

\subsection{Statistics of Flux Spectra}

In Table \ref{dab}
we list statistics of the flux for the spectra from the
simulations. The statistics are from each pixel in all $N^2$ spectra through
each box in the $z$ direction.  The $\Delta $ values are the mean flux that
we should add onto the current value to
obtain that in the next larger box.
The overall trend of the mean flux $\bar{F}$ with box size
is hard to discern because the changes are small compared to
the measurement errors (discussed below).  The underlying trend is apparently
for larger mean flux in larger boxes, but this seems to reverse for
the largest 4 boxes, where the flux decreases in the larger boxes.

\begin{table}
\caption{\label{dab}Statistics of the flux in all spectra from the
simulations. We
list the mean flux, the error on the mean, the change in the mean to
obtain the flux in the next larger box
and the normalized variance of the transmitted flux.
These statistics all refer to the flux in each pixel in the box.
The column Mode($\bar{F}_L)$ is
different and refers to the mean flux per sight line, ($\bar{F}_L$),
rather than per pixel. The Mode is the most common of the mean flux values
in bins of 0.0005.
}
\begin{tabular}{lclccc}
\hline
Name & $\bar{F}$ & error & $\Delta $ & $Var(F/\bar{F})$ & Mode($\bar{F}_L$)\cr
\hline
A    &  0.8714 & 0.0039 &  .....  & 0.051 & 0.8875 \cr 
A2   &  0.8721 & 0.0059 & -0.0007 & 0.058 & 0.9075 \cr
A3   &  0.8734 & 0.0113 & -0.0013 & 0.060 & 0.9325 \cr
A4   &  0.8741 & 0.0168 & -0.0007 & 0.067 & 0.9375 \cr
A6   &  0.8612 & 0.0332 &  0.0129 & 0.073 & 0.9575 \cr
A7   &  0.8335 & 0.0531 &  0.0277 & 0.095 & 0.9725 \cr
A2kp &  0.8716 & 0.0059 & -0.0002 & 0.058 & 0.9075 \cr 
A3kp &  0.8707 & 0.0118 &  0.0009 & 0.058 & 0.9275 \cr
A4kp &  0.8701 & 0.0181 &  0.0006 & 0.063 & 0.9325 \cr

\end{tabular}
\end{table}

We estimated the errors on the mean flux values in Table \ref{dab}
from the standard deviation of the  mean flux values for
tiles across the $xy$ face of each box. We use tiles of continuous area
because the spectra in adjacent sight lines are highly correlated, and
hence the error on the mean flux value is much larger than the standard
deviation divided by the square root of the number of samples. We use
a 4x4 tiling for A and A2,  3x3 for A3 and A4 and 2x2 for A6 and A7.
These choices are a compromise between having enough tiles to give a small
random error and having tiles large enough  to reduce the inter-tile
correlations.

The error given by tiling captures some of the variation
in the mean density on large scales across the boxes. However it is less
than the external error that we would want to use when we compare to real
spectra because it misses all of the variation that we would see if we
started each box with different random phases, and we allowed each mode
to have a random amplitude, and it misses the variation due to all
modes larger than the box. On the other hand, this error from tiling is larger
than the smallest change that we can consider indicative of a trend when
we compare a series of boxes. This is because the statistics are
evaluated from the whole of each box and all boxes use the same
amplitudes and phases for their modes.

In Table \ref{dab} we also list the variance of the flux $F/\bar{F}$
evaluated for all sight lines through the box, where $\bar{F}$ is the
mean flux in the box, and not that in each sight line. We have
$Var(F/\bar{F})=  N^{-3}\sum( (F / \bar{F}) -1)^2 $,
where the sum is over all $N^3$ pixels from all spectra though one side of
the box. The variance of the flux in each pixel is then
$Var(F) = \bar{F}^{2} \times Var(F/\bar{F})$.
The $Var(F/\bar{F})$ quantity
decreases systematically with increasing box size, because, we will now see,
the fraction of pixels with flux $<0.97$ is up to a factor of two smaller in the
larger boxes.

\begin{figure}
\includegraphics[width=84mm]{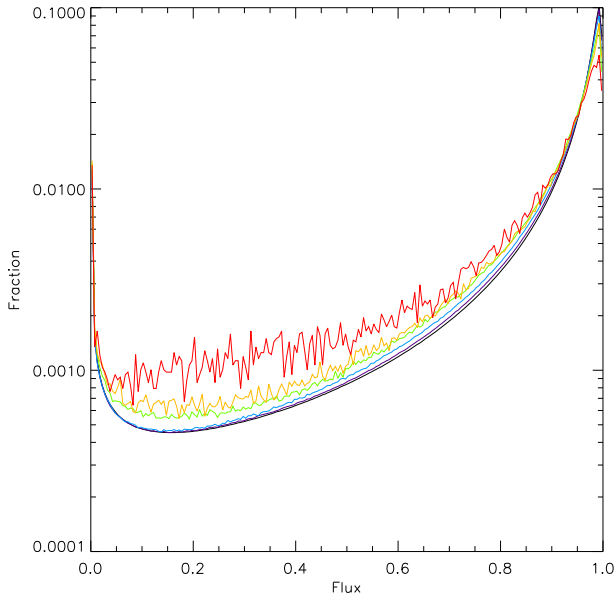}
\caption{
The distribution of the flux per 5~\kms\ pixel in the spectra from the A series
simulations. We use 200 bins for the flux, each of size 0.005.
The distribution from the larger boxes are lower on the plot for most
fluxes ($0.2 <$ flux $< 0.9$).
}
\label{fluxpdf}
\end{figure}

\begin{figure}
\includegraphics[width=84mm]{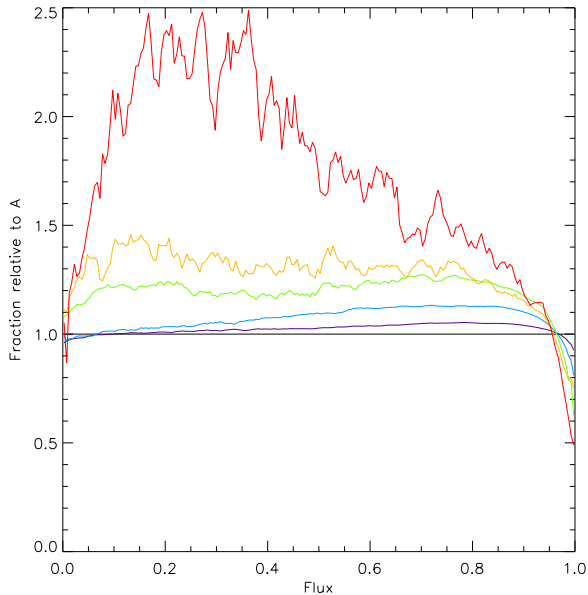}
\caption{
Distribution of the flux per pixel for A series. We show the fraction
of pixels divided by those for simulation A. The results for larger boxes
are nearer to the horizontal line at 1.0 for most flux values.
}
\label{fluxpdfratio}
\end{figure}

In Figure \ref{fluxpdf}
we show the distribution of the flux per pixel for all spectra through
each A series simulation, also called the flux pdf. 
In Figure \ref{fluxpdfratio}
we show the same value, divided by the fractions for simulation A.
The simulations all have approximately the same frequency
of pixels with a flux of 0.96 -- 0.97. The larger simulations have higher
frequencies above 0.97, and lower below. Our impression that the spectra
from the smallest boxes have more absorption is confirmed; they do have
a much larger fraction of pixels with $0.05 < \rm Flux < 0.9$. We
also confirm that the larger simulations have more pixels with very high flux.
We see smaller changes between the larger simulations, indicating convergence
by the size of box A. If this trend continues, then the fractions for
simulation A will be
within approximately 5\% of those for a much larger simulation.

We see that the differences
between the simulations decrease to less than 10\% for fluxes
below 0.02; they all have the same
fraction of their spectra occupied by the bottoms of saturated
absorption lines. This is reasonable because Figs. \ref{pdfdelta}
and \ref{pdfdeltaa} show
they all have approximately the same frequency (per Mpc$^3$) of
high density regions. The largest differences between the boxes
are for intermediate flux levels from the sides of saturated lines,
or the bottoms of nearly saturated lines, both of which are a
small fraction of the pixels.

In Table \ref{dab}
we also list Mode($\bar{F_L}$)
the mode of the mean flux values, $\bar{F}_L$, with one mean per sight line.
The modes are the most common mean fluxes when we use bins of 0.0005.
In contrast with
the mean flux per pixel, these modes per sight line show a systematic
decrease with increasing box size.
These modes are all much less than the mode of the flux in individual pixels,
which is 0.990 to 0.995 for all boxes, as seen in Fig. \ref{fluxpdf}.

\subsection{Statistics of the Lines}

In this section we quantify the types of lines seen in the simulations.
We obtain line statistics by fitting Voigt profiles as described in \citet{zhang97}.
As in \citet{tytler04b} and J05 (\S 5.1, 6.2) we consider only lines with
$12.5 < \lnhi < 14.5$~\cmm . We also limit our discussion to
lines with central optical depths $\tau > 0.05$, which is a new 
constraint for this paper.

\subsubsection{Line Widths: $b$-values}

\begin{figure}
\includegraphics[width=84mm]{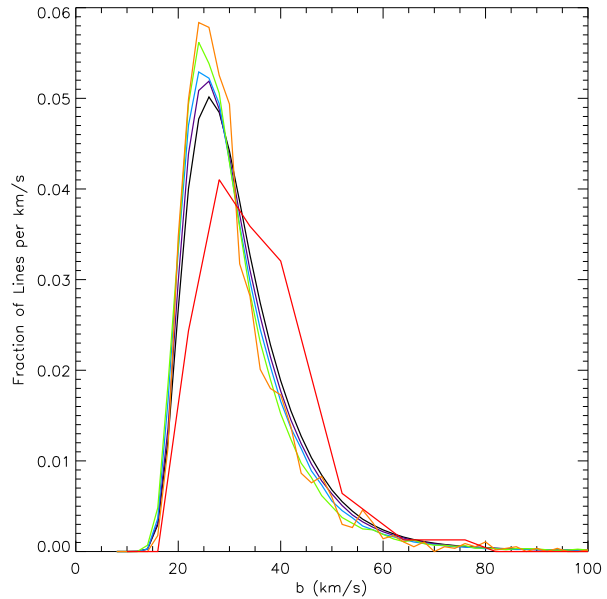}
\caption{
The distribution of the $b$-values for the lines with $12.5 < \lnhi\ < 14.5$
(\cmm ) and $\tau > 0.05$ from the A series simulations. We show the fraction
of all lines that are in 2~\kms\ wide bins (centered at 20, 22, 24 \kms\ etc.
), or 6~\kms\ for A7. The corresponding best estimate $b_{\sigma}$
values are shown in Table \ref{tablebsig}. The results from the larger boxes
are lower at $b = 25$~\kms , with the exception of the lowest curve which is
from the smallest box.
}
\label{bparamA}
\end{figure}

In Figure \ref{bparamA}
we show the distribution of the $b$-values. The
distributions show small but clearly systematic changes with box size.
In detail, the larger boxes have wider lines, fewer narrow lines with
$b < 28$~\kms , and
more lines with $b > 28$~\kms , and a slightly broader distribution.
The smallest box A7 is
an exception to this, presumably because the fundamental mode in this box is
nonlinear at ${z = 2}$.
The change in typical line width can come from a combination of three
factors: larger absorbing regions, giving more Hubble flow across a
line; larger peculiar velocities from the increase in large scale power;
and higher temperatures also from the increase in velocities \citep{theuns99,
bryan99}.

\begin{figure}
\includegraphics[width=84mm]{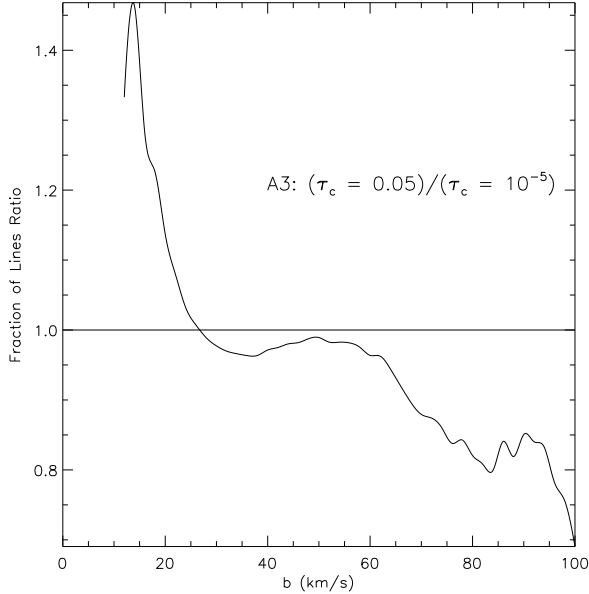}
\caption{
The b-parameter distribution for the A3 box. The vertical axis is
the ratio of the fraction of lines in two samples,  the sample
with $\tau > 0.05$ divided by the fraction for the sample with $\tau > 10^{-5}$.
We find all lines with \lnhi $> 12.5$ (\cmm ) and plot the
ratio of the fraction of lines which have $\tau > 0.05$ with the fraction
of lines which have $\tau > 10^{-5}$ as a function of the $b$-value of the line.
}
\label{fig.bparam-2}
\end{figure}

In Fig. \ref{fig.bparam-2}
we see that the $b$-value distribution is
sensitive to the minimum line central optical depth $\tau $. We use a
sample with $12.5 < $ \lnhi\ $< 14.5$~\cmm\ and $\tau > 0.05$.
If instead we use a sample with $\tau > 10^{-5}$ we see a different
$b-$value distribution that has a larger fraction of lines with
$b < 27$~\kms\ and a smaller fraction of lines with larger $b$-values.
Our sample is the subset of the total which lacks broad shallow lines.
The shape of the distribution in the figure comes from the simultaneous
requirement that \lnhi $> 12.5$ (\cmm ) and $\tau > 0.05$. Narrow lines
with \lnhi $> 12.5$ always have $\tau > 0.05$ and hence they are not
effected  by the 0.05 limit. Very broad lines often have \lnhi $> 12.5$
and $\tau < 0.05$.

In \citet[Fig. 18]{tytler04b} we showed that \citet{hui99c} function gave an excellent fit
to the distribution of $b$-values from a simulation B, which had 35~kpc
cell size, half of that used here.
The Hui-Rutledge function has only one parameter, the \bsig\ that
describes the typical line width.

In Table \ref{tablebsig}
we give estimates for the \bsig\ values that best fit the distributions of the
lines from each simulation.  We use the maximum likelihood method since it
treats the individual $b$-values, and not the binned values that we show in
 the plots. We make two improvement on J05. First, we now fit only lines
with $b < 40$ \kms\ because we want \bsig\ to describe the most common
lines, and not the rare broad features that are hard to see in real 
spectra because of photon noise, flux calibration problems and uncertain continua. 
When we included all $b$-values in J05 the
\bsig\ was larger by about 0.5 \kms . Second, we completely sample the boxes,
where as J05 had few spectra, all of which began at the lowest density part
of that box.  The values that we give here differ from those that we
gave in J05 for the same simulations for these reasons. We estimate the errors
using the tiles, as we did for the for the mean flux, and again the
same comment apply; external errors are larger, and we can be interested
in differences between boxes that are less than the external errors.
We did not estimate the errors for B, B2 and B3 but we give the value of
0.8~\kms\ from A4 because these boxes are all the same size and contain the
same number of \lya\ lines of a given \nhi .

\begin{table}
\caption{\label{tablebsig}
Estimates of the $b_{\sigma}$ parameter. The $\Delta $ values are
the value of \bsig\ in the row above minus the value in the current row,
except for A2kp which we subtract from A.
}
\begin{tabular}{lccc}
\hline
Name & $b_{\sigma}$ & error & $\Delta $ \cr
 & (\kms ) & (\kms ) & (\kms )\cr
\hline
A    &  26.7 & 0.2 & ... \cr 
A2   &  25.6 & 0.4 & 1.1 \cr
A3   &  24.8 & 0.5 & 0.8 \cr
A4   &  24.1 & 0.8 & 0.7 \cr
A6   &  22.4 & 1.2 & 1.7 \cr
A7   &  ...  & ... & ... \cr
A2kp &  26.2 & 0.4 & 0.5 \cr 
A3kp &  26.3 & 0.5 & $-$0.1 \cr
A4kp &  26.3 & 0.7 & 0.0 \cr
B2   &  22.3 & 0.8 & ...\cr
B    &  22.1 & 0.8 & 0.2\cr
A4   &  24.1 & 0.8 & $-$2.0\cr
B3   &  28.3 & 0.8 & $-$4.2\cr
\end{tabular}
\end{table}

The mode of the $b$-value distribution is $b_{peak} =0.9457 $\bsig .
We do not list \bsig\ for A7 because this box is barely large enough to
contain a single complete line, and we do not obtain fits to $b$-values.
The $b$-values show a simple trend: a systematic increase with increasing box
size, consistent with the change in shape of the pdf of the $b$-values. We
discuss the convergence behavior of this statistic in \S 10.

\subsubsection{Column Density Distribution: $f(N)$}
\label{secfn}

\begin{figure}
\includegraphics[width=84mm]{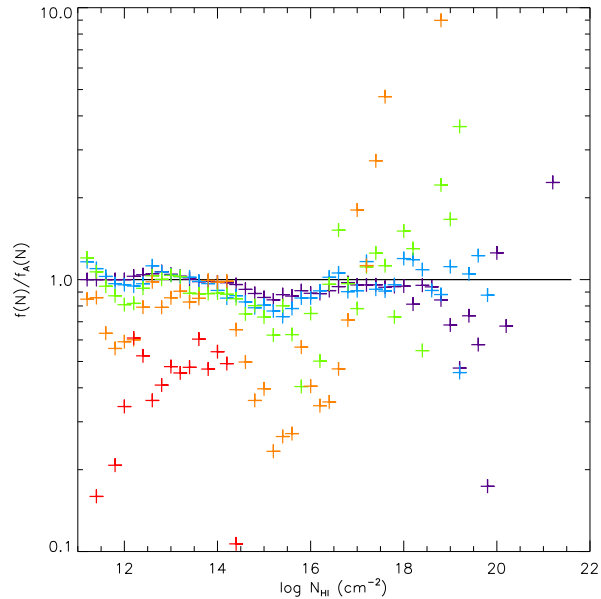}
\caption{The column density distribution f(N) relative to the A box.
}
\label{fig.columns}
\end{figure}

In Figure \ref{fig.columns}
we show the distribution of the H~I
column densities of the lines, relative to the values for box A.
Here the function $f(N)$ is the
differential distribution of lines, per (linear) \cmm , and per unit absorption
distance $X$. The coordinate $X(z)$ is defined such that the density of absorbers
per unit $X$ should be independent of $X$ and $z$. The number density of
non-evolving objects per unit redshift along a line of sight is given by
\citep[Eqn. 3]{tytler81}

\begin{equation}
\label{numdens}
N(z) = N_o Y^2 (H(z)/H_o)^{-1}
\end{equation}

where $Y \equiv (1+z)$ and we define the function

\begin{eqnarray*}
H(z)/H_o&\equiv&E(z) \\
      &=&[\Omega_M Y^3 + (1-\Omega_M-\Omega_\Lambda) Y^2 + \Omega_\Lambda]^{1/2}
\nonumber
\end{eqnarray*}

Setting ${X(z=0) = 0}$, $X$ can then be defined \citep{tytler82} using
${N(X) = N(z)\,dz/dX = constant}$, which gives

\begin{eqnarray}
\label{Xdef}
X(z)&=&\int_0^z Y^2 E^{-1}(z)\,dz
\nonumber \\
    &=&\int_0^z Y[Y(1+z\Omega_m)-z(2+z)\Omega_\Lambda]^{-1/2}\,dz
\end{eqnarray}

Until this decade it was common to use models with
\ol = 0 and $q_0 = 0$ or 1/2. We now
use the cosmological parameters that we gave in \S \ref{secenzo},
\ol = 0.73 and \om = 0.27, which  at $z=2$ give $dX/dz = 3.17801$.
For $q_0 = 0$ the value of $dX/dz = 3$ is similar, but for $q_0 = 1/2$ we have
the significantly different value $dX/dz = 3^{1/2} = 1.73$.
A single sight line through the A box then covers $\delta X = (dX/dz)\delta z =0.163571$
where $\delta z = L_v (1+z)/c = 0.0514693$
and $L_v$ is the velocity span of one sight line, 5143.37 \kms .

Larger boxes have larger  maximum column densities.
The smallest box A7 has no lines with \lnhi $> 15$~\cmm\ while
the larger boxes have a higher density of lines with \lnhi $< 17$~\cmm.
 Indeed the trends are
rather complex with e.g., box A showing a noticeably higher density of
systems with $14.5 <$ \lnhi $~< 16$~\cmm\ than all the other boxes.
We see strong correlations between the $f(N)$ values for similar $N$ values
because adjacent sight lines sample almost the same absorbing gas and hence
almost the same column densities. We also see
large deviations when \nhi\ changes by about a factor of ten.
Together these features  make it hard to assess the errors and rate of
convergence.

\begin{figure}
\includegraphics[width=3.4in]{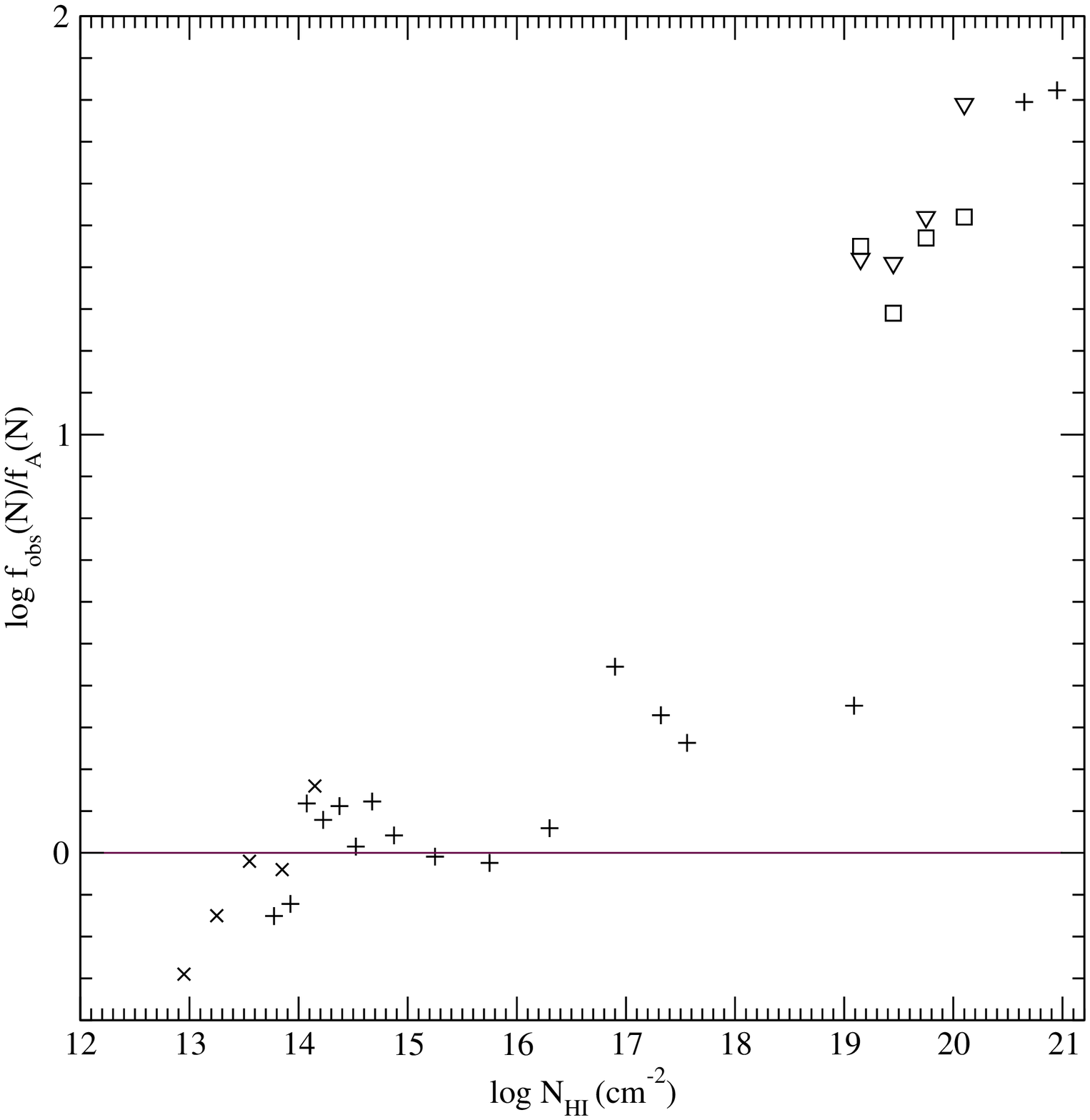} 
\caption{The column density distribution at $z=2$
from observational data relative to the
box A. We use linear interpolation on the $f(N)$ values for box A from Table \ref{tabfNofA} to find values at the \lnhi\ given in data tables.
The observations points are from \citet{kim97} (x), \citet{petitjean93} (+)
and \citet{omeara07a} (squares: MIKE, triangles: UVES).
We make no changes to the $f(N)$ values from \citet[Table 5 \& Fig. 7]{omeara07a} because they
use \ol $=0.7$ while we use 0.73, which is a small difference, and they find no
evidence for rapid evolution. If instead, the number of these lines per unit $z$
were to evolve as rapidly as $N(z) \propto (1+z)^{2.6}$ then $f(N,X) \propto (1+z)^{2.1}$ (\citet{omeara07a} find that the power drops by
approximately 0.5 for $z > 2$ when we change from $z$ to $X$)
and the points drop by 0.23 for MIKE and 0.33 for UVES.
We make two types of corrections to all the other points.
We define $f(N,X)$ using $X(z)$ and $dX/dz$ as in Eqn. \ref{Xdef} for
our assumed cosmology.
Both \citet{kim97} and \citet{petitjean93} use $X_p(z) = 0.5[(1+z)^2-1]$ for
\ol $=0$ and  $q_0 = 0$ giving $dX_p/dz=(1+z)$. We calculate $f(N,X) = f(N,X_{p})(dX_p/dz)/(dX/dz)$,
at the mean redshift of each sample, where $f(N,X_p)$ are the published values.
For \citet{kim97} we take the $f(N,X_p)$ values from Table 2, with a mean
$z=2.31$ from two QSOs, and we multiply these values by a factor $d(X_p)/dX =0.980 $.
For \citet{petitjean93} we take the $f(N,X_p)$ values from Table 2. We assume a mean
$z=2.8$, the mean \zem\ from Table 1, weighted by the number of lines and
evaluated for \lya\ at a rest wavelength of 1120~\AA . We multiply
these $f(N, X_p)$ values by a factor $d(X_p)/dX = 1.037$.
For \lnhi $<17.2$~\cmm\
we make corrections for redshift evolution assuming $f(N)$ per unit $z$ evolves as $N(z) \propto (1+z)^{2.8}$ \citep{kirkman05a,kirkman07a}, and hence $f(N,X) \propto (1+z)^{2.3}$. We then multiply the $f(N,X)$ values from \citet{kim97} by 0.798
and those from \citet{petitjean93} by 0.581.
For $17.2 < $ \lnhi\ $< 19.1$~\cmm\ we assume that the $N(z) \propto (1+z)^{2.8}$
\citep{sargent89a,lanzetta91a,stenglerlarrea95} and a mean redshift of 3.0.
These evolution correction factors could have  large errors.
The point from \citet{petitjean93} at \lnhi 19.09~\cmm\ is for a large range
for $17.68 < $ \lnhi $<20.5$~\cmm . We plot the point at the mean of the  ,
but lower \lnhi\ lines tend to be more common, so the point should be plotted at some
(unknown) lower \nhi value.
}
\label{fig.columnsdata}
\end{figure}

In Figure \ref{fig.columnsdata}
we compare observed values for the column density
distribution to the values from box A. We list values for box A in
Table \ref{tabfNofA}.
We have attempted to correct the data to $z=2$ and our 
 $X$ definition.
 
\begin{table}
\caption{\label{tabfNofA}
The Column Density Distribution for simulation A.
$f(N)$ is lines with line central optical depth $\tau > 10^{-5}$
per \cmm\ per unit X, the absorption distance
from Eqn. \ref{Xdef}. The lines are counted in bins of width
0.2 in \lnhi\ (e.g. 11.4 -- 11.6) , and we report the value
at the listed bin centers (e.g. 11.5).
When we estimate $f(N)$ in bins of width \lnhi $=0.5$ instead,
we find differences
of approximately 0.02 at \lnhi $\simeq 13$~\cmm , and 0.05 at
$19 <$ \lnhi $<20$~\cmm .
}
\begin{tabular}{cccc}
\hline
\lnhi\ & $\log f(N)$ & \lnhi\ & $\log f(N)$\cr
(\cmm ) & (cm$^{2}$ X$^{-1}$) & (\cmm ) & (cm$^{2}$ X$^{-1}$)\cr
\hline
     11.5  & $-$9.80489  &    16.3  & $-$16.9895 \cr
     11.7  & $-$9.91356  &    16.5  & $-$17.2893 \cr
     11.9  & $-$10.0239  &    16.7  & $-$17.5894 \cr
     12.1  & $-$10.1645  &    16.9  & $-$17.9246 \cr
     12.3  & $-$10.3565  &    17.1  & $-$18.2563 \cr
     12.5  & $-$10.5504  &    17.3  & $-$18.5748 \cr
     12.7  & $-$10.8223  &    17.5  & $-$18.9155 \cr
     12.9  & $-$11.1150  &    17.7  & $-$19.1734 \cr
     13.1  & $-$11.4108  &    17.9  & $-$19.4402 \cr
     13.3  & $-$11.7170  &    18.1  & $-$19.7179 \cr
     13.5  & $-$12.0511  &    18.3  & $-$20.0306 \cr
     13.7  & $-$12.3695  &    18.5  & $-$20.3788 \cr
     13.9  & $-$12.7416  &    18.7  & $-$20.7238 \cr
     14.1  & $-$13.1143  &    18.9  & $-$21.1386 \cr
     14.3  & $-$13.4898  &    19.1  & $-$21.5316 \cr
     14.5  & $-$13.9236  &    19.3  & $-$21.8351 \cr
     14.7  & $-$14.2540  &    19.5  & $-$22.1258 \cr
     14.9  & $-$14.6523  &    19.7  & $-$22.4385 \cr
     15.1  & $-$15.0434  &    19.9  & $-$22.7923 \cr
     15.3  & $-$15.4000  &    20.1  & $-$23.1537 \cr
     15.5  & $-$15.7948  &    20.3  & $-$23.5523 \cr
     15.7  & $-$16.0774  &    20.5  & $-$24.0321 \cr
     15.9  & $-$16.3919  &    20.7  & $-$24.3957 \cr
     16.1  & $-$16.7003  &    20.9  & $-$24.8255 \cr

\end{tabular}
\end{table}

Numerical simulations have often found too few systems with high
\nhi\ values
\citep{katz96a, gnedin98b, gardner01a,dodorico06a}.
Simulations need both sufficient volume
to contain the long wavelengths modes to get enough CDM halos
\citep{bagla05a}, and they need high enough resolution to make the clumps of
gas that cause Lyman limit systems (LLS). In recent work,
\citet{kohler07a} are able to reproduce
both the mean flux in the forest and the column density distribution of
LLS at  $z=4$ in $4h^{-1}$~Mpc boxes with $2h^{-1}$~kpc resolution. They
also obtained approximately the real number of LLS per unit $z$.

Our simulations lack absorption systems with large column densities. 
We lack a factor of 1.2 for $14 <$ \lnhi\ $ < 15$~\cmm .
We see an increasing lack at \lnhi $>17$~\cmm , reaching a factor of
approximately 30 by \lnhi $=19$~\cmm ,
and a factor of 70 for Damped \lya\ lines (DLAs) with \lnhi $\simeq 21$~\cmm .
The point shown as a plus from \citet{petitjean93} is unreliable
at \lnhi $\simeq 19$~\cmm .
We are also uncertain about the errors in the $f(N)$ measurements,
especially since we have not checked that the $f(N)$ values are consistent
with the total mean absorption that we assume at $z=2$. Our simulation A
has approximately the correct total absorption, and hence the excess lines with
\lnhi $<14$~\cmm\ should make approximately the same total absorption as the
lack with \lnhi $>14$~\cmm .

The lack at the higher columns, \lnhi $> 17$~\cmm\ is due to insufficient
numerical resolution in collapsed halos which give rise to this absorption
and the lack of self-shielding against the UV background radiation.
The lack is large enough that we must clearly remove high column lines
from both simulations and real spectra when we want to make quantitative
comparisons, as first pointed out by \citet{tytler04b} and \citet{jena05a}.

\section{1D Power Spectra of the Flux}

Following \citet{croft97b}
and others, we define the flux contrast as
\begin{equation}
f(u) = (F/\bar{F}) -1,
\label{fluxc}
\end{equation}
where $u$ is velocity and
$\bar{F}$ is the mean of the flux from all spectra through a box,
listed in Table \ref{dab}.
While our signal looks like that used by
\citet{croft98} and \citet{mcdonald00a}, there is an important difference.
They both take $\bar{F}$ to be the mean flux in each spectrum.
We discuss this alternative choice below in \S \ref{sect.signal}.

Since the mode of the flux in each pixel is larger
than the mean flux, $f(u)$ is typically larger than zero.
This definition resembles that of \dc -1, except that \dc\
involves division by the mean density of CDM that is
a parameter input into the simulations and identical for all.
In contrast, the mean flux is not known until spectra are made, and it
varies from simulation to simulation and with $z$.

We compute the Fourier transform of the flux contrast using
$f(u)$ in place of \dc -1 in Eqn. \ref{eqnft}.
We measure the (one-dimensional) flux power of each sight line in the
$z$ direction, and we present the average of the power from all sight lines.
We have explicitly checked that we obtain the same power as does McDonald
from the same spectrum.

\begin{figure}
\includegraphics[width=84mm]{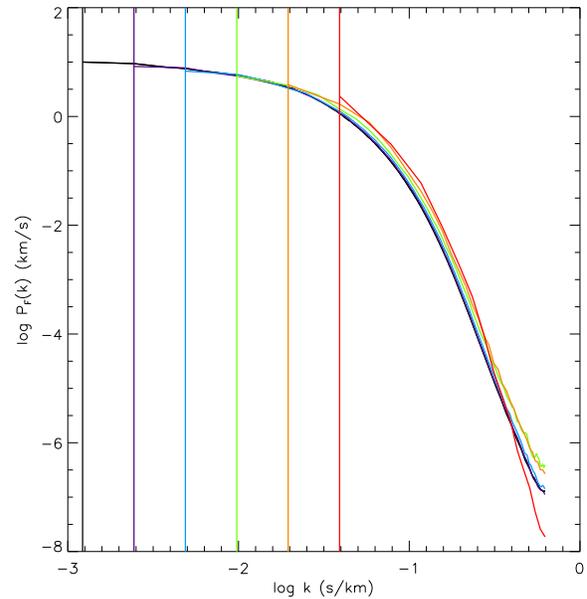}
\caption{The power spectrum of the 1D flux contrast (Eqn. \ref{fluxc}, flux
divided by mean flux in that simulation box)
from A-series boxes for all $N^{2}$ sight lines along the z-direction.
The $k$ is in s/km and the smallest boxes gives the smallest amount of power
at log $k = -1$.
}
\label{fpow}
\end{figure}

\begin{figure}
\includegraphics[width=84mm]{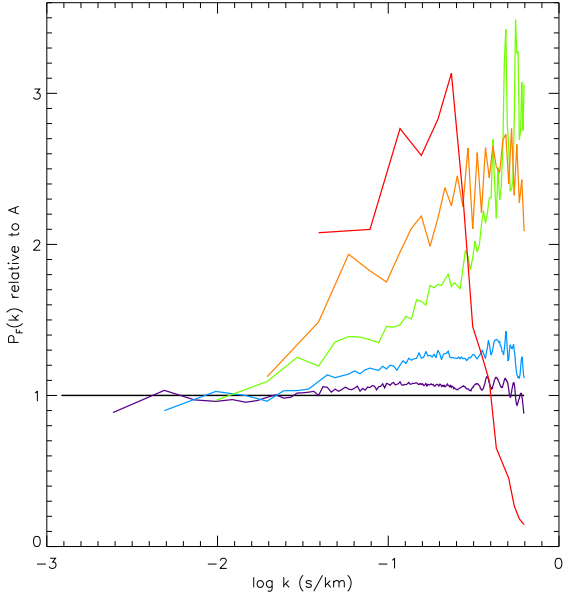}
\caption{
The power spectrum as in Fig. \ref{fpow} but now divided by the power
from simulation A, which is the line at vertical value 1.
}
\label{fpowa}
\end{figure}

In Figure \ref{fpow}
we show the power spectrum of the flux, $P_F$, in this case from
all spectra parallel to the $z$ axis in each simulation. We tabulate values in
Tables \ref{fluxpower}
and
\ref{fluxpowerA},
and in Figure \ref{fpowa}
we show the power divided by
that in simulation A.  In contrast to the CDM power, the differences between
simulations are small, and of the opposite sign. In general, the larger boxes
have smaller flux power, the opposite of the trend that we saw in Fig.  \ref{fig.pdmspc}
for the power of the CDM.
On the largest scale sampled by a box, the power at log $k < -2$ s/km in the
A2 and A3 boxes is slightly less than that in the largest A box.
This might be simply the effect of the larger modes in the larger boxes.
On intermediate scales $-1.5 < \rm{log}~ k  < -0.6$ there is a systematic decrease in
power with increasing box size, with larger changes on smaller scales (larger
$k$). However, on the smallest scales, log $k > -0.5$ s/km,
corresponding to sine wavelengths $\lambda < 20 $~\kms\ (4 cells in the
simulation), the trends change direction. The differences between the
simulations become less, and A2 returns to approximately the same power as A.
The largest deviations in the ratios of the power of the flux are seen
at log $k > -0.7$ (s/km), corresponding to changes in the shapes of the
narrow lines.

\begin{table*}
\caption{\label{fluxpower} The log of the power of the flux in the A series
simulations. The subscripts are values of log (k) (s/km), and the R are the
ratios of the power (not the log) to the values in A at the same $k$.
}
\begin{tabular}{lllllllll}
\hline
Box & log $P_{-2.5}$ & log $P_{-2}$  &  log $P_{-1.5}$  &
log $P_{-1}$ & $R_{-2.5}$ & $R_{-2}$ & $R_{-1.5}$ & $R_{-1}$ \cr
\hline
A    &  0.933 &  0.757 &  0.239 &  -1.310 & 1.00 &   1.00 &   1.00 &  1.00  \cr
A2   &  0.910 &  0.739 &  0.246 &  -1.279 & 0.95 &   0.96 &   1.02 &  1.08  \cr
A3   &  ..... &  0.767 &  0.253 &  -1.233 & .....&   1.02 &   1.03 &  1.20  \cr
A4   &  ..... &  0.742 &  0.325 &  -1.144 & .....&   0.97 &   1.22 &  1.47  \cr
A6   &  ..... &  ..... &  0.336 &  -1.060 & .....&   .... &   1.25 &  1.78  \cr
A7   &  ..... &  ..... &  ..... &  -0.945 & .....&   .... &   .... &  2.32  \cr
A2KP &  0.919 &  0.747 &  0.244 &  -1.325 & 0.97 &   0.98 &   1.01 &  0.97  \cr
A3KP &  ..... &  0.784 &  0.255 &  -1.314 & .... &   1.07 &   1.04 &  0.99  \cr
A4KP &  ..... &  0.778 &  0.342 &  -1.238 & .... &   1.05 &   1.27 &  1.18  \cr

\end{tabular}
\end{table*}

\begin{table}
\caption{\label{fluxpowerA} The 1D power spectrum of the flux in simulation A.
We include all wavenumbers below $k \le 0.01$ (s/km). We then pick eight
points from each log $k$ decade and we force the last point to be the
Nyquist frequency. We computed the error on the power from 4x4
tiles across the face of the box. The error is the
standard error on the mean power from all the tiles. The $P_e$ and its error
refer to the evolving spectra that we discuss in \S \ref{secevol}. We show
powers of ten in parenthesis: $3.522(-8) = 3.522 \times 10^{-8}$.
}
\begin{tabular}{lllllllll}
\hline
 log $k $ & $P(k)$ & $P_e(k)$ (km/s) & error & error$_e$ \cr
\hline
     -2.9134 &  10.016 &  10.021   & 0.441 & 0.441 \cr
     -2.6124 &   9.371 &   9.377   & 0.409 & 0.411 \cr
     -2.4363 &   8.134 &   8.135   & 0.381 & 0.382 \cr
     -2.3114 &   7.576 &   7.576   & 0.363 & 0.364 \cr
     -2.2144 &   6.865 &   6.872   & 0.311 & 0.311 \cr
     -2.1353 &   6.446 &   6.447   & 0.283 & 0.284 \cr
     -2.0683 &   6.053 &   6.057   & 0.270 & 0.270 \cr
     -2.0103 &   5.790 &   5.794   & 0.250 & 0.251 \cr
     -1.9592 &   5.407 &   5.412   & 0.236 & 0.237 \cr
     -1.6830 &   3.264 &   3.260   & 0.124 & 0.123 \cr
     -1.5155 &   1.855 &   1.849   & 0.651(-1) & 0.648(-1) \cr
     -1.3949 &   1.068 &   1.068   &  0.315(-1) & 0.314(-1) \cr
     -1.3006 &  6.243(-1)  &  6.256(-1) & 0.170(-1) & 0.171(-1) \cr
     -1.2232 &  3.748(-1)  &  3.749(-1) & 0.106(-1) & 0.107(-1) \cr
     -1.1575 &  2.269(-1)  &  2.273(-1) & 0.600(-2) & 0.620(-2) \cr
     -1.1005 &  1.376(-1)  &  1.376(-1) & 0.375(-2) & 0.380(-2) \cr
     -1.0501 &  8.435(-2)  &  8.398(-2) & 0.226(-2) & 0.229(-2) \cr
     -1.0049 &  5.185(-2)  &  5.171(-2) & 0.149(-2) & 0.149(-2) \cr
     -0.8565 &  7.716(-3)  &  7.725(-3) & 0.266(-3) & 0.272(-3) \cr
     -0.6630 &  2.878(-4)  &  2.904(-4) & 0.137(-4) & 0.140(-4) \cr
     -0.5296 &  2.210(-5)  &  2.215(-5) & 0.121(-5) & 0.127(-5) \cr
     -0.4277 &  3.272(-6)  &  3.633(-6) & 0.196(-6) & 0.269(-6) \cr
     -0.3452 &  7.803(-7)  &  9.018(-7) & 0.451(-7) & 0.954(-7) \cr
     -0.2759 &  2.357(-7)  &  3.350(-7) & 1.341(-8) & 5.048(-8) \cr
     -0.2162 &  1.296(-7)  &  2.193(-7) & 7.166(-9) & 3.727(-8) \cr
     -0.2041 &  1.286(-7)  &  2.073(-7) & 7.110(-9) &  3.522(-8) \cr

\end{tabular}
\end{table}

The large changes in the power of the CDM with box size, contrasted with
the small changes in the power of the flux implies that the bias will change
rapidly with box size, approximately as does the CDM.
Attention must be given to the appropriate smoothing of the fields to
reduce the sensitivity of the CDM power to the box size \citep{mcdonald02}.

\citet{mcdonald03} shows how the power of the flux changes when he increases
 his box size from 28.2 to 56.3 to 112.7~Mpc, while keeping the cell size constant at
 220~kpc. He also sees that the power is lower in the smaller boxes

We see that the change in power with box-size is consistent with the
simultaneous change in the $b$-value distribution. \citet{viel03c}
showed quantitatively how power of the flux on small scales
responds to changes in line $b$-values.
For $-2 <$ log $ k < -0.2$ \citet{viel03c} saw essentially the same
power from randomly placed \lya\ lines as from real spectra.
They also showed that making all $b$-values larger by a factor of two decreased
the power at $-1.5 < $ log $k < -0.7 $. Our larger boxes have larger $b$-values
and they show decreased power on these scales.
If all other factors are unchanged, we would expect higher
temperatures for the IGM in larger boxes, which we confirm in \S 6.2.

\subsection{Autocorrelation of the Flux}

Although containing exactly the same information as the power, the
autocorrelation can better illustrate characteristics of the signal that are
hard to see in the power, such as correlations over large scales. While a
power spectrum is a complete statistical description of a
random Gaussian field, the flux distribution is far from Gaussian, hence
neither the power spectrum nor the autocorrelation will be a complete
description.

The autocorrelation of the flux for a given velocity shift $\delta u$ can
be computed directly from a transmission spectrum as
$\xi_F(\delta u) = <(F(u)-\bar{F}) (F(u+\delta u)-\bar{F}) >$, where
$\bar{F}$ is the mean flux in a box from Table \ref{dab}. The
brackets refer to an average across the pixels of the spectrum.
We choose to
obtain the autocorrelation from
\begin{equation}
\xi_F (\delta u) = { \bar{F}^2 \over 2\pi } \int  P(k)e^{jk \delta u}dk
\end{equation}
where $P(k)$ is the power of the flux contrast
and we can use the line of sight averaged power directly because we have
divided by the mean flux of each cube.
The results we report here, as with the flux power, are the average
autocorrelation profiles at each velocity shift from all lines of sight.

For the power spectrum, the longest non-zero mode is one wave in the box.
By analogy, for a sight line parallel to the box edges,
the largest lag in a box is $u = L_v/2$.
 We expect the autocorrelation to decrease up to scales of $L_v/2$ and then
to rise again, since a lag of $u$ in one direction is simultaneously
a lag of $L_v-u$ in the reverse direction. This is in contrast with real
spectra where the autocorrelation will continue to decrease with increasing lag.
The number of samples of each lag drops from $N$ for lags of one pixel to
N/2 for lags of L/2. When we shift a spectrum by some lag, we loop around the
periodic boundary conditions, making the first and last pixels adjacent, so that
all shifted spectra have the same length.

\begin{figure}
\includegraphics[width=84mm]{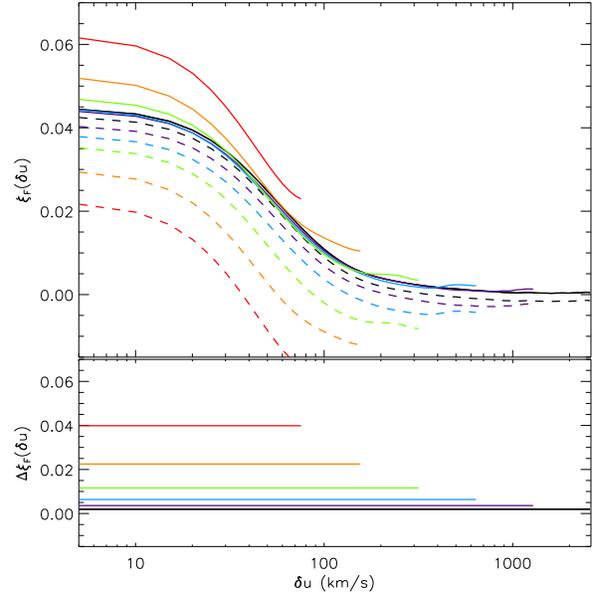}
\caption{
How the autocorrelation of the flux depends on box box size and
the definition of the signal. Autocorrelation vertically again velocity lag horizontally.
The solid curves were calculated
using the mean flux of the box.
The dotted curves use instead the mean flux of each sight line,
$\bar{F}_L$. We show the difference of the two, from Eqn. \ref{eqndelauto},
in the bottom panel. The signals from the larger boxes extend farthest to the right.
}
\label{autodef}
\end{figure}

In Figure \ref{autodef}
we show the autocorrelation of the flux calculated using the mean of the flux
from all spectra through each box.
The autocorrelation of the flux falls with increasing lags, and it falls
to lower values in larger boxes. The larger boxes also have smaller
correlation for most velocity lags.
However, the autocorrelation does not drop to zero on
the largest scales in a box, rather it stops falling at some low plateau
value.

\begin{figure}
\includegraphics[width=84mm]{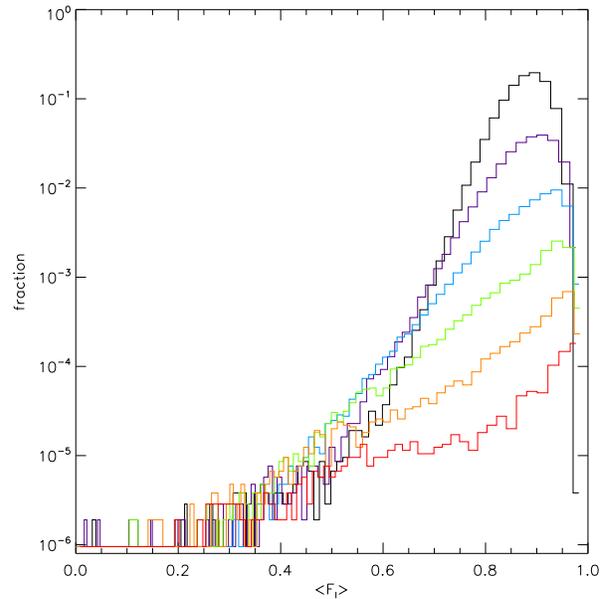}
\caption{
The distribution of the mean flux per sight line, $\bar{F}_L$, for all
simulations in the A series. The vertical scale shows the number of sight
lines, divided by the number of sight lines in the A box. This division lowers
the distributions from the smaller boxes (lower histograms at $F_L = 0.8$)
making it easier to see each distribution.
}
\label{fmapn}
\end{figure}

If we define the autocorrelation using the mean flux
in each sight line $\bar{F}_L$, instead of the mean of the box, then the
autocorrelations are all reduced.
We show these autocorrelation functions as dashed lines in Fig.  \ref{autodef}.
The amount by which the two autocorrelation calculations differ is equal to
the variance of the line of sight mean flux values about their mean, which is
the mean flux of each cube that we give in Table \ref{dab},
\begin{equation}
\Delta \xi _{F}(\delta u) = \frac{1}{N^{2}}
\sum_{L=1}^{N^{2}}(\bar{F}-\bar{F}_{L})^{2} \equiv Var(\bar{F}_{L}),
\label{eqndelauto}
\end{equation}
where $N^2$ is the number of sight lines we use to sample each $N^3$ box.
The smaller boxes have larger sight-line to sight-line variance, as can be seen
in Fig. \ref{fmapn} and hence their autocorrelation (and other statistics) is
decreased the most then we use the mean flux per sight line.

\section{Velocity Field, Baryon Temperature and Baryon Density}
\label{tempdens}

Having seen how the statistical properties of the \lyaf\ depend on box size we now
return to the examine the changes in the velocity field, and the
baryon temperature and baryon density in the
simulations, since these fields control how the \lyaf\ changes.

\subsection{How gas velocity changes with box size}

\begin{figure}
   \includegraphics[width=84mm]{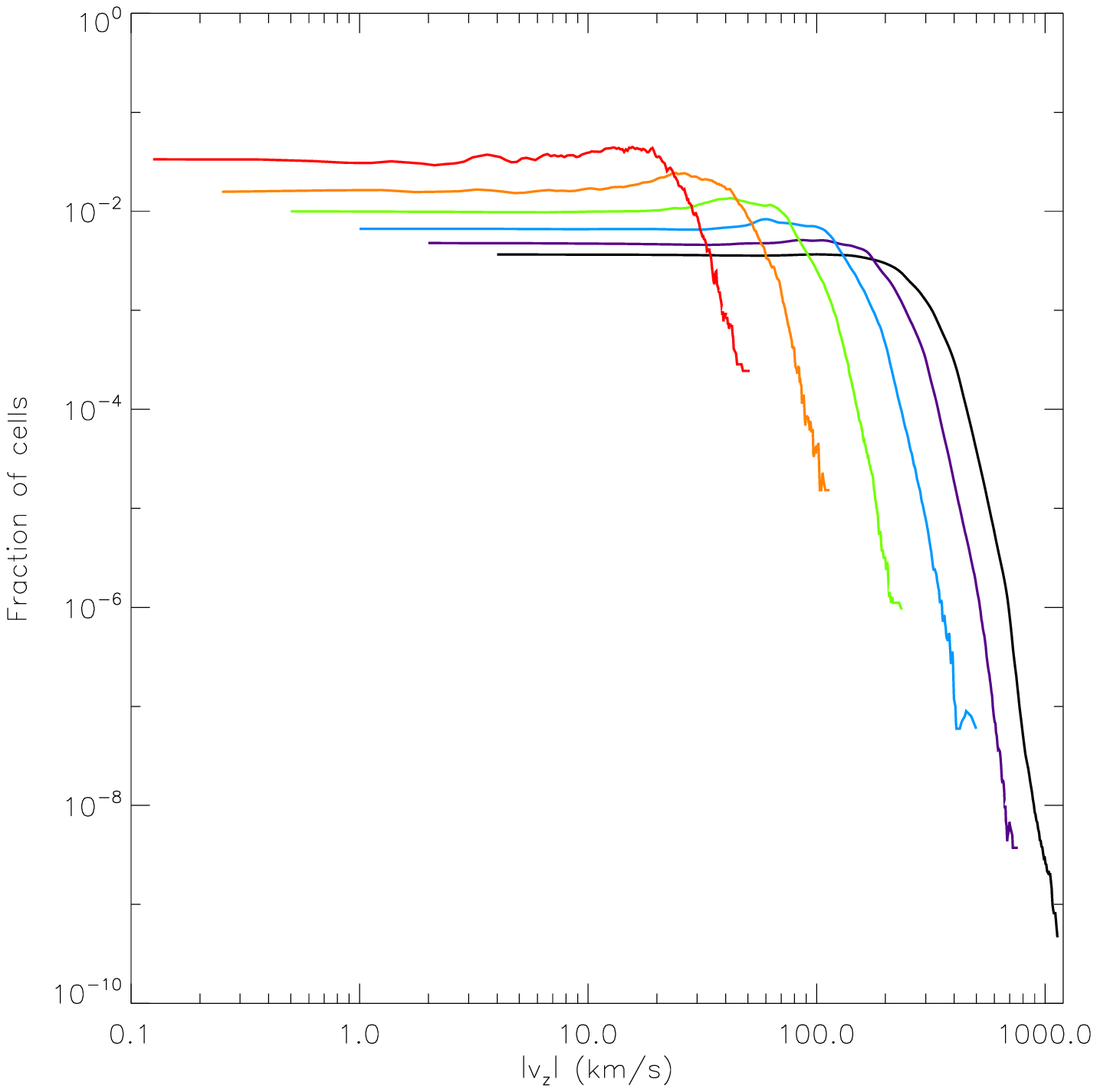}
   \caption{\label{velbox}
The change in the velocity field with box size. The fraction of cells, on a log scale,
as a function of the modulus of the gas velocity in the $z$ direction. We sample
the fractions in steps of 8~\kms\ for A
4~\kms\ for A2, 2~\kms\ for A3, 1~\kms\ for A4, 0.5~\kms\ A6 and 0.25~\kms\ for A7,
and we report the fraction of cells per 1~\kms , for all simulations, which accounts
for why the total area under the curves falls with increasing box size.
   }
\end{figure}

In Fig. \ref{velbox}
we see that the velocity of the cells increases dramatically with
increasing box size. Velocities $> 160$~\kms\ are common in the three
largest boxes but non-existent
in the two smallest boxes.  In Table \ref{tabtemp}
we list the minimum, median and maximum baryon velocity in each box.
These velocities increase by factors of 1.23 -- 2.59 as we double the box size,
with the largest factors applying to the maximum velocity for the smallest pair
of boxes. The maximum changes the most because this is sensitive to the rare
high density regions. However, all cells show a systematic increase in velocity,
as illustrated by the factor of 1.34 increase in the median velocity
going from the A2 to A box.

\begin{table}
\caption{\label{tabtemp} Statistics of the proper velocity (km/s) for the baryons
in the cells in the A series simulations.
}
\begin{tabular}{lrrr}
\hline
Box & mean & median  & max  \cr
\hline
A  &   136.4 &    138.2 &  1153.9\cr
A2 &   110.8 &    103.0 &   764.3\cr
A3 &   74.9  &    70.9  &   515.1\cr
A4 &   47.2  &    45.0  &   277.2\cr
A6 &   27.8  &    26.7  &   153.3\cr
A7 &   14.1  &    13.8  &    59.2\cr
\end{tabular}
\end{table}

\subsection{How gas temperature and density change with box size }

\begin{figure}
   \includegraphics[width=94mm,height=94mm,angle=-90]{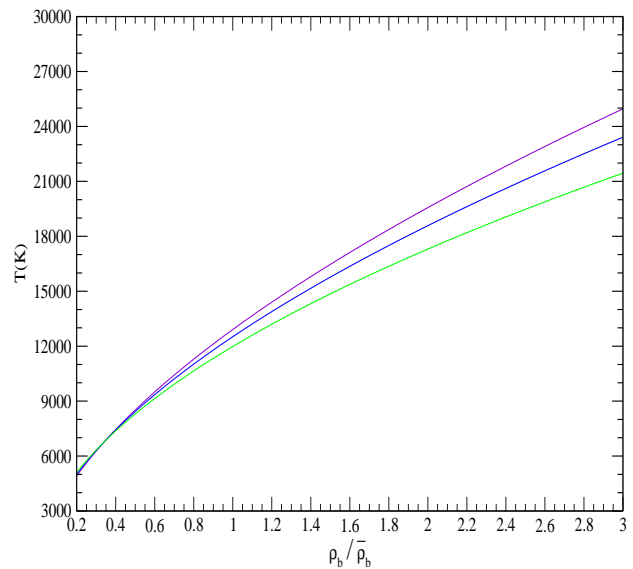} 
   \caption{\label{trhofits}
Power law fits to the most common temperature at a given baryon overdensity,
from J05 Table 9. The fits shown,
from top to bottom on the right, are for the A2, A3 and A4 simulations.
   }
\end{figure}

In \citet[Fig. 19]{tytler04b} we showed the temperature of cells in simulation
B as a function of their baryon density. Simulation B has the same parameters
as the A series used here, but with cells that are half the size.
We fit a broken power law to
the ridge line that specifies the most common T at a given density, but noted
that these fits were not very satisfactory in shape.
In Table 9 of J05 we fit single power laws
$T(\rho ) = T_0(\rho _b / \bar{\rho _b} )^{\alpha }$
to $0.2 < \rho _b / \bar{ \rho _b}  < 3$ in many simulations,
where $ \bar { \rho _b} $ is the cosmological density of baryons.
We found values of $T_0 =$ 11,982, 12,561 and 12,910~K for A4, A3 and A2,
showing an increase in temperature at a given density with box size.
We also saw a systematic increase in the index $\alpha $ with box size
which corresponds to a larger difference in temperature at higher
densities, and near identical temperatures at $\rho _b/ \bar{ \rho _b} = 0.35$.
In Figure \ref{trhofits}
we show these fits for simulations A4, A3 and A2.

\begin{figure}
\includegraphics[height=84mm,width=84mm]{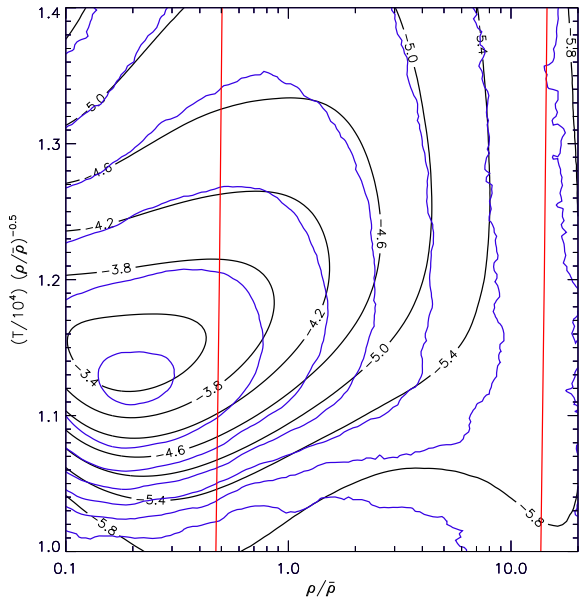}
\caption{\label{tdenscat}
Contour plot of the temperature of cells against their baryon overdensity.
Blue and black contours correspond to A4 and A boxes respectively.
We generate the contours from the scatter plot between
the grid values of $T/(\rho_{b}/\bar{\rho_{b}})^{0.5}$ and
$\rho_{b}/\bar{\rho_{b}}$
by binning the x-axis and y-axis in logarithmic intervals
of 0.001 and 0.02 respectively and computing the number of points in each
2D mesh. The iso-level values show the fraction of points (in logarithmic
units) contained within each contour relative to the total number
of points in the simulation contained within the boundaries of each axis.
The nearly vertical line on the left shows the typical
cells responsible for \lya\ lines with \lnhi $=12.5$~\cmm\ while the line
on the right is for 14.5~\cmm .
 }
\end{figure}

Fig. \ref{tdenscat}
is a contour plot of the temperature T of cells as a function of baryon overdensity
 for simulations A and A4.
The vertical axis is $T/(\rho _b / \bar{ \rho _b})^{0.5}$ to remove much of the tendency for
T to increase of with density.
We are most interested in the densities that produces the \lyaf .
We found in \S \ref{sec.cdmden} that \citet[Eqn. 10]{schaye01a} suggests that
the \lyaf\ lines with \lnhi = 12.5 -- 14.5~\cmm\ that we use to study the
$b$-value distribution typically come from baryon overdensities 0.5 -- 15.7.
On Fig. \ref{tdenscat} we show lines of constant \lnhi ,
\begin{equation}
T=10^4 K \left( {8.82 \times 10^{12} \over \nhi\ ~\cmm }\right)^{3.846}
\left({ \rho \over \bar{\rho }} \right)^{5.769},
\end{equation}
obtained from \citet[Eqn. 10]{schaye01a}.
We see, with close inspection, that, at overdensities above $\simeq 0.5$
the contours of the larger box have on average shifted to higher temperatures,
particularly in the regions closer to the frequency peak of this 2D distribution.
The gas that makes the \lyaf\ absorption is clearly hotter in the larger box.

\begin{figure}
  \includegraphics[width=84mm]{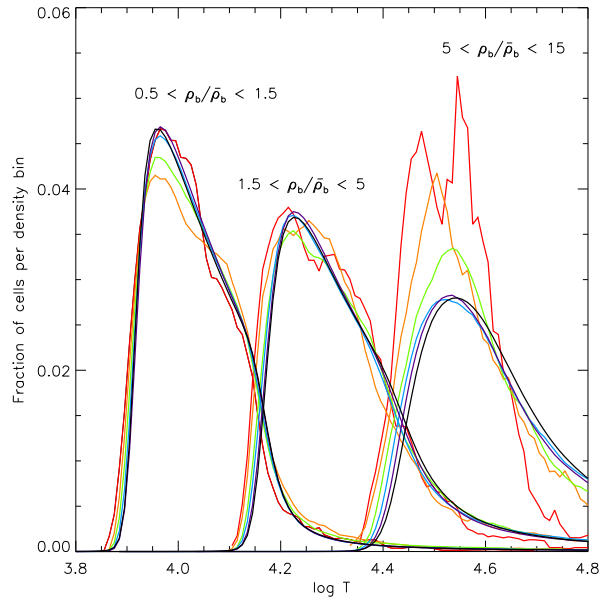}
\caption{\label{temppdf3}
The pdf of the temperature (in K) of cells for three contiguous ranges of baryon
overdensity, the ranges that are responsible for \lya\ lines with
\lnhi $>12.5$~\cmm\ (the histograms on the left for the lower overdensities) to $<14.5$~\cmm\ (histograms on the right for the higher overdensities).
We show the fraction of all cells in the density range, and we sample the
temperature in steps of $\log T = 0.01$.
 }
\end{figure}

In Fig. \ref{temppdf3}
we show the pdf of the temperature per cell for three
different baryon overdensity ranges, 0.5 -- 1.5, 1.5 -- 5 and 5 -- 15. For each density
range, we show six distributions, one for each box size. The temperature
pdf shows very little change with box size for low overdensities
0.5 -- 1.5, but the  intermediate and especially
the higher densities the larger boxes have systematically higher temperatures.
This tendency of increasing temperature with box size at higher
but not lower overdensities confirms
the power law fits to the most common temperatures
 from J05 that we showed in Fig. \ref{trhofits}.

\begin{figure}
   \includegraphics[width=84mm]{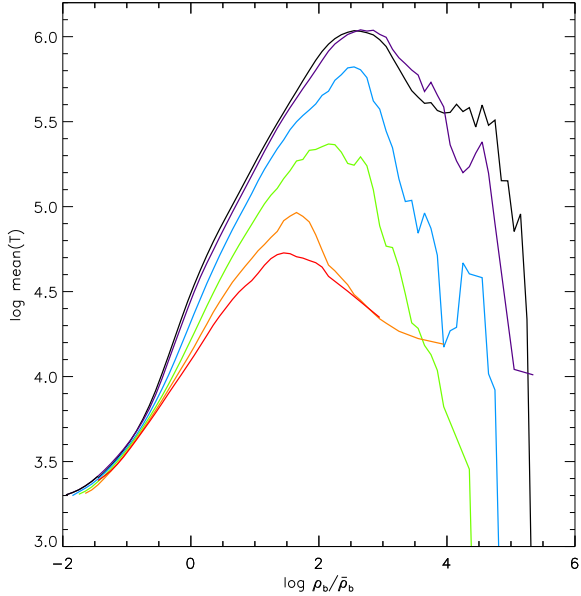}
   \caption{\label{mntempvsdenbox}
The mean temperature (K) of all cells as a function of log baryon overdensity, evaluated in
log overdensity intervals of 0.1. At $-0.5 <$ log $\rho _b / \bar{\rho }_b < 2.5$ the
order of the curves is that of increasing box size towards the top.
   }
\end{figure}

In Fig. \ref{mntempvsdenbox}
we show the mean temperature of cells as a function of baryon overdensity
for the A series simulations. We see a dramatic increase in the
mean temperature in the larger boxes especially
at higher overdensities. The percentage increase in the mean temperature
with box size decreases at lower overdensities, as we just saw for the much more
restricted range of densities in \ref{temppdf3}.

\begin{figure}
   \includegraphics[width=84mm]{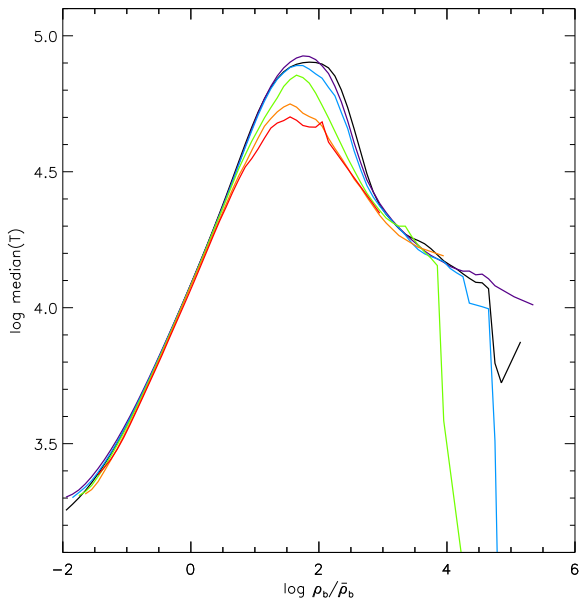}
   \caption{\label{mdtempvsdenbox}
The median temperature (50\% are hotter) of all cells as a function of log baryon
overdensity, sampled in bins of log overdensity = 0.1. The order of the curves is
that of increasing box size to the top at log $\rho _b / \bar{\rho }_b = 2.2$.
   }
\end{figure}

Fig. \ref{mdtempvsdenbox}
is like Fig. \ref{mntempvsdenbox} but now showing the temperature which is exceeded by
50\% of cells, the median.
We again see that the larger boxes are hotter but the differences are much smaller, especially
at the low densities of the \lyaf . The much larger increase in the mean temperature comes from
relatively few cells that have undergone shock heating to temperatures much larger than the median
and well above the temperature at which there is sufficient H~I to make \lya\ lines.

\begin{figure}
   \includegraphics[width=84mm]{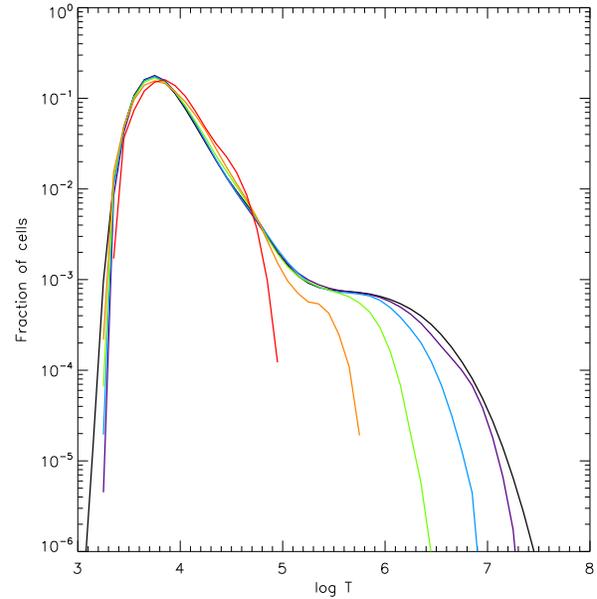}
   \caption{\label{gastemppdfbox}
The fraction of cells as a function of baryon temperature (K), sampled
in bins of $\log T = 0.1$.
The larger boxes extend farther to the right.
   }
\end{figure}

In Fig. \ref{gastemppdfbox}
we see that relatively few cells are attaining much higher temperatures in the 
larger boxes. The temperatures above $10^5$~K come from shocks which are rarer and weaker
in the
smaller boxes because the velocities and peak densities are lower.

\begin{figure}
   \includegraphics[width=84mm]{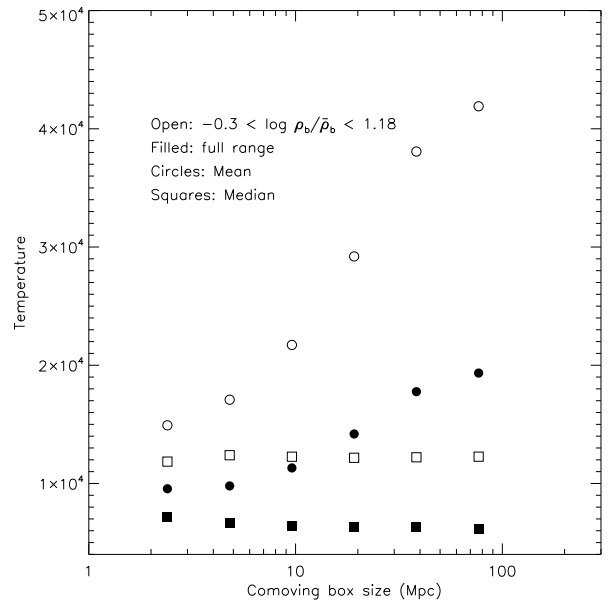}
   \caption{\label{glbtempbox}
The mean (circles) and median (squares) baryon temperature (K) as a function of
box size. We evaluate these statistics in bins of 0.1 in log T,
for both the entire box, and for the densities that produce the \lyaf .
   }
\end{figure}

Fig. \ref{glbtempbox}
summaries the changes in temperature with box size, and shows that the trends
of relevance to the \lyaf\ are only revealed by specific statistical measures.

\begin{figure}
   \includegraphics[width=84mm]{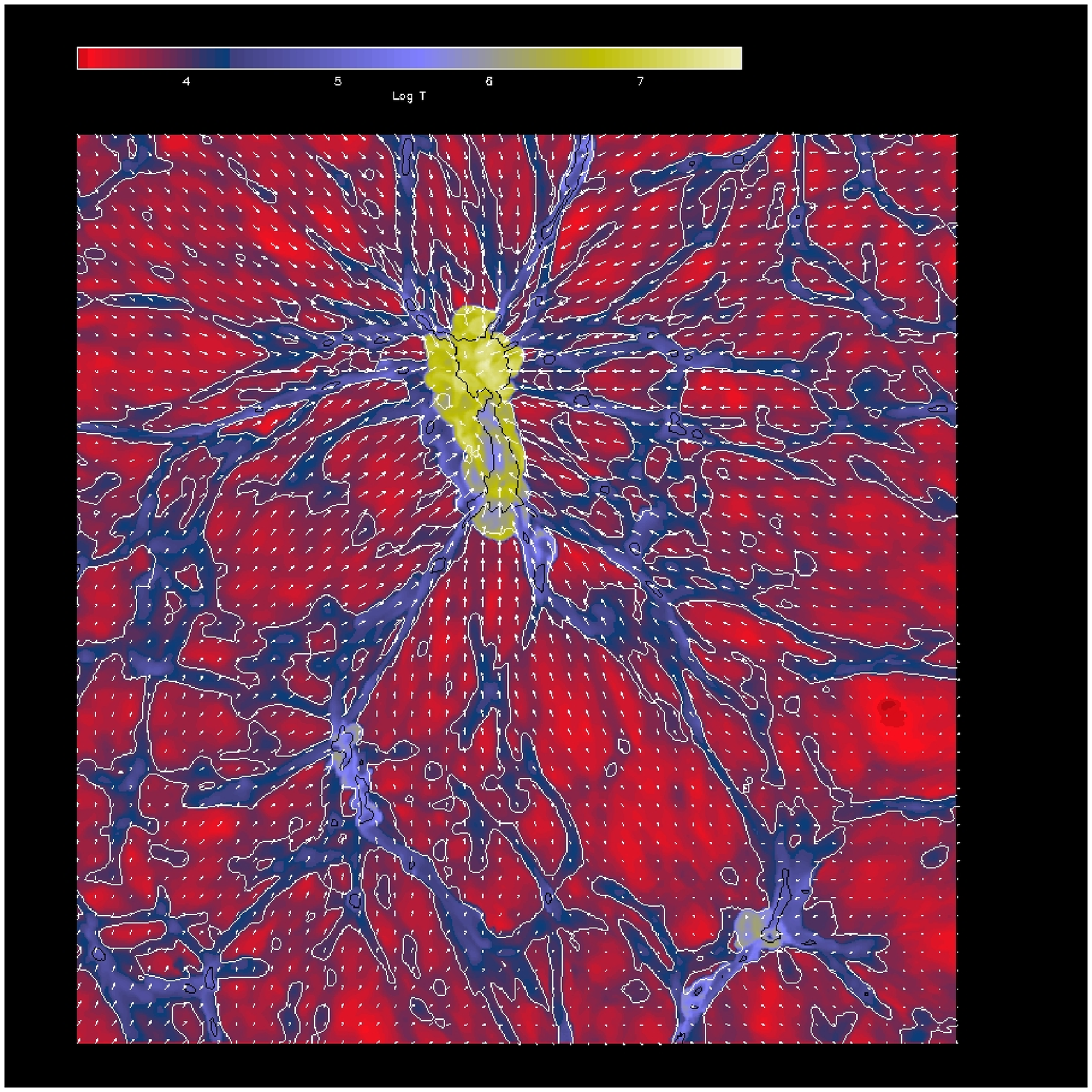}
   \caption{\label{dentempimage}
A 30x30 Mpc section of the A box, one cell thick and containing the highest
density cell in the whole box. The color scale shows the baryon temperature,
with the red to blue boundary at the typical temperature for a \lya\
lines with \lnhi $=12.5$~\cmm . The red regions are cooler and the
yellow and whitish regions are hotter than the blue regions.
We show two contours, one for the baryon overdensity of 0.5
and the other for 15, the range responsible for many \lyaf\ lines.
The arrows show the baryon velocity with a length linearly proportional to
the amplitude of the velocity. The typical arrow in the upper right is 280 \kms .
   }
\end{figure}

In Fig. \ref{dentempimage}
we show a slice of the A box one cell thick. We added two contours one of which
shows the minimum typical over density for the IGM and the other
the upper overdensity. The arrows show the velocity of the cells. While most cells
making \lya\ absorption are surrounded by cooler gas, those that are flowing into the
highest density regions are next to hotter gas.

\begin{figure}
\includegraphics[width=84mm]{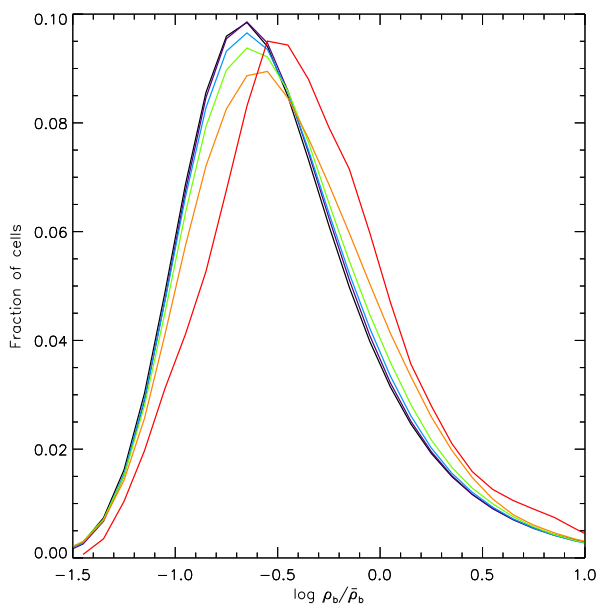}
\caption{
The effect of box size on the pdf of the log baryon overdensity in each cell.
At $\log \rho _b / \bar{\rho }_b = 0$ the curves for larger boxes are lower on the plot.
The \lyaf\ typically comes from $-0.3 < \log \rho _b / \bar{\rho }_b < 1.2$,
all to the right of the peak, where the larger boxes have systematically lower
fractions of their cells.
}
\label{gasdenpdfboxlin}
\end{figure}

\begin{figure}
\includegraphics[width=84mm]{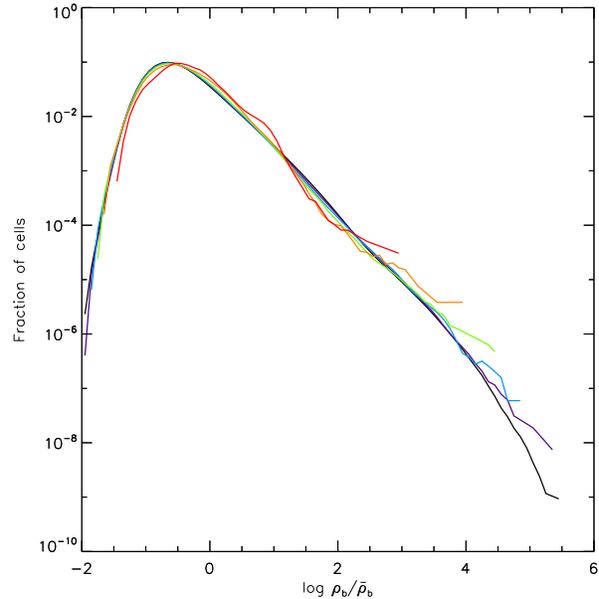}
\caption{
As Fig. \ref{gasdenpdfboxlin} but with the log of the fraction of cells on the
vertical axis to show the relative changes.
The curves for larger boxes extend farther to the right and are lower on the plot at
$\rho _b / \bar{\rho }_b = 0$.
}
\label{gasdenpdfbox}
\end{figure}

In Figs. \ref{gasdenpdfboxlin} and \ref{gasdenpdfbox}
we show the pdf of the baryon density per cell for different box size.
We see that most of the pdf moves to lower density in the larger boxes.
In the density range responsible for typical \lyaf\ lines with \lnhi\ 12.5 -- 14.5~\cmm\
there are systematically fewer cells in the larger boxes, which can explain why we
saw in Table \ref{dab}
less absorption in  the larger boxes.
If the density distribution drops by a constant factor for all
densities relevant to the \lyaf , then the $f(N)$ will remain unchanged in shape.
The log vertical scale in Fig. \ref{gasdenpdfbox} shows that this is approximately
true, but in detail there is a slightly larger relative decrease in the number of
low density cells. The larger boxes then have relatively more lines with higher
\lnhi\ values, as we have already seen in Fig. \ref{fig.columns} for columns $13 <$ \lnhi $< 15$~\cmm . Since lines with higher \nhi\ values tend to be wider, since they
come from higher densities where the gas is hotter, consistent with the larger
 \bsig\ values in the larger boxes.

\section{Simulations with Nearly Constant Mean Flux and b-values }
\label{fixedout}

We ran a second series of simulations using input parameters that we
adjusted to make the mean flux and \bsig\ values approximately constant,
at the values for the simulation A.
We adjusted the intensity of the UVB \gammah\ and the amount of
heating per He~II ionization, \gammahe .
We determined these parameters iteratively using scaling relations similar
to those described in J05. At redshift $z=2$ the ionizing background were
multiplied by the factors \gammah , listed in Table \ref{sim_table}.
For these KP simulations, we also augmented the UVB by additional factors
to make the mean flux at those redshifts closer to the values reported in
Keck HIRES spectra by \citet{kirkman05a}.  At $z \leq 2$ we multiplied the
fluxes by 1.
At $z=2$ -- 3 we multiply by $1.3(z-2)$, and
at $z>3$ we multiplied by 1.3.

In Table \ref{dab} we see that the mean fluxes are indeed similar to that of A,
although the modes are less so.
In Table \ref{tablebsig}
we see that the \bsig\ values are all similar and between those for A
and A2.
We could have iterated further to improve the agreement, but felt that this
was not necessary for this work.

\begin{figure}
\includegraphics[width=84mm]{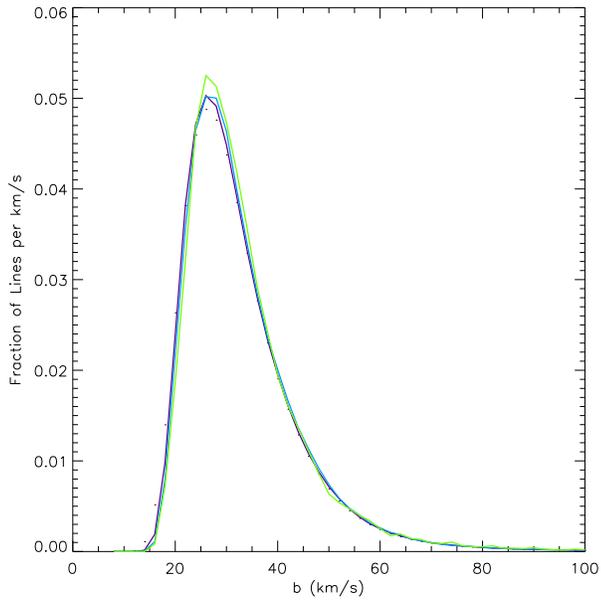}
\caption{
The $b$-value distribution for boxes in the KP-series, which should be
compared to Fig. \ref{bparamA} for the A series simulations. Just to the right of the peak,
the curves are from the top A4kp (green), A3kp (blue) and A2kp (violet).
}
\label{bparamAkp}
\end{figure}

In Fig. \ref{bparamAkp}
we show the b-value distribution for the KP series.
The three are more similar to each other and to A than were the similar sized
simulations from the A series, as we expect.

\begin{figure}
\includegraphics[width=84mm]{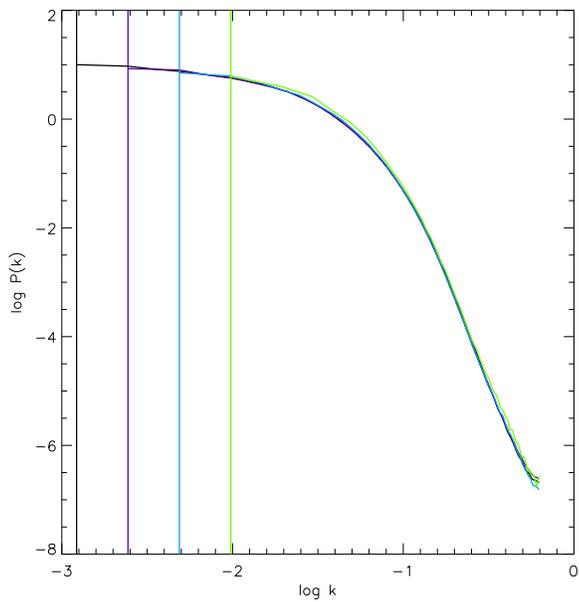}
\caption{
The power spectrum of the 1D flux contrast (Eqn. \ref{fluxc}) from KP series,
A2kp (violet), A3kp (blue) and A4kp (green) together with A (black).
The curves from the larger boxes extend farther to the left.
Compare to Fig. \ref{fpow} for the entire A series.
}
\label{fpowerkp}
\end{figure}

\begin{figure}
\includegraphics[width=84mm]{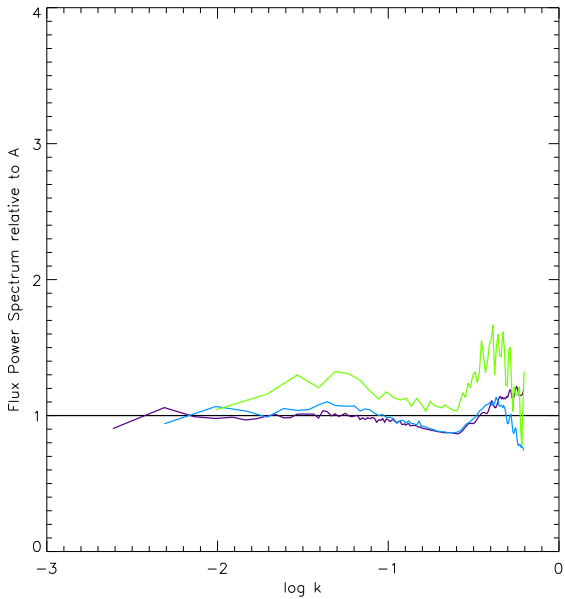}
\caption{
The power spectrum of the 1D flux contrast (Eqn. \ref{fluxc}) from KP series
divided by the power from A. From bottom to top at log $k=-0.5$ s/km we
show  A2kp (violet), A3kp (blue) and A4kp (green). Compare to Fig. \ref{fpowa}
which is on nearly the same vertical scale.
}
\label{fluxpowratiokp}
\end{figure}

In Fig. \ref{fpowerkp}
we show for the power of the flux contrast for the KP simulations,
and in Fig. \ref{fluxpowratiokp}
we show the same divided by the power from A.
Comparing to Fig.  \ref{fpowa} we see that the power in the KP series is
factors of 2 -- 3 closer to the power in A than were the
simulations of the same size
in the A series, though the differences are less reduced on the smallest scales.

\clearpage

\begin{figure}
\includegraphics[width=84mm]{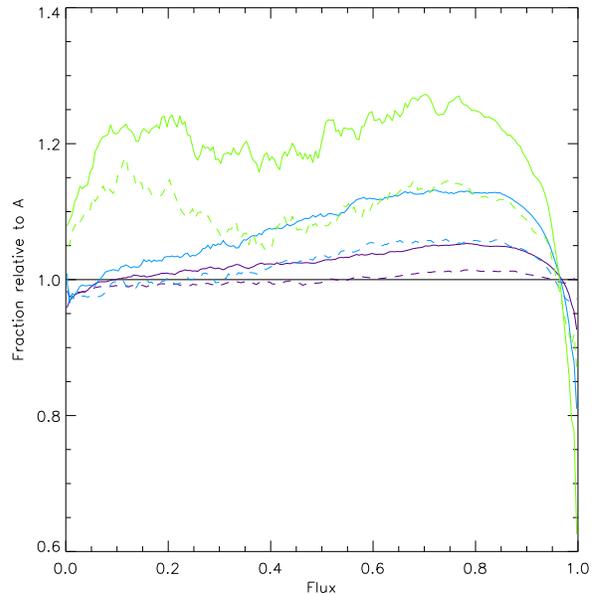}
\caption{
Distribution of the flux per pixel for KP and A series. We show the fraction
of pixels divided by those for simulation A. We show simulations A2, A3 and A4
(solid lines) from top to bottom at Flux $=0.7$, and below,
A2kp, A3kp and A4kp (dashed lines). Compare to Fig. \ref{fluxpdfratio}
for the A series.
}
\label{fluxpdfratiokp}
\end{figure}

In Fig. \ref{fluxpdfratiokp}
we see that the distribution of the flux in the KP series is
significantly closer to A than are the A series
simulation of the same size. The difference is about a factor of two for
A3 and A4, such that A4KP is similar to A3, and A3KP is similar to A2.
The improvement seems larger for the larger boxes. For A2 the frequency of
$Flux = 0.8$ is 1.05 times the frequency in A, while for A2KP this is 1.02.
There are even larger improvements at fluxes above 0.97.

In general, we see that the adjustments in the \gammah\ and \gammahe\ that
we made for the KP series allow a given KP simulations to have approximately 
\lyaf\ statistics of an A series simulation that is twice the size.
For some applications, we can then save a factor of 8 in computing resources.
We can use larger \gammahe\ values, corresponding to more heat input,
to partially compensate for the effects of limited box size. We simultaneously
need smaller \gammah\ values to maintain the same mean flux.

\section{Signal Definition and Normalization}
\label{sect.signal}

The division by the mean flux can introduce significant ambiguity, because
there are many ways to select the mean, and the mean is a function
of $z$. There are two main ways of defining the mean flux; global and local.

In this paper we use global definitions for the mean flux that come close to
approximating the true mean flux at each $z$.
We have divided spectra by the mean flux from the whole of each simulation box.
We could alternatively have taken an estimate of the mean flux from a calibrated
measurement \citep{tytler04b,kirkman05a,kirkman07a}. When we use real spectra
we must remove the metal lines and the strong \lya\ lines of LLS and
DLAs because they add significant absorption to the \lyaf\ that is not from the
low density IGM and that will be missing from simulations.

In contrast, \citet{hui01,kim04a,mcdonald06a} and others have used local
measures of the mean flux. They divide
each spectrum by its own mean flux, since their goal is to avoid continuum
fitting or to reduce the errors in the continuum level.
\citet{kim04a} [Fig. 2] obtained similar power spectra at
$k > 0.002$ s/km when they divided real spectra by either fitted continua or
the mean flux.

We do not advocate division by the local mean flux for several reasons.
We need to know the lengths of each spectrum to make a precise
comparison with other data or simulations.
For extremely long artificial spectra, division by the mean flux in
individual spectra is not very different to dividing by the overall mean
flux of the whole sample of many
spectra, but for the short spectra, including those from our boxes,
the differences are huge.

In real spectra the mean flux varies significantly from spectrum to spectrum
due to large scale structure. In
\citet{tytler04b} (Fig. 13, 16, Table 4) and we found that at $z=1.9$ the
standard deviation of the mean absorption in 121~\AA\ segments
from the low density IGM alone
is about 1/3 of the mean amount of absorption. In addition, the metal lines
and strong \lya\ lines and the low density IGM all
contribute similar amounts to the variation in the total amount of absorption.
Hence, when we divide by the mean flux in each spectrum, we are
removing much of the large scale structure signal, and introducing undesired
correlations with the metal lines and strong \lya\ lines,
with no guarantee that we are removing any errors in the continuum level.
Indeed, the continuum level errors of most interest are on the short
scales of the flux calibration errors and the emission line shapes, and not
necessarily correlated with the mean flux across the whole of a spectrum.

\begin{figure}
\includegraphics[width=84mm]{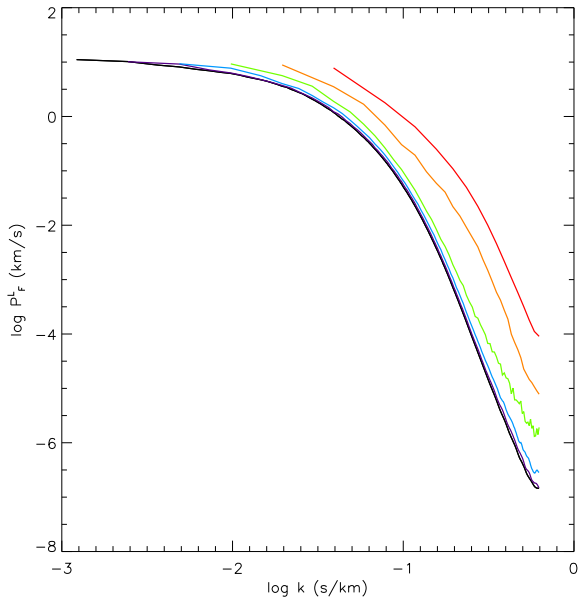}
\caption{\label{fpanl} The 1D power spectra of the flux where the signal is
divided by the mean flux in each sight line, and not the mean of the box.
Compare to Fig. \ref{fpow} that is identical except that there
we divided the flux in each spectrum by the mean flux in the simulation box.
}
\end{figure}

\begin{figure}
\includegraphics[width=84mm]{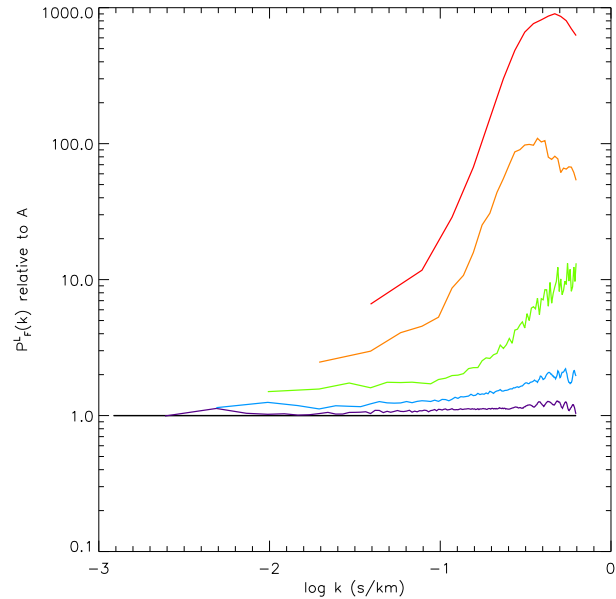}
\caption{\label{fpanlr} The 1D power spectra of the flux where we divided the
flux in each spectrum by the mean flux in that spectrum.
As Fig. \ref{fpanl} but now we divide the power by that in the A box. 
Compare to Fig. \ref{fpowa} where we divided each spectrum by the mean flux in the whole box.
}
\end{figure}

In Figure \ref{fpanl}
we show the power spectra of the flux obtained when we
divide the flux in each spectrum by the mean flux of that spectrum, $P^L_F$.
We show in Fig. \ref{fpanlr}
the power of the flux, divided by the mean flux in a sight line,
and then divided by the same quantity for box A.
The power in the larger boxes is little changed from in
Fig. \ref{fpow} where we divided by the mean flux in the whole box,
but the power in the smaller boxes is raised.

\section{How Resolution Changes the Simulated IGM}
\label{cellres}

\begin{figure}
\includegraphics[width=84mm]{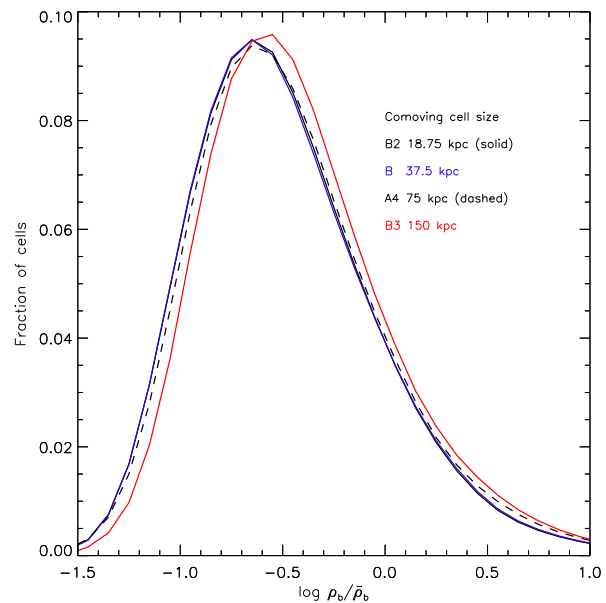}
\caption{\label{gasdenpdfcelllin} The pdf of the log baryon overdensity per cell for the
B series simulations which differ in only their cell size. We show the
fraction of cells sampled in intervals of $\log \rho _b/\bar{\rho}_b = 0.1$.
}
\end{figure}

\begin{figure}
\includegraphics[width=84mm]{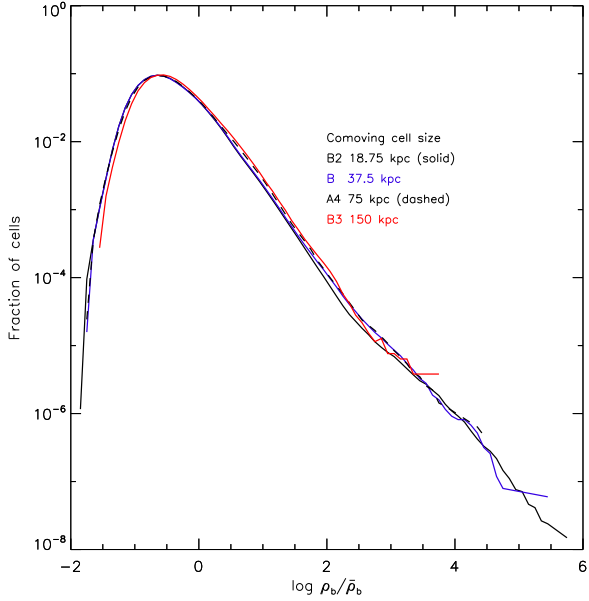}
\caption{\label{gasdenpdfcelllog} As Fig. \ref{gasdenpdfcelllin} but with
a log scale vertically. We sample in bins of size 0.1 in the log baryon overdensity.
Simulations with
smaller cells contain larger densities and extend farther to the right.
}
\end{figure}

In Figs. \ref{gasdenpdfcelllin} and \ref{gasdenpdfcelllog}
we show how the pdf of the baryon overdensity per cell varies with the cell size.
For the common densities shown in Fig. \ref{gasdenpdfcelllin} we see that simulations
with smaller cells have systematically lower densities. These changes are larger for
cell sizes 150 to 75 to 37.5~kpc, but the changes are too small to see
when the cell size drops from 37.5 to 18.75~kpc. The explanation for this trend
to lower densities is given in Fig. \ref{gasdenpdfcelllog}
where we see that simulations with smaller cells contain a few cells with much
larger densities. These cells contain the baryons that is depleted from the bulk of the volume.

\begin{figure}
\includegraphics[width=84mm]{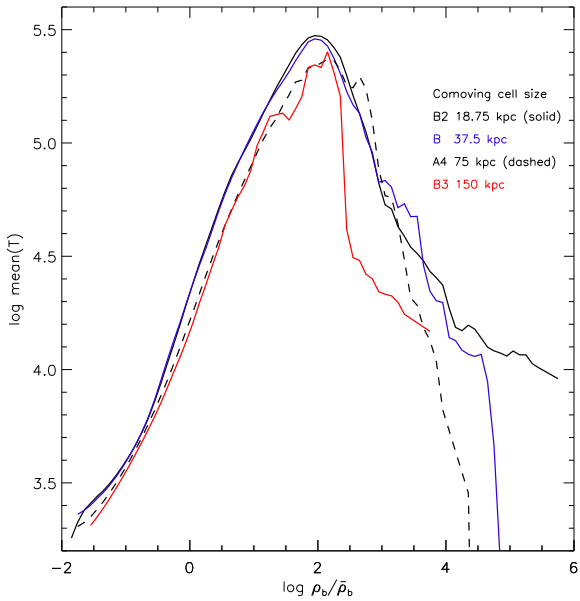}
\caption{\label{mntempvsden} The mean temperature of cells as a function of the
log baryon overdensity, sampled in bins of log overdensity 0.1.
}
\end{figure}

In Fig. \ref{mntempvsden}
we see the mean temperature of cells at a given baryon overdensity increases with
decreasing cell size. The increase is minimal when we decrease the cells from
37.5 to 18.5~kpc, suggesting that 37.5~kpc is small enough for the current work.

\begin{figure}
\includegraphics[width=84mm]{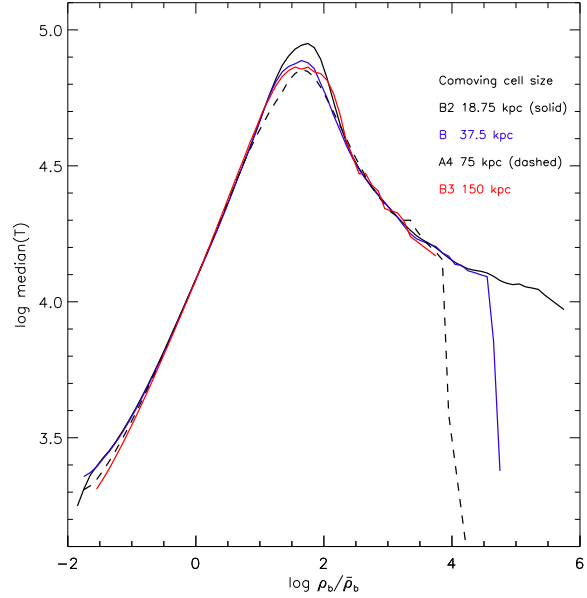}
\caption{\label{mdtempvsden} As Fig. \ref{mntempvsden} but showing the
temperature that is exceeded in 50\% of cells, the median,
sampled in bins of log overdensity = 0.1
}
\end{figure}

In Fig. \ref{mdtempvsden}
we show the median instead of the mean temperature. The changes are now much smaller,
except near log overdensity = 1.7 where we see the peak temperatures and
no sign of convergence as we decrease the cell size.

\begin{figure}
\includegraphics[width=84mm]{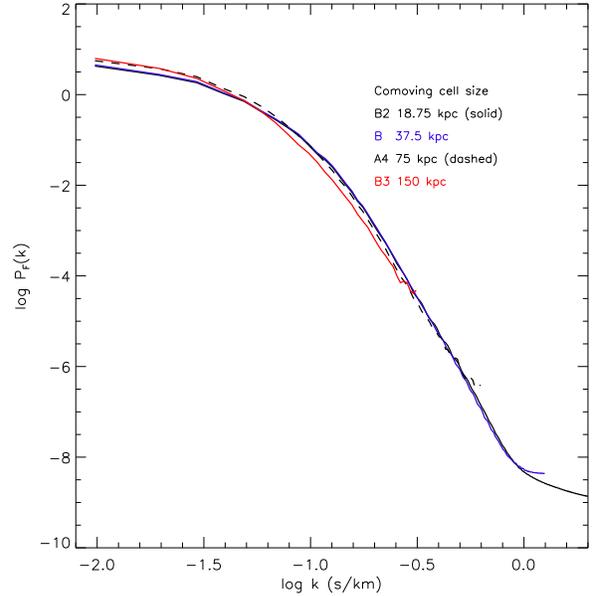}
\caption{\label{fpowerb} The 1D power spectra of the flux of the
B series simulations which differ in only their cell size. We divided the
flux in each spectrum by the mean flux in that simulation box.
We terminate the power spectra at the Nyquist frequency for that cell
size:
B3 (150~kpc cells,  log $k=-0.5$, red line),
A4 (75~kpc, log $k =-0.2$, dashed black line),
B  (37.7~kpc, log $k =+0.1$ s/km, blue line),
B2 (18.75~kpc, log $k =0.4$~s/km, black line).
}
\end{figure}

In Fig. \ref{fpowerb}
we show how the power of the flux depends on the cell size.
With smaller cells there is less power on the largest scales
and more on small scales. The boxes with smaller cells begin with
more power in total because their power spectra extend to smaller scales.
Their structure becomes non-linear on small scales earlier and this
encourages the growth of power on small scales at the expense of large ones.

\begin{figure}
\includegraphics[width=84mm]{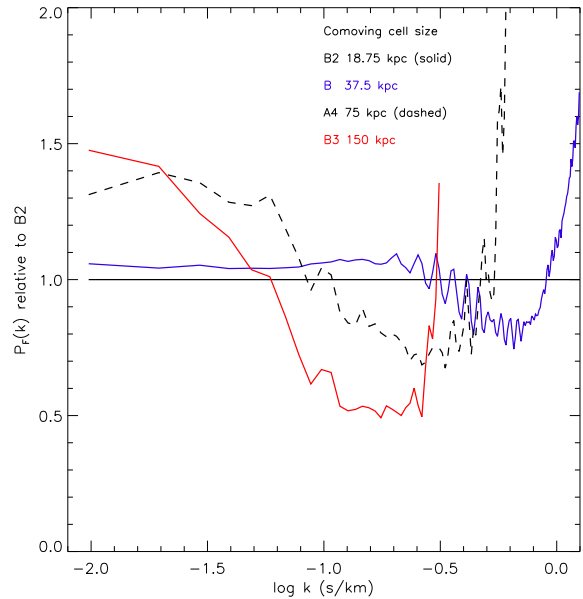}
\caption{\label{fpowerbrat} As Fig. \ref{fpowerb} but showing the
ratio of the power of the flux to the power from the B2 simulation.
On the far left the simulations are from the top
B3 (150~kpc cells, red line),
A4 (75~kpc, dashed black line),
B  (37.7~kpc, blue line),
B2 (18.75~kpc, black line at 1.0).
}
\end{figure}

In Fig. \ref{fpowerbrat}
we show the ratio of the power of the flux to
the power from the B2  simulation that has the smallest cells.
In general, the boxes with smaller cells have smaller flux power on the
largest scales (small $k$) and more power on the smallest scales.
We see that the maximum $k$ value at which the power is larger than in B2
shifts systematically to higher values with the smaller cells: from
$-1.32$~s/km (B3, 150~kpc cells) to
$-1.0$~s/km  (A4, 75~kpc cells) to
$-0.5$~s/km  (B, 37.5~kpc cells).
We see that factor by which the power is larger than in B2 on large scales
(small $k$) is approximately constant over a range of $k$ values,
at approximately 1.35 for A4 and
1.07 for B. This suggests rapid convergence as the cell size decreases
below 18.75~kpc. However, the convergence on smaller scales is much less rapid.
The ratio of the power to that in B2 is a minimum on scales near a factor of two
larger than the Nyquist frequency. These minimum values for the power 
ratios are large and approach 1.0 slowly as we decrease the cell size: from
0.52 (B3) to
0.75 (A4) to
0.83 (B).
This behavior suggests that cells smaller than 10~kpc
will be needed to get the power at log $k \simeq 0$~s/km
to within a factor of 0.9 of the value in a simulation
with much smaller cells.

One other feature of the power spectra of the flux
is more troubling. We see that the power
increases steeply on the smallest scales, just above the Nyquist frequency.
We saw similar behavior in Fig. \ref{fpow} for the A series.
Early in this investigation we saw much larger versions of
these upturns in power which were caused by errors in the
generation of spectra that lead to discontinuities in the flux. We continued
 searching
for errors and found no more. However, the behavior is clearly not physical, because
simulations with smaller cells do show that the power ratios continue to decline
smoothly on decreasing scales. We should not use the power from these simulations
on scales within log $k = 0.2$~s/km of the Nyquist frequency.

\citet[Fig. 6]{mcdonald03} shows how the power of the flux varied for three hydro-PM
simulations (no shocks) in 6.25~Mpc boxes with cell sizes of 24.4, 48.8 and 97.6~kpc.
While we both see
that the largest cell size gives results very different from intermediate sizes
(50 -- 75~kpc), we do not see much else in common between our results.
This confirms the point made by \citet{mcdonald03},
that the results of resolution studies
depend on the nature of the small scale force calculations and physics.

\begin{figure}
\includegraphics[width=84mm]{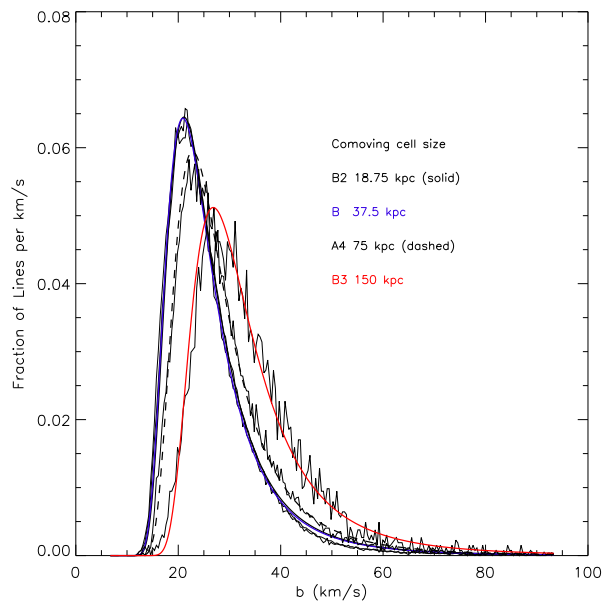}
\caption{\label{bres} The distribution of the $b$-values for \lya\ lines
with $12.5 < \lnhi\ < 14.5$ (\cmm ) and line central optical
depth $\tau > 0.05$ in
B series simulations. The jagged thin lines are the distributions of the values
and the smooth curves are the Hui-Rutledge fits to each pdf.
From the right, at a fraction of 0.004, the simulations are
B3 (150~kpc cells, red line),
A4 (75~kpc, dashed black line),
B2 (18.75~kpc, black) and
B  (37.7~kpc, blue line)
which is out of order and largely hidden under B2.
}
\end{figure}

In Fig. \ref{bres}
we see that the $b$-value distributions moves to significantly smaller velocities
with smaller cells, except for B2 which has slightly larger velocities
than B, reversing the trend. In Table \ref{tablebsig} we list the \bsig\ values
for the Hui-Rutledge fitting formula. The $\Delta $ column
shows that the \bsig\ value drops 4.2~\kms\ from 150 to 75~kpc cells, and then 2.0~\kms\ going to 37.5~kpc, but it increases by 0.2~\kms\ going to 18.75~\kms\ cells.
Since the internal error in the measurement is about 0.8~\kms, 35~kpc cells seem to give a fair estimate of the \bsig\ that would apply to a simulation with much smaller cells. The fitting function gives  an excellent representation of the $b$-value
distributions. In detail we see systematic differences between these distributions
and the function, e.g. the function is too high around the most common
$b$-values (especially for the larger cell sizes) and has too many lines with $b > 40$ (for B2, B) or $> 50$~\kms\ (for B3). As for the A-series, we use only lines with $b < 40$~\kms\ when we estimate the \bsig\ values.

\begin{figure}
\includegraphics[width=84mm]{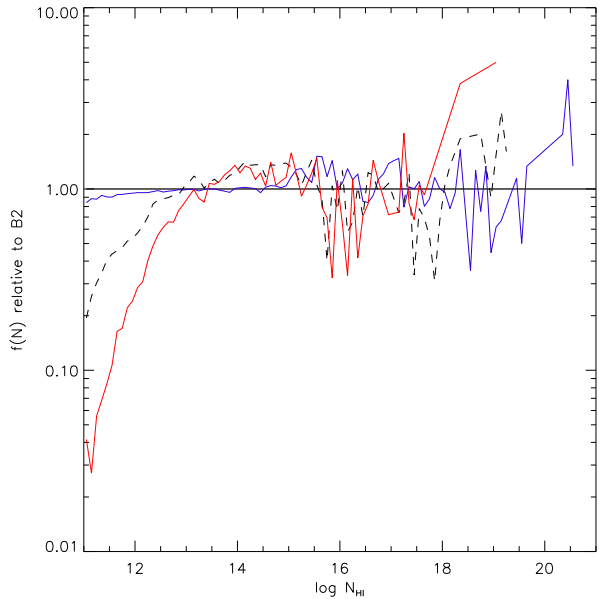}
\caption{\label{resfhnratio} How the column density distribution depends on
the cell size of a simulation. We show the $f(N)$, measured using all lines with
line central optical depth $\tau > 10^{-5}$. Here only we evaluate $f(N)$ per unit $z$
rather than $X$, and we divided by the $f(N)$ from the B2 simulation.
We give values for evaluated in bins of width 0.1.
On the far left the simulations are from the bottom
B3 (150~kpc cells, red line),
A4 (75~kpc, dashed black line),
B  (37.7~kpc, blue line),
B2 (18.75~kpc, black line at 1.0).
}
\end{figure}

In Fig. \ref{resfhnratio}
we see that the simulations with smaller cells have
factors of several more \lya\ lines with
the lowest column densities \lnhi $< 13$~\cmm .
However, the small cells also give about 20\% fewer lines
with $13<$ \lnhi\ $<17$~
\cmm\ where the precise range depends on the cell size. Hence simulations with
smaller cells are slightly farther from data than was simulation A (75~kpc cells)
that has too few lines of high \lnhi\ (Fig. \ref{fig.columnsdata}).

\section{Convergence and Comparison with Data}
\label{seccon}

In Table \ref{tabconverge}
we summarize how various statistics change as we increase the size of the
simulation box. In Table \ref {tabdataerr}
we compare the difference between the values we measure in the A and A2 
boxes to the likely errors from measurements of data.

We see very small changes in the mean flux with increasing box size, in
part because the amount of absorption is small compared to the mean flux.
The changes are better seen in the amount of absorption itself. 
The rate of convergence suggest that the mean flux in A is within
0.0007  of the value expected in a much larger simulation. This is about a
factor of 14 smaller than the measurement error of
approximately 0.01, from sample size,
continuum level errors and difficulties removing
absorption from metal lines and strong \Lya\ lines down to some fixed \nhi\ value
\citep{tytler04c, kirkman07a,kim07b}. The mean flux in simulation A is essentially
identical to that from Eqn. 10 of J05, scaled to $z=2$, and we expect
this to remain true in a much larger box.

\begin{table}
\caption{\label{tabconverge} The Convergence of Statistics of the A series.
The ratios are of quantities from smaller boxes A4, A3 or A2 to the value in
the largest box A. When a range is indicated, we list the
largest value in that range.
}
\begin{tabular}{llll}
\hline
Quantity & A4/A & A3/A  & A2/A  \cr
\hline
Flux mean & 1.0031 & 1.0023 & 1.0008\cr
Absorption $=1-$ flux mean & 0.9790 & 0.9844 & 0.9946\cr
Flux pdf (F=0.995--1.0) & 0.62 & 0.81 & 0.93 \cr
Flux pdf (F=0.8) & 1.24 & 1.13  & 1.05  \cr
Typical line width \bsig\ (\kms ) & 0.903 & 0.929  & 0.959 \cr
$f(N)$ \lnhi $=12.5-14.5$ & 0.83 & 0.85 & 1.10\cr
$f(N)$ \lnhi $=15$ & 0.73 & 0.81 & 0.84\cr
Flux P(k=0.01) & 0.966 & 1.023 & 0.959\cr
Flux P(k=0.1) & 1.465 & 1.194 & 1.074\cr
Frequency of CDM density & 0.93 & 1.03 & 0.98\cr
Mode CDM density & 0.2 & 0.2 & 0.50\cr
\end{tabular}
\end{table}

\begin{table}
\caption{\label{tabdataerr} Comparison of Statistics from Simulation A to
those from Data. The column headed ``A" lists the
value of the quantity in box A.
The column A2-A lists the value of the parameter in
box A2 minus the value in box A.
The errors on the data values are guessed, not precise, values.
}
\begin{tabular}{llllll}
\hline
Quantity &  A & A2-A & Data & $\sigma(\rm data)$\cr
\hline
Flux mean &  0.8714 & $-0.0007$ & 0.869 & 0.01\cr
\bsig\ (\kms ) &  26.7 & 1.1 & 23.6 & 1\cr
$log f(\lnhi $=14.3) & $-13.49$ & $-0.03$ & $-13.39$ & 0.2\cr
Flux P(k=0.01) & 5.8  & 0.23 & 7 & 1\cr
Flux P(k=0.1) & 0.049 & $-0.004$ & 0.13 & 0.05\cr
\end{tabular}
\end{table}
For the flux pdf, Figure \ref{fluxpdfratio} indicates that simulation A
will be within about 5\% of the frequencies for a much larger simulation.

For \bsig\ the error from the simulation box size is comparable
to that for data. The $\Delta $  values in Table \ref{tablebsig}
do not show much evidence for convergence, since the change in
\bsig\ from A2 to A is larger than the change from A3 to A2,
and from A4 to A3.
This slow convergence can be traced back to the effects of the long modes
of the CDM density fluctuations that change the sizes of the absorbing
regions, the velocities and the temperatures.

We also see that the \bsig\ for A is significantly larger than for
data \citep{kim01,jena05a} and the difference will be still
larger in a larger box. 
 To better match data, we need a simulation with less heat
input (smaller \gammahe ) or larger \sig\ (Figs. 21 and 38 of J05), which is
a surprise since the value we are using, \sig $=0.9$, is large compared to
the WMAP 3-year suggestion. We \citep{tytler04b,jena05a} and
others \citep{viel06b,seljak05,viel07a}
have previously noted that the \lyaf\ data prefer much larger \sig\ values than
does the CMB anisotropy. \citet{slosar07a} use \lyaf , Supernovae and galaxy
clustering data with WMAP 3-year data to estimate $n=0.965 \pm 0.012$ and
\sig $=0.85 \pm 0.02$, compared to \sig $=0.80 \pm 0.03$ without the \lyaf .

The changes that we will need
to make to the simulations match the  column density distribution of data
will also change the $b$-value distribution and the \bsig\ value.
The minimum $b$-value in the \lyaf\ increases as \nhi\ increases
\citep{kirkman97a} until we reach \lnhi $ = 15$~\cmm\ after which the
values start declining in our HIRES spectra and in
simulations \citep[Figs. 3, 5]{misawa04}. Since simulations need fewer
lines with \lnhi $< 14$~\cmm\ and, in compensation to conserve the total absorption,
 more at $14 <$ \lnhi\ $< 15$~\cmm\
(Fig. \ref{fig.columnsdata}), the mean $b$ values will increase,
exacerbating the differences with data.

The lack of high column lines is the second conspicuous difference
between our simulations and data. As noted previously, this is due to
insufficient spatial resolution and lack of self-shielding in collapsed
dense halos.
In Fig.  \ref{fig.columnsdata} we saw that our simulations
have too many lines with \lnhi $< 14$~\cmm , a slight lack of lines with
\lnhi $ = 14 - 15$~\cmm , and a large lack with \lnhi $>17$~\cmm .
This lack of high column lines will reduce the power to below that in real
spectra that include such lines.
We also noted that our sight lines that are parallel to the box sides
are too short to contain the full damping wings of DLAs.
Fig.  \ref{fig.columns}
shows convergence as the box size increases and suggests that
the $f(N)$ values from simulation A for \lnhi $=12.5 -14.5$~\cmm\
are within about 10\% of the values we would obtain from a much larger box.

\begin{figure}
   \includegraphics[width=84mm]{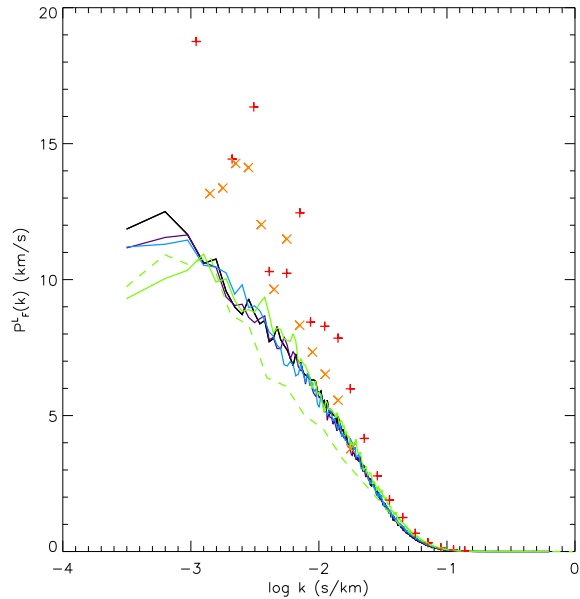}
   \caption{\label{powerdatlin}
Comparison of the power of the flux in the \lyaf\ with that for our
simulations. Lines show the power of spectra from simulations A (black),
A2 (violet), A3 (blue) and A4 (green), our usual color scheme.
The dashed green line is for simulation B2 from J05.
We use flux spectra that travelled in random
directions for a distance of $\delta z = 0.2$ to approximate the lengths
of real spectra (Appendix C), 
but without evolution of the IGM (Appendix B).
Before we obtained the power, we divided all spectra by the 
mean flux in that spectrum, to match what was done with the data. 
In simulation A, the largest box, the longest mode parallel to a box edge has
log$k > -2.91$~s/km.
Values of the power for box A are listed under $P_e$ in Table 12.
We show the power from \citet{mcdonald06a} linearly extrapolated to $z=2$
(orange x) and the PJ05 power (J05 \S 6.4) from 6 HIRES and UVES spectra
(red +) with metal lines masked.
   }
\end{figure}

\begin{figure}
   \includegraphics[width=84mm]{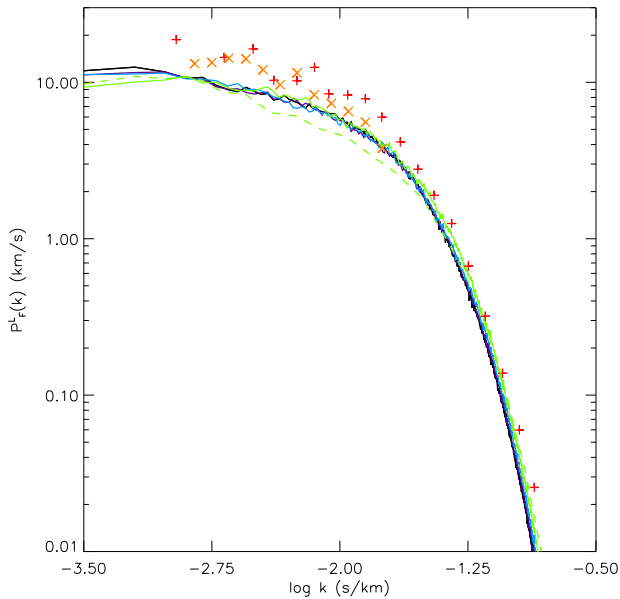}
   \caption{\label{powerdatlog}
As Fig. \ref{powerdatlin} but on a log scale.
   }
\end{figure}

In Figs. \ref{powerdatlin} and \ref{powerdatlog}
we compare the power of the flux of the \lyaf\ in data to that in our
simulations.  The power from the simulations is less than in the data at all
$k$ values. The power in the simulations is too low by about 20\%
 at $-1.6 < $log $k < -1.1$~s/km rising to about 50\% on large scales at 
log $k < -2$~s/km.
 
We are most concerned about the missing power on large scales.
There we have SDSS measurements that we trust more than those from J05 on small scales,
and there should be no problems from residual metal lines in the real spectra at
large scales. The values that we give for the errors on the
power on the data in Table \ref {tabdataerr} are guesses based on the spread
between different measurement values. We note that the differences between the
simulation and data seem less at large $k$ since only a small change in $k$ would
be needed to align the two. However, the errors on $k$ are very small, and hence
we are interested in the vertical change in the power and not a horizontal shift in
$k$.

We had expected the power of the simulation to be smaller than in data on small scales
(large $k$) because the $b$-values in the simulations are larger than in data.
The sense of the differences are consistent: larger $b$ values correspond to less
power at log~$k > -1.5$~s/km \citep{viel03c}. We also knew that we lacked large
scale power when we began this investigation and we had hoped to understand this 
difference, but we have not.

A major conclusion of this paper is that a much larger box will not
bring the power from the
simulations up to that in the data since we saw in Fig. \ref{fpowa} that
the effects of doubling the box size are ten times smaller than the amount of
missing power.

We have also shown that improving the resolution of a simulation by reducing the
cell size makes the simulation more different from data at large scales.
In Fig. \ref{fpowerbrat} we saw that when we decrease the cell size, from
75~kpc to 18.75~kpc we decrease the power in the simulation at log~$k < -1$~s/km,
by 30 -- 40\% at the largest scales. Using these small cells, the power in the 
simulation is then about a factor of two ($1.5 \times 1.35$) below that in data.
Simultaneously, the
power increases for the largest few $k$ values, which brings the simulation 
closer to the data.
The power from the B2 simulation is the lowest of all in
Figs. \ref{powerdatlin} and \ref{powerdatlog}  and yet it has the
same input parameters and box size as A4 and 4 times smaller cells than
the A series. In J05 we noted that B2 has a lower \bsig\ value 
(corresponding to higher small scale power) but higher mean flux
(lower power) than the A-series. Hence to better match data we should re-run B2
using a lower \gammah\ to increase the \lya\ absorption and this will
increase the power, making the difference from the data less than a factor of 
two.

We know that our simulations have too many low column density lines and too few
with high column densities. When we used cells 4 times smaller, these 
differences got worse, as did the difference in the power.
We also noted that a four times reduction in cell size does not correct the
large lack of lines with \lnhi $>17$~\cmm , lines that we hope are excluded from
the data. \citet{kohler07a} showed that they obtain the correct number of such LLS 
using 2~kpc cells. We are curious whether 2~kpc cells might also match the
entire column density distribution and perhaps the power.

\subsection{What Cell Size do we Need?}
\label{convres}

We know from our analysis of the KP series of simulations in \S \ref{fixedout}
that we can mimic much of the effects on the \lyaf\
of doubling the size of a box by instead increasing
the \gammahe\ parameter that increases the heat input per He~II ionization.
We must simultaneously decrease the rate of H~I ionizations by
reducing the \gammah\ to bring the amount of H~I back to the level that
gives the observed mean flux value.
Small simulation boxes are too cold compared to large ones.

In Figs. \ref{gasdenpdfcelllin} and \ref{gasdenpdfcelllog}
we saw that there was very little change in the pdf of the baryon density per
cell, for typical densities, going from 37.5 to 18.75~kpc cells, implying 
that 37.5~kpc is acceptable for our work.

In Fig. \ref{mdtempvsden}
we see no convergence by even 18.75~kpc for the median temperature
at log baryon overdensity near 1.7. Smaller cells are leading to higher
median temperatures. Fig. \ref{mntempvsden} however shows that
mean temperatures are converged by 18.75~kpc, again
suggesting that 37.5~kpc is small enough for the current work.

In Fig. \ref{fpowerbrat}
we saw that changing the cell size from 75~kpc to 18.75~kpc has a complex effect on the
power spectrum. The power drops by 30 -- 40\% on
large scales ($\log k < -0.7$~s/km)
but then increases by up to 30\% before falling again on the
smallest-scales near the Nyquist frequency.
However, changing the cell size from 37.5 to 18.75~kpc has a much smaller
effect, reducing the power by only 5\% at log$k < -0.7$~s/km.
This implies that 37.5~kpc is small enough for all but the highest
precision work.

We see a similar rapid convergence for the $b$-values. In Fig. \ref{bres}
we saw that the $b$-value distribution changes noticeably from 150 to 75 to
37.5~kpc cells, but the change going to 18.75~kpc is barely detectable.

 In Fig. \ref{resfhnratio}
we saw that  reducing the cell size from 37.5~kpc to 18.75~kpc made the column 
density distribution increase by about 5\% at \lnhi $=12$~\cmm\ and 
decrease by 20\% at $15 <$ \lnhi $<17$~\cmm .

In summary, a cell size of 37.5~kpc is adequate at this time, but
slightly smaller cells (or correction factors) will be needed for the highest
accuracy work.  We see rapid convergence as the cell size drops below
37.5~kpc. We will
definably need to apply corrections if we use cells of 75~kpc or larger.

\subsection{How Large a Box do we Need?}

We summarise the amount of convergence by
listing the statistical parameters in the order of increasing difference
of their A2/A ratios from unity. The mean flux (1.0008), amount of absorption
(0.995), and the frequency of the CDM density (0.98) are the most converged
quantities.
Then follow the \bsig\ (0.96),
the flux pdf (1.05, 0.93) and
the power of the flux (0.96, 1.07).
The $f(N)$ (1.10, 0.84) is less converged and the mode
of the CDM density (0.50) shows no sign of convergence in our boxes.

Some of the CDM statistics converge while others do not.
In Fig.  \ref{pdfdeltaa}
we see convergence in the frequencies of the CDM densities.
We also see convergence with the power of the CDM density in
Fig.  \ref{fig.pdmspc}, however the mode of the CDM density distribution
in Fig. \ref{fig.cdm_pdf_updated} shows no sign of converging as the
box size increases. We discussed how this was caused by the rare very high
density regions in the larger boxes.
These high density regions are not in the
low density IGM and yet they dominate many of the CDM statistics,
including the power and the pdf.

To first order, the values in Table \ref{tabconverge}
show that doubling the size of a box typically halves the difference of a
parameter value from its value in the largest box.
If this trend were to continue unchanged, we know from the sum of the geometric
series 1/2 + 1/4 + 1/8... that the value in a very large box would
be approximately the value in A plus the difference from A2 to A.
In practice we can expect the series to converge more quickly as the box
size increases past several hundred Mpc to include the peak in the matter
power \citep{bagla05a}.
Hence the values of the parameters in a very large box would be similar to
the value in A plus the value in the column A2-A in Table \ref {tabdataerr}.

We can estimate the change in \bsig\ if we had run
simulation A with a resolution of $18.75$~kpc instead of 75~kpc,
using the scaling relations given
in J05; the value of \bsig\ would change from 26.7~\kms\ to 25.1~\kms .
If the $b$-values converge with box size
as do the other statistics, which we have not
established, then we might guess that a much larger box, many hundreds of Mpc
in size, with 18.75~kpc cells would give
\bsig $=25.1+1.1 = 26.2$~\kms\ that is 2.6~\kms\ larger than the data.

This comparison with the measurement errors for data suggests that
the box size is a relevant but probably not the dominant error for our
largest box. The only exception is the line widths where the difference between
the values from our two largest boxes is similar to the measurement
error. We would like factor of several larger boxes to
reduce this uncertainty. Since the $b$-values are closely related to the
small scale power, we would expect that larger boxes will also bring useful
improvements in the accuracy of the small-scale power.

We can most easily detect the increase in the
size of a simulation box in data on the smallest scales, the \lya\ line widths.
This is primarily because  it is easier to make high accuracy measurement
of small scale features of the \lyaf\ than of the large scale trends,
such as the power of the flux.

\section{Physical Explanations}

We have deliberately treated the simulations like observational data. We
concentrated on reporting how statistics that describe the IGM and
the \lyaf\ change with box size. We have resisted the temptation to follow
additional side line investigations that might reveal the physical
causes of these effects.
We now give a short discussion of the possible physical
explanations for the changes that we see with box size.

The larger boxes differ from smaller ones in only two ways; they contain
more cells and they contain longer mode perturbations that do not fit in the
smaller boxes.
The extra modes add
more power in total to the simulations, leading  to larger velocities and more
kinetic energy. We will see flows that are coherent on
larger distances, and larger velocity differences on a given distance scale.
Changes in the (negative) gravitational potential energy
are more complex.
Potential wells become deeper where long modes are positive density
fluctuations, and shallower elsewhere, giving a cancellation to first order.
We saw in Figure \ref{pdfdeltaa}
that non-linear effects lead to rather complex changes in the
CDM density per cell. We saw a decrease in the number of cells
near and above the most common density and an increase in the number of lower
density cells.

We hope to explain, using the longer modes alone,
all the changes that we see with box size. We see changes in both the
elemental density, velocity and temperature fields
and in the various \lyaf\ statistics that are composites of these fields.

We saw in Fig.  \ref{fig.pdmspc} that by $z=2$ the longer modes have
evolved to add power on all scales, but especially the largest ones.
This is a Richardson-Kolmogorov cascade \citep{kritsuk07a}
of energy from large to small scales, which is a non-linear effect.
The study of \citet{bagla05a} shows the growth in the number of halos
as a result of this cascade, where they use
a fixed box size and truncate the initial power to sub-box scales.
We do not know if these results are sensitive to
the finite box size, the box shape and the periodic boundary conditions.

\subsection{Why the gas causing the \lyaf\ is hotter in larger boxes}

The gas in the IGM that causes the \lyaf\ is hotter in the larger simulation
boxes because of the enhanced heating by shocks.
We now discuss why we believe this, and not an alternative explanation involving
reduced adiabatic cooling.

We have discussed how the extra longer modes evolve to give more power in
the CDM and baryon density fluctuations on all scales, and hence 
larger peculiar velocities.
The most obvious explanation is that the increased density and velocity
perturbations lead to faster collisions of the gas in given cells, 
and more cells that experience a collision of a given velocity.
This thermalisation effect applies to the warm-hot IGM at $z \simeq 0$
that is not seen in H~I absorption, but it is less clear if the effect
is also important for the IGM that produces the \lyaf\ $z =2$.
Most cells in the IGM have not been in any collisions by $z=2$, and hence we 
require that the heat from collisions
 spreads far beyond the cells that have contained shocks \citep{cen99,dave01}.

We now discuss an alternative explanation, that the larger boxes are 
hotter because the
gas that makes the \lyaf\ has cooled less than in the smaller boxes.
The IGM is heated when H~I, He~I and He~II are ionized. The temperature
drops in time due to Hubble expansion. Lower density regions expand faster
and cool more, leading to the well known increases in temperature
with density
for the gas causing the \lyaf\ \citep{hui97}, as illustrated for two of
our simulations in \citet[Fig. 19]{tytler04b} and Fig. 34 of J05.
The longer modes in larger boxes may give rise to higher
densities that reduce the adiabatic cooling from Hubble expansion.
The gas that makes \lyaf\ lines
is hotter in larger boxes because it expands less and cools less.

However, this can not be the entire explanation.
When we add the longer modes we make some densities higher and others
lower. We expect the two to give opposing effects
that will tend to cancel to first order.
For this mechanism to give a net heating of the IGM
we require that the long modes make a larger increase in the number of 
hotter cells
than in the number of cooler cells. The asymmetry favouring heating
comes from the distribution of the number of cells as a function of
density.
There are more cells at lower densities, and hence when the
long modes adjust all densities, there is
a net flow of cells to higher density.
We can see the pdf of cells as a function of density in \citet[Fig. 34]{jena05a}
where the most common baryon density is near 0.2 of the mean density,
below the typical density that leads to lines with \lnhi = 12.5~\cmm .
This asymmetry is a version of the Malmquist bias, in which we see a net
increase in the number of objects detected in a flux limited sample
\citep{gonzalez97}.
The effect depends on the steepness of the pdf of the density per cell.

We saw in Fig. \ref{gasdenpdfboxlin} that larger boxes had a smaller fraction of their
cells in the density range responsible for the \lyaf .  The temperature depends on the
relative number of cells with different densities. In Fig. \ref{gasdenpdfbox} we saw
that there is a larger decrease in the number of cells with log $\delta _b \simeq 0$
compared to cells with log $\delta _b \simeq 1$, except for the smallest box. Since the
lower densities correspond to lower temperatures, we then expect the larger boxes to
have fewer cells with lower temperatures. The trend is in the direction needed 
to make the larger boxes have hotter gas, and wider \lya\ lines,
but the effects we see on the plots we just discussed all seem too small
to offer a credible explanation.

In Figure \ref{fig.columns} we saw that in larger boxes the column density distribution
has relatively more lines near the higher end of the 12.5 -- 14.5~\cmm\ range that we
use when we measure \bsig . Such lines tend to have larger $b$-values. We do
not know if the $b$-value pdf changes at a given \nhi ,
or if we can we explain the larger \bsig\ entirely in
terms of the change in $f(N)$. 

In summary we do not find any convincing evidence for the second explanation,
that the \lyaf\
lines are wider in larger boxes because they come from gas that cooled less. Rather
we prefer the first explanation, that the larger boxes are hotter because the
velocities are larger giving more and stronger shocks.

\subsection{Why the \lyaf\ lines are wider in larger boxes}

\lya\ line widths are set by three components \citep{bryan99}, thermal from
the gas temperature, peculiar velocities
from hydrodynamical motions and Hubble from the
cosmological expansion \citep{sargent80}.
We have shown in Figs. \ref{velbox} and \ref{mntempvsdenbox} that the
gas temperature and peculiar velocities increase with box size for fixed
photoionization and cosmological parameters. Hubble broadening depends on the size of
the line forming regions, which in turn depends on the balance between the previous
two gas attributes. High temperature leads to thermal expansion and wider
lines, while increased peculiar
velocities are indicative of high compressions that tend to
decrease the size. \citet{bryan99} determined that in the column density
range relevant to the b-parameter distribution, $10^{12.5}-10^{14.5}$ cm$^{-2}$,
thermal broadening increasingly dominates over Hubble broadening at
larger column densities.
Therefore we conclude the \lyaf\ lines are wider in the
larger boxes because there is more thermal broadening, which is more
important than the reduction in $b$-values from the larger peculiar velocities.

\subsection{How  might we make simulations with the appropriate temperatures}

We discuss several potential ways of making the IGM in our simulations cooler
and closer to the temperature required by \lyaf\ data.

First, we should mention that there remains a slight possibility that there is
no problem with the IGM temperature,
and rather a mis-match in the comparison of the simulation and data because 
the $b$-value distributions are not determined using the same codes.
We consider this unlikely in part because the power spectrum
measurements tell a consistent story.
The power spectrum is measured independantly, from mostly different spectra,
and the higher small-scale 
power in the data compared to the simulations is consistent
with the smaller $b$-values in the data compared to simulations.

We have seen that increasing the resolution of our simulations does
lead to a cooler IGM with smaller $b$-values. However, we can not run the
ideal simulation.
A simulation with 18.75~kpc cells (like B2) in a 76.8~Mpc box (like A) would have
 $8192^3$ grid cells, a factor of $4^3$ too large for the supercomputers we use.
An adaptive mesh refinement (AMR) simulation with increased resolution
at the densities of interest is also not practical. 
We are interested in the
filamentary structure at overdensities above 1.
A simulation of the 76.8~Mpc box would  require
three levels of refinement, each decreasing the cell size by a factor of 2,
to have an effective resolution equal to that of the B2 simulation.
However, the volume fraction of regions with densities above the cosmic 
mean is large, requiring a prohibitive number of refinement subgrids.

The temperature of the IGM at ionization is set in large part
by the mean energy of the photons
that cause the ionization. Softer spectra have steeply declining numbers
of photons with
increasing energy above the ionization threshold. We could make
the temperature lower by using an ionizing spectrum that was softer than that
specified by Haardt and Madau,
at either 1 Rydberg for H~I or 4 Ry for He~II or for both.

Early results from simulations that include the effects of
 radiation transfer suggest that this makes the IGM hotter
and exacerbates the difference with data.
\citet{paschos07}
have carried out an approximate treatment of the helium ionization due
to discrete QSO sources and have found out that the
temperature of the IGM at the cosmic mean density can be on average 68\%
above in an optically thin simulation. This increases the widths of \lya\
lines by only about 1.3 \kms\ at $z=2.5$,
because other factors (Hubble flow, turbulence) dominate the line widths.
\citet{bolton04a} also found an increase in the IGM temperature in their radiative
transfer calculation.

We could also decrease the temperature by decreasing the He/H abundance ratio.
For our Haardt and Madau UVB spectrum, more heat is input per baryon
when the baryons are in $^4$He rather than in $^1$H.
Since He is ionized later than H, if we decrease the number of baryons in He
we decrease the temperature at $z=2$.
However, constraints on the primordial He abundance from standard big bang nucleosynthesis,
from CMB anisotropy constraints on the baryon density and from
observations of the He abundance in extragalactic H~II regions all make this
suggestion a long shot.

\begin{figure}
\includegraphics[width=84mm]{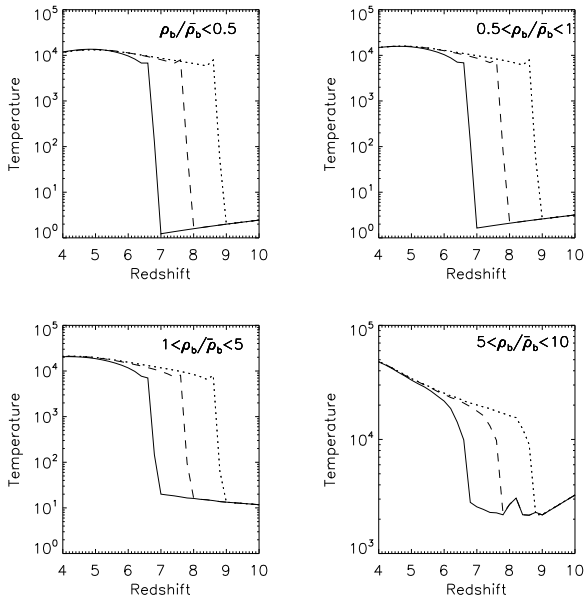}
\caption{The mean temperature (in K) of the IGM as a function of the epoch
of reionization.
 The four panels from the upper left show the mean temperatures for increasing ranges
 of density. We use a different scale at the highest densities in the lower right
 because they are at higher temperatures.
}
\label{fig.tempevl}
\end{figure}

The IGM might be cooler at $z=2$ if we ionize it earlier since the temperature of
the IGM is set in part by the
amount of expansion following the ionization. We now show that the effect is 
goes in the opposite direction, and is much too
small to be relevant at $z=2$. Again, we could make the ionization of either
 H~I or He~II or both occur earlier. We do this by increasing the intensity of the
UVB at early times. Integrated over time from the earliest $z$ we then need
more photons to reach a given ionization at $z=2$
because recombinations are faster when the IGM is denser.
However, we find that although cosmic expansion tends to cool down the IGM, it is the 
rapid increase of the UV
background intensity with decreasing redshift, as computed by Haardt \& Madau,
that causes the IGM to heat up, rather than cool down.
We demonstrate this in Fig. \ref{fig.tempevl} where we show how the mean temperature of the IGM changes with the
epoch of reionization. These results are from three simulations in 100 $h^{-1}$ Mpc
boxes each with $128^3$ cells. We use the Haardt \& Madau
ionizing spectrum described in \S \ref{secenzo}.
We use a very steep function to increase the flux from zero to the normal intensity
at some high $z$.
When we initiate the ionizing flux at redshifts
7, 8 or 9, the mean temperature is slightly higher when we ionize earlier, and
the changes are insignificant at $z < 5$.
These results differ from
\citet[Fig. 2]{miraldaescude94} who found a steep decrease in temperature with decreasing
$z$
because they used an ionizing spectrum that was continuously decreasing in intensity over time,
which is not now favored by data.

Lastly, we can change the \lya\ line widths by changing the
amplitude (\sig ) and shape
($n_s$) of the primordial power spectrum.
Fig 21 of J05 shows that a larger \sig\ to match the large-scale power of data
will also make the $b$-values smaller, as needed to match data.

We know from J05 that simulation A4 has approximately the correct $b$-values.
The temperature of the IGM at $z=2$, and at
the mean baryon density, may be close to 12,000~K
if \sig $=0.9$ and we neglect radiative transfer.
We expect a higher temperature is needed to match data if \sig $> 0.9$.

\subsection{Why do the simulations lack large scale power?}

We have seen that relative to \lyaf\ data
our simulations lack power on all scales, especially the largest
ones and their lines are too wide. We also noted that the simulations
have a different $f(N)$ distribution from the data, especially a lack
of lines with \lnhi $> 17$~\cmm .

We discussed in \S \ref{secfn} whether simulations with more
absorption from lines with \lnhi $>14$~\cmm\ and especially $>17$,
would have more power on large scales. We doubt that this will be
enough to match data since we have attemeted to exclude lines with \lnhi 
$>17$~\cmm\ from the data, and the changes required
in the column density distribution at
smaller \lnhi\ values are not large factors.

There are at least 
three other parameters that could increase the power in the simulations:
the amplitude (\sig ) and slope ($n_s$) of the initial power, and the
temperature of the gas (\gammahe ). We would have to simultaneously adjust the
\gammah\ to maintain the observed amount of H~I absorption, which does match data.
Our simulations use relatively high
values for both \sig = 0.9 and $n_s = 1.0$, and a best guess for the
heating by the UVB.

The power in the simulations could be below the data because of errors in the data.
The power of the data presented in J05 (PJ05) include only \lya\ lines
(no metals) with \lnhi $< 17.2$~\cmm , hence they should be directly
comparable to the simulations, but they may still contain some metal lines.
In J05 we noted that
the PJ05 power spectrum values could be too large, because
they are 30\% larger than our estimates of the power from \citet{kim04e}
after we removed the power from metal lines. However, the SDSS measurement of power
should be as reliable as any current measurement, and the SDSS power is very
similar to that from J05.

 The most obvious way to increase the large-scale power in the simulations
 is to increase \sig\  above the large value of 0.9 that we used. We saw in Fig 21 of 
 J05 that larger \sig\ values also give smaller $b$-values, as required to better match data.
 We expect that larger \sig\ also gives larger small scale power, because the lines are narrower. 
 However, 
 \citet[Fig. 10a]{mcdonald02} found the opposite, perhaps because his simulations
 use simpler representations of small scale physics.

We should also investigate changing the power spectrum tilt or shape.
\citet[Fig. 10a]{mcdonald02} suggests that decreasing the slope from $n=1.0$ to
$n=0.95$ will increase the large-scale flux power in simulations by about 6\%, which is the desired
direction of change, but much less than the factor of 1.5 -- 2
increase needed to match data.

\section{Summary of the Comparison with Data}

Our simulations differ from data in at least three ways:
their $b$-values are too large,
they have too many lines with \lnhi $<14$~\cmm\ and too few with larger \nhi ,
and most conspicuously, the power spectra of their flux has too low an amplitude.
We have have found that increasing the box size does not help while decreasing the
cell size, or adding radiative transfer both make the critical differences larger. 
A \sig $>0.9$ is one way in which these simulations might match data. 

While is is early to reach a conclusion, we do feel that there is a real
difference between our simulations of the IGM and the data we are using.
It is  premature to conclude that we are adding too much
heat, or that we need \sig $> 0.9$, because we do not yet understand why
we do not match the power spectrum of the flux, and we need to be more
careful when we excise high column density \lya\ lines and metals from both
data and simulations. However, it is looking increasingly difficult to make
simulations that match the $b$-values, mean flux and flux power of the \lyaf\
at $z=2$
using popular values for the astrophysical and cosmological parameters.

\section*{Acknowledgments}
This work was supported by NSF grant AST 0507717. Simulations were produced
using the facilities of the NCSA SDSC supercomputer centers with LRAC
allocation MCA98N020.
We thank Brian O'Shea for help with ENZO and we
are extremely grateful to Pat McDonald for generously sharing his
considerable experience and insights.
We also acknowledge helpful discussion with
Alexei Kritsuk,
Avi Loeb,
Mordecai Mac-Low,
Joop Schaye and
Paul Shapiro.

\bibliographystyle{mn2e}
\bibliography{ms29dt}

\begin{thebibliography}{}

\bibitem[\protect\citeauthoryear{{Abel}, {Anninos}, {Zhang} \& {Norman}}{{Abel}
  et~al.}{1997}]{abel97a}
{Abel} T.,  {Anninos} P.,  {Zhang} Y.,    {Norman} M.~L.,  1997, New Astronomy,
  2, 181

\bibitem[\protect\citeauthoryear{Anninos, Zhang, Abel \& Norman}{Anninos
  et~al.}{1997}]{anninos97a}
Anninos P.,  Zhang Y.,  Abel T.,    Norman M.~L.,  1997, {New Astronomy}, 2,
  209

\bibitem[\protect\citeauthoryear{{Bagla} \& {Ray}}{{Bagla} \&
  {Ray}}{2005}]{bagla05a}
{Bagla} J.~S.,  {Ray} S.,  2005, \mnras, 358, 1076

\bibitem[\protect\citeauthoryear{{Barkana} \& {Loeb}}{{Barkana} \&
  {Loeb}}{2004}]{barkana04a}
{Barkana} R.,  {Loeb} A.,  2004, \apj, 609, 474

\bibitem[\protect\citeauthoryear{{Bodenheimer}, {Laughlin}, {R{\'o}zyczka} \&
  {Yorke}}{{Bodenheimer} et~al.}{2007}]{bodenheimer07a}
{Bodenheimer} P.,  {Laughlin} G.~P.,  {R{\'o}zyczka} M.,    {Yorke} H.~W.,
  eds, 2007, {Numerical Methods in Astrophysics: An Introduction}

\bibitem[\protect\citeauthoryear{{Bolton}, {Meiksin} \& {White}}{{Bolton}
  et~al.}{2004}]{bolton04a}
{Bolton} J.,  {Meiksin} A.,    {White} M.,  2004, \mnras, 348, L43

\bibitem[\protect\citeauthoryear{{Bracewell}}{{Bracewell}}{1986}]{bracewell}
{Bracewell} R.,  1986, The Fourier Transform and its Applications.
McGraw-Hill

\bibitem[\protect\citeauthoryear{{Bryan}, {Machacek}, {Anninos} \&
  {Norman}}{{Bryan} et~al.}{1999}]{bryan99}
{Bryan} G.~L.,  {Machacek} M.,  {Anninos} P.,    {Norman} M.~L.,  1999, \apj,
  517, 13

\bibitem[\protect\citeauthoryear{{Bryan} \& {Norman}}{{Bryan} \&
  {Norman}}{1997}]{bryan97a}
{Bryan} G.~L.,  {Norman} M.~L.,  1997, in {Clarke} D.~A.,  {West} M.~J.,  eds,
  Computational Astrophysics; 12th Kingston Meeting on Theoretical Astrophysics
  Vol.~123 of Astronomical Society of the Pacific Conference Series,
  {Simulating X-Ray Clusters with Adaptive Mesh Refinement}.
p.~363

\bibitem[\protect\citeauthoryear{{Bryan}, {Norman}, {Stone}, {Cen} \&
  {Ostriker}}{{Bryan} et~al.}{1995}]{bryan95a}
{Bryan} G.~L.,  {Norman} M.~L.,  {Stone} J.~M.,  {Cen} R.,    {Ostriker} J.~P.,
   1995, Computer Physics Communications, 89, 149

\bibitem[\protect\citeauthoryear{{Cen} \& {Ostriker}}{{Cen} \&
  {Ostriker}}{1999}]{cen99}
{Cen} R.,  {Ostriker} J.~P.,  1999, \apj, 514, 1

\bibitem[\protect\citeauthoryear{Croft \& Gaztanaga}{Croft \&
  Gaztanaga}{1997}]{croft97b}
Croft R.~A.~C.,  Gaztanaga E.,  1997, \mnras, 285, 793

\bibitem[\protect\citeauthoryear{{Croft}, {Weinberg}, {Katz} \&
  {Hernquist}}{{Croft} et~al.}{1998}]{croft98}
{Croft} R.~A.~C.,  {Weinberg} D.~H.,  {Katz} N.,    {Hernquist} L.,  1998,
  \apj, 495, 44

\bibitem[\protect\citeauthoryear{{Dav{\' e}} \& {Tripp}}{{Dav{\' e}} \&
  {Tripp}}{2001}]{dave01}
{Dav{\' e}} R.,  {Tripp} T.~M.,  2001, \apj, 553, 528

\bibitem[\protect\citeauthoryear{{D'Odorico}, {Viel}, {Saitta}, {Cristiani},
  {Bianchi}, {Boyle}, {Lopez}, {Maza} \& {Outram}}{{D'Odorico}
  et~al.}{2006}]{dodorico06a}
{D'Odorico} V.,  {Viel} M.,  {Saitta} F.,  {Cristiani} S.,  {Bianchi} S.,
  {Boyle} B.,  {Lopez} S.,  {Maza} J.,    {Outram} P.,  2006, \mnras, 372, 1333

\bibitem[\protect\citeauthoryear{{Eisenstein} \& {Hu}}{{Eisenstein} \&
  {Hu}}{1999}]{eisenstein99}
{Eisenstein} D.~J.,  {Hu} W.,  1999, \apj, 511, 5

\bibitem[\protect\citeauthoryear{{Gardner}, {Katz}, {Hernquist} \&
  {Weinberg}}{{Gardner} et~al.}{2001}]{gardner01a}
{Gardner} J.~P.,  {Katz} N.,  {Hernquist} L.,    {Weinberg} D.~H.,  2001, \apj,
  559, 131

\bibitem[\protect\citeauthoryear{{Gnedin}}{{Gnedin}}{1998}]{gnedin98b}
{Gnedin} N.~Y.,  1998, \mnras, 299, 392

\bibitem[\protect\citeauthoryear{{Gnedin}, {Baker}, {Bethell}, {Drosback},
  {Harford}, {Hicks}, {Jensen}, {Keeney}, {Kelso}, {Neyrinck}, {Pollack} \&
  {van Vliet}}{{Gnedin} et~al.}{2003}]{gnedin03a}
{Gnedin} N.~Y.,  {Baker} E.~J.,  {Bethell} T.~J.,  {Drosback} M.~M.,  {Harford}
  A.~G.,  {Hicks} A.~K.,  {Jensen} A.~G.,  {Keeney} B.~A.,  {Kelso} C.~M.,
  {Neyrinck} M.~C.,  {Pollack} S.~E.,    {van Vliet} T.~P.,  2003, \apj, 583,
  525

\bibitem[\protect\citeauthoryear{{Gonzalez} \& {Faber}}{{Gonzalez} \&
  {Faber}}{1997}]{gonzalez97}
{Gonzalez} A.~H.,  {Faber} S.~M.,  1997, \apj, 485, 80

\bibitem[\protect\citeauthoryear{Haardt \& Madau}{Haardt \&
  Madau}{2001}]{haardt01a}
Haardt F.,  Madau P.,  2001, in Clusters of galaxies and the high redshift
  universe observed in X-rays, Recent results of {XMM-Newton and Chandra},
  {XXXVIth Rencontres de Moriond}, {XXIst Moriond Astrophysics Meeting, March
  10-17, 2001, Savoie France}. Edited by {D.M. Neumann \& J.T.T. Van} Modelling
  the {UV/X-ray} cosmic background with {CUBA}

\bibitem[\protect\citeauthoryear{{Heitmann}, {Lukic}, {Fasel}, {Habib},
  {Warren}, {White}, {Ahrens}, {Ankeny}, {Armstrong}, {O'Shea}, {Ricker},
  {Springel}, {Stadel} \& {Trac}}{{Heitmann} et~al.}{2007}]{heitmann07a}
{Heitmann} K.,  {Lukic} Z.,  {Fasel} P.,  {Habib} S.,  {Warren} M.~S.,  {White}
  M.,  {Ahrens} J.,  {Ankeny} L.,  {Armstrong} R.,  {O'Shea} B.,  {Ricker}
  P.~M.,  {Springel} V.,  {Stadel} J.,    {Trac} H.,  2007, ArXiv e-prints, 706

\bibitem[\protect\citeauthoryear{{Heitmann}, {Luki{\'c}}, {Habib} \&
  {Ricker}}{{Heitmann} et~al.}{2006}]{heitmann06a}
{Heitmann} K.,  {Luki{\'c}} Z.,  {Habib} S.,    {Ricker} P.~M.,  2006, \apj,
  642, L85

\bibitem[\protect\citeauthoryear{{Hockney} \& {Eastwood}}{{Hockney} \&
  {Eastwood}}{1988}]{hockney88}
{Hockney} R.~W.,  {Eastwood} J.~W.,  1988, {Computer simulation using
  particles}.
Bristol: Hilger, 1988

\bibitem[\protect\citeauthoryear{{Hui}, {Burles}, {Seljak}, {Rutledge},
  {Magnier} \& {Tytler}}{{Hui} et~al.}{2001}]{hui01}
{Hui} L.,  {Burles} S.,  {Seljak} U.,  {Rutledge} R.~E.,  {Magnier} E.,
  {Tytler} D.,  2001, \apj, 552, 15

\bibitem[\protect\citeauthoryear{{Hui} \& {Gnedin}}{{Hui} \&
  {Gnedin}}{1997}]{hui97}
{Hui} L.,  {Gnedin} N.~Y.,  1997, \mnras, 292, 27

\bibitem[\protect\citeauthoryear{{Hui} \& {Rutledge}}{{Hui} \&
  {Rutledge}}{1999}]{hui99c}
{Hui} L.,  {Rutledge} R.~E.,  1999, \apj, 517, 541

\bibitem[\protect\citeauthoryear{Janknecht, Reimers, Lopez \& Tytler}{Janknecht
  et~al.}{2006}]{janknecht06a}
Janknecht E.,  Reimers D.,  Lopez S.,    Tytler D.,  2006, \aa, 458, 427

\bibitem[\protect\citeauthoryear{{Jena}, {Norman}, {Tytler}, {Kirkman},
  {Suzuki}, {Chapman}, {Melis}, {Paschos}, {O'Shea}, {So}, {Lubin}, {Lin},
  {Reimers}, {Janknecht} \& {Fechner}}{{Jena} et~al.}{2005}]{jena05a}
{Jena} T.,  {Norman} M.~L.,  {Tytler} D.,  {Kirkman} D.,  {Suzuki} N.,
  {Chapman} A.,  {Melis} C.,  {Paschos} P.,  {O'Shea} B.,  {So} G.,  {Lubin}
  D.,  {Lin} W.,  {Reimers} D.,  {Janknecht} E.,    {Fechner} C.,  2005,
  \mnras, 361, 70

\bibitem[\protect\citeauthoryear{{Katz}, {Weinberg}, {Hernquist} \&
  {Miralda-Escude}}{{Katz} et~al.}{1996}]{katz96a}
{Katz} N.,  {Weinberg} D.~H.,  {Hernquist} L.,    {Miralda-Escude} J.,  1996,
  \apj, 457, L57

\bibitem[\protect\citeauthoryear{{Kauffmann} \& {Melott}}{{Kauffmann} \&
  {Melott}}{1992}]{kauffmann92}
{Kauffmann} G.,  {Melott} A.~L.,  1992, \apj, 393, 415

\bibitem[\protect\citeauthoryear{{Kim}, {Bolton}, {Viel}, {Haehnelt} \&
  {Carswell}}{{Kim} et~al.}{2007}]{kim07b}
{Kim} T.~.,  {Bolton} J.~S.,  {Viel} M.,  {Haehnelt} M.~G.,    {Carswell}
  R.~F.,  2007, ArXiv e-prints 0711.1862, 711

\bibitem[\protect\citeauthoryear{{Kim}, {Hu}, {Cowie} \& {Songaila}}{{Kim}
  et~al.}{1997}]{kim97}
{Kim} T.,  {Hu} E.~M.,  {Cowie} L.~L.,    {Songaila} A.,  1997, \aj, 114, 1

\bibitem[\protect\citeauthoryear{{Kim}, {Cristiani} \& {D'Odorico}}{{Kim}
  et~al.}{2001}]{kim01}
{Kim} T.-S.,  {Cristiani} S.,    {D'Odorico} S.,  2001, \aap, 373, 757

\bibitem[\protect\citeauthoryear{{Kim}, {Viel}, {Haehnelt}, {Carswell} \&
  {Cristiani}}{{Kim} et~al.}{2004a}]{kim04e}
{Kim} T.-S.,  {Viel} M.,  {Haehnelt} M.~G.,  {Carswell} B.,    {Cristiani} S.,
  2004a, \mnras, 351, 1471

\bibitem[\protect\citeauthoryear{{Kim}, {Viel}, {Haehnelt}, {Carswell} \&
  {Cristiani}}{{Kim} et~al.}{2004b}]{kim04a}
{Kim} T.-S.,  {Viel} M.,  {Haehnelt} M.~G.,  {Carswell} R.~F.,    {Cristiani}
  S.,  2004b, \mnras, 347, 355

\bibitem[\protect\citeauthoryear{{Kirkman} \& {Tytler}}{{Kirkman} \&
  {Tytler}}{1997}]{kirkman97a}
{Kirkman} D.,  {Tytler} D.,  1997, \apj, 484, 672

\bibitem[\protect\citeauthoryear{Kirkman, Tytler, Lubin \& Charlton}{Kirkman
  et~al.}{2007}]{kirkman07a}
Kirkman D.,  Tytler D.,  Lubin D.,    Charlton J.,  2007, \mnras, 376, 1227

\bibitem[\protect\citeauthoryear{Kirkman, Tytler, Suzuki, Melis, Hollywood,
  James, So, Lubin, Jena, Norman \& Paschos}{Kirkman et~al.}{2005}]{kirkman05a}
Kirkman D.,  Tytler D.,  Suzuki N.,  Melis C.,  Hollywood S.,  James K.,  So
  G.,  Lubin D.,  Jena T.,  Norman M.~L.,    Paschos P.,  2005, \mnras, 360,
  1373

\bibitem[\protect\citeauthoryear{{Kohler} \& {Gnedin}}{{Kohler} \&
  {Gnedin}}{2007}]{kohler07a}
{Kohler} K.,  {Gnedin} N.~Y.,  2007, \apj, 655, 685

\bibitem[\protect\citeauthoryear{{Kritsuk}, {Padoan}, {Wagner} \&
  {Norman}}{{Kritsuk} et~al.}{2007}]{kritsuk07a}
{Kritsuk} A.~G.,  {Padoan} P.,  {Wagner} R.,    {Norman} M.~L.,  2007, ArXiv
  e-prints, 706

\bibitem[\protect\citeauthoryear{Lacey \& Cole}{Lacey \& Cole}{1994}]{lacey94}
Lacey C.,  Cole S.,  1994, \mnras, 271, 676

\bibitem[\protect\citeauthoryear{{Lanzetta}}{{Lanzetta}}{1991}]{lanzetta91a}
{Lanzetta} K.~M.,  1991, \apj, 375, 1

\bibitem[\protect\citeauthoryear{{Lukic}, {Heitmann}, {Habib}, {Bashinsky} \&
  {Ricker}}{{Lukic} et~al.}{2007}]{lukic07a}
{Lukic} Z.,  {Heitmann} K.,  {Habib} S.,  {Bashinsky} S.,    {Ricker} P.~M.,
  2007, ArXiv Astrophysics e-prints

\bibitem[\protect\citeauthoryear{{Mandelbaum}, {McDonald}, {Seljak} \&
  {Cen}}{{Mandelbaum} et~al.}{2003}]{mandelbaum03}
{Mandelbaum} R.,  {McDonald} P.,  {Seljak} U.,    {Cen} R.,  2003, \mnras, 344,
  776

\bibitem[\protect\citeauthoryear{{McDonald}}{{McDonald}}{2003}]{mcdonald03}
{McDonald} P.,  2003, \apj, 585, 34

\bibitem[\protect\citeauthoryear{{McDonald} \& {Miralda-Escud{\'
  e}}}{{McDonald} \& {Miralda-Escud{\' e}}}{2001}]{mcdonald01b}
{McDonald} P.,  {Miralda-Escud{\' e}} J.,  2001, \apj, 549, L11

\bibitem[\protect\citeauthoryear{{McDonald}, {Miralda-Escud{\' e}} \&
  {Cen}}{{McDonald} et~al.}{2002}]{mcdonald02}
{McDonald} P.,  {Miralda-Escud{\' e}} J.,    {Cen} R.,  2002, \apj, 580, 42

\bibitem[\protect\citeauthoryear{{McDonald}, {Miralda-Escud{\' e}}, {Rauch},
  {Sargent}, {Barlow}, {Cen} \& {Ostriker}}{{McDonald}
  et~al.}{2000}]{mcdonald00a}
{McDonald} P.,  {Miralda-Escud{\' e}} J.,  {Rauch} M.,  {Sargent} W.~L.~W.,
  {Barlow} T.~A.,  {Cen} R.,    {Ostriker} J.~P.,  2000, \apj, 543, 1

\bibitem[\protect\citeauthoryear{{McDonald}, {Seljak}, {Burles}, {}
  et~al.,}{{McDonald} et~al.}{2006}]{mcdonald06a}
{McDonald} P.,  {Seljak} U.,  {Burles} S.,  {}   et~al., 2006, \apjs, 163, 80

\bibitem[\protect\citeauthoryear{{Meiksin} \& {White}}{{Meiksin} \&
  {White}}{2004}]{meiksin04a}
{Meiksin} A.,  {White} M.,  2004, \mnras, 350, 1107

\bibitem[\protect\citeauthoryear{{Miralda-Escud\'e} \& Rees}{{Miralda-Escud\'e}
  \& Rees}{1994}]{miraldaescude94}
{Miralda-Escud\'e} J.,  Rees M.,  1994, \mnras, 266, 343

\bibitem[\protect\citeauthoryear{{Misawa}, {Tytler}, {Iye} M.~{Paschos},
  {Norman}, {Kirkman}, {O'Meara}, {Suzuki} \& {Kashikawa}}{{Misawa}
  et~al.}{2004}]{misawa04}
{Misawa} T.,  {Tytler} D.,  {Iye} M.~{Paschos} P.,  {Norman} M.,  {Kirkman} D.,
   {O'Meara} J.,  {Suzuki} N.,    {Kashikawa} N.,  2004, \aj , astroph/0303476,
  128, 2954

\bibitem[\protect\citeauthoryear{{Neto}, {Gao}, {Bett}, {Cole}, {Navarro},
  {Frenk}, {White}, {Springel} \& {Jenkins}}{{Neto} et~al.}{2007}]{neto07a}
{Neto} A.~F.,  {Gao} L.,  {Bett} P.,  {Cole} S.,  {Navarro} J.~F.,  {Frenk}
  C.~S.,  {White} S.~D.~M.,  {Springel} V.,    {Jenkins} A.,  2007, ArXiv
  e-prints, 706

\bibitem[\protect\citeauthoryear{{Norman} \& {Bryan}}{{Norman} \&
  {Bryan}}{1999}]{norman99}
{Norman} M.~L.,  {Bryan} G.~L.,  1999, in ASSL Vol. 240: Numerical Astrophysics
  {Cosmological Adaptive Mesh Refinement}.
p.~19

\bibitem[\protect\citeauthoryear{{Norman}, {Bryan}, {Harkness}, {Bordner},
  {Reynolds}, {O'Shea} \& {Wagner}}{{Norman} et~al.}{2007}]{norman07a}
{Norman} M.~L.,  {Bryan} G.~L.,  {Harkness} R.,  {Bordner} J.,  {Reynolds} D.,
  {O'Shea} B.,    {Wagner} R.,  2007, ArXiv e-prints 0705.1556, 705

\bibitem[\protect\citeauthoryear{{O'Meara}, {Prochaska}, {Burles}, {Prochter},
  {Bernstein} \& {Burgess}}{{O'Meara} et~al.}{2007}]{omeara07a}
{O'Meara} J.~M.,  {Prochaska} J.~X.,  {Burles} S.,  {Prochter} G.,  {Bernstein}
  R.~A.,    {Burgess} K.~M.,  2007, \apj, 656, 666

\bibitem[\protect\citeauthoryear{O'Shea, Bryan, Bordner, Norman, Abel, Harkness
  \& Kritsuk}{O'Shea et~al.}{2004}]{oshea04a}
O'Shea B.~W.,  Bryan G.,  Bordner J.,  Norman M.~L.,  Abel T.,  Harkness R.,
  Kritsuk A.,  2004, in Adaptive Mesh Refinement -- Theory and Applications.
  Eds. T. Plewa, T. Linde \& V.G. Weirs, Springer Lecture Notes in
  Computational Science and Engineering.~ Intoducing {ENZO}, an {AMR} cosmology
  application

\bibitem[\protect\citeauthoryear{{O'Shea}, {Nagamine}, {Springel}, {Hernquist}
  \& {Norman}}{{O'Shea} et~al.}{2005}]{oshea05a}
{O'Shea} B.~W.,  {Nagamine} K.,  {Springel} V.,  {Hernquist} L.,    {Norman}
  M.~L.,  2005, \apjs, 160, 1

\bibitem[\protect\citeauthoryear{Paschos, Norman, Bordner \& Harkness}{Paschos
  et~al.}{2007}]{paschos07}
Paschos P.,  Norman M.~L.,  Bordner J.~O.,    Harkness R.,  2007, ArXiv
  e-prints 0711.1904

\bibitem[\protect\citeauthoryear{{Pen}}{{Pen}}{1997}]{pen97a}
{Pen} U.-L.,  1997, \apj, 490, L127

\bibitem[\protect\citeauthoryear{{Petitjean}, {Webb}, {Rauch}, {Carswell} \&
  {Lanzetta}}{{Petitjean} et~al.}{1993}]{petitjean93}
{Petitjean} P.,  {Webb} J.~K.,  {Rauch} M.,  {Carswell} R.~F.,    {Lanzetta}
  K.,  1993, \mnras, 262, 499

\bibitem[\protect\citeauthoryear{{Regan}, {Haehnelt} \& {Viel}}{{Regan}
  et~al.}{2007}]{regan07a}
{Regan} J.~A.,  {Haehnelt} M.~G.,    {Viel} M.,  2007, \mnras, 374, 196

\bibitem[\protect\citeauthoryear{Sargent, Young, Boksenberg \& Tytler}{Sargent
  et~al.}{1980}]{sargent80}
Sargent W.,  Young P.,  Boksenberg A.,    Tytler D.,  1980, \apjs, 42, 41

\bibitem[\protect\citeauthoryear{{Sargent}, {Steidel} \&
  {Boksenberg}}{{Sargent} et~al.}{1989}]{sargent89a}
{Sargent} W.~L.~W.,  {Steidel} C.~C.,    {Boksenberg} A.,  1989, \apjs, 69, 703

\bibitem[\protect\citeauthoryear{{Schaye}}{{Schaye}}{2001}]{schaye01a}
{Schaye} J.,  2001, \apj, 559, 507

\bibitem[\protect\citeauthoryear{Schaye, Aguirre, Kim, Theuns, Rauch \&
  Sargent}{Schaye et~al.}{2003}]{schaye03a}
Schaye J.,  Aguirre A.,  Kim T.-S.,  Theuns T.,  Rauch M.,    Sargent W. L.~W.,
   2003, \apj, 596, 768

\bibitem[\protect\citeauthoryear{{Seljak}, {Makarov}, {McDonald} \& {and
  others}}{{Seljak} et~al.}{2005}]{seljak05}
{Seljak} U.,  {Makarov} A.,  {McDonald} P.,    {and others} 2005, PhysRevD, 71,
  3515

\bibitem[\protect\citeauthoryear{{Sirko}}{{Sirko}}{2007}]{sirko07a}
{Sirko} E.,  2007, PhD thesis, Princeton University

\bibitem[\protect\citeauthoryear{{Slosar}, {McDonald} \& {Seljak}}{{Slosar}
  et~al.}{2007}]{slosar07a}
{Slosar} A.,  {McDonald} P.,    {Seljak} U.,  2007, New Astronomy Review, 51,
  327

\bibitem[\protect\citeauthoryear{{Stengler-Larrea}, {Boksenberg}, {Steidel},
  {Sargent}, {Bahcall}, {Bergeron}, {Hartig}, {Jannuzi}, {Kirhakos}, {Savage},
  {Schneider}, {Turnshek} \& {Weymann}}{{Stengler-Larrea}
  et~al.}{1995}]{stenglerlarrea95}
{Stengler-Larrea} E.~A.,  {Boksenberg} A.,  {Steidel} C.~C.,  {Sargent}
  W.~L.~W.,  {Bahcall} J.~N.,  {Bergeron} J.,  {Hartig} G.~F.,  {Jannuzi}
  B.~T.,  {Kirhakos} S.,  {Savage} B.~D.,  {Schneider} D.~P.,  {Turnshek}
  D.~A.,    {Weymann} R.~J.,  1995, \apj, 444, 64

\bibitem[\protect\citeauthoryear{{Theuns}, {Leonard}, {Schaye} \&
  {Efstathiou}}{{Theuns} et~al.}{1999}]{theuns99}
{Theuns} T.,  {Leonard} A.,  {Schaye} J.,    {Efstathiou} G.,  1999, \mnras,
  303, L58

\bibitem[\protect\citeauthoryear{{Tytler}}{{Tytler}}{1981}]{tytler81}
{Tytler} D.,  1981, Nature, 291, 289

\bibitem[\protect\citeauthoryear{{Tytler}}{{Tytler}}{1982}]{tytler82}
{Tytler} D.,  1982, Nature, 298, 427

\bibitem[\protect\citeauthoryear{Tytler, Kirkman, O'Meara, Suzuki, Orin, Lubin,
  Paschos, Jena, Lin \& Norman}{Tytler et~al.}{2004}]{tytler04b}
Tytler D.,  Kirkman D.,  O'Meara J.,  Suzuki N.,  Orin A.,  Lubin D.,  Paschos
  P.,  Jena T.,  Lin W.-C.,    Norman M.,  2004, \apj, 617, 1

\bibitem[\protect\citeauthoryear{{Tytler}, {Kirkman}, {O'Meara}, {Suzuki},
  {Orin}, {Lubin}, {Paschos}, {Jena}, {Lin}, {Norman} \& {Meiksin}}{{Tytler}
  et~al.}{2004}]{tytler04c}
{Tytler} D.,  {Kirkman} D.,  {O'Meara} J.~M.,  {Suzuki} N.,  {Orin} A.,
  {Lubin} D.,  {Paschos} P.,  {Jena} T.,  {Lin} W.,  {Norman} M.~L.,
  {Meiksin} A.,  2004, \apj, 617, 1

\bibitem[\protect\citeauthoryear{{Viel}, {Becker}, {Bolton}, {Haehnelt},
  {Rauch} \& {Sargent}}{{Viel} et~al.}{2007}]{viel07a}
{Viel} M.,  {Becker} G.~D.,  {Bolton} J.~S.,  {Haehnelt} M.~G.,  {Rauch} M.,
  {Sargent} W.~L.~W.,  2007, ArXiv e-prints 0709.0131

\bibitem[\protect\citeauthoryear{{Viel}, {Haehnelt}, {Carswell} \&
  {Kim}}{{Viel} et~al.}{2003}]{viel03c}
{Viel} M.,  {Haehnelt} M.,  {Carswell} R.,    {Kim} T.,  2003, astro-ph/0308078
  (relevant Fig missing from published version)

\bibitem[\protect\citeauthoryear{{Viel}, {Haehnelt} \& {Lewis}}{{Viel}
  et~al.}{2006}]{viel06b}
{Viel} M.,  {Haehnelt} M.~G.,    {Lewis} A.,  2006, \mnras, 370, L51

\bibitem[\protect\citeauthoryear{{Weymann}, {Jannuzi}, {Lu}, {Bahcall},
  {Bergeron}, {Boksenberg}, {Hartig}, {Kirhakos}, {Sargent}, {Savage},
  {Schneider}, {Turnshek} \& {Wolfe}}{{Weymann} et~al.}{1998}]{weymann98}
{Weymann} R.~J.,  {Jannuzi} B.~T.,  {Lu} L.,  {Bahcall} J.~N.,  {Bergeron} J.,
  {Boksenberg} A.,  {Hartig} G.~F.,  {Kirhakos} S.,  {Sargent} W.~L.~W.,
  {Savage} B.~D.,  {Schneider} D.~P.,  {Turnshek} D.~A.,    {Wolfe} A.~M.,
  1998, \apj, 506, 1

\bibitem[\protect\citeauthoryear{{Zhang}, {Anninos}, {Norman} \&
  {Meiksin}}{{Zhang} et~al.}{1997}]{zhang97}
{Zhang} Y.,  {Anninos} P.,  {Norman} M.~L.,    {Meiksin} A.,  1997, \apj, 485,
  496

\end{thebibliography}

\appendix
\section{How we make Flux Spectra}

We calculate the optical depth $\tau(v) $ using equations in \S 4 of J05,
and the flux from $F = exp ( -\tau )$, where $F= 1.0$ in the absence of
absorption, and $F=0$ at the center of saturated absorption lines.
The spectra from all simulations are smooth functions of $v$, with no
discreteness from the cell size, because to determine
$\tau $ at each position we integrate over at least $\pm 600$ cells
to the left and right of the absorbing cell. At $\approx 5$ km/s
velocity resolution at z=2 this corresponds to an integration range of at
least 6000 km/s. The number of cells involved in the integration is
even larger when the absorption is from a high density region
that makes a DLA line with extensive wings.

The bulk of the spectra that we present here, unlike those in J05, were made
at a fixed redshift.
These spectra are frozen in time, as if made by light
travelling infinitely fast at the chosen $z.$ We convert from the Mpc per grid
cell into velocity $v$ in \kms\ of a spectrum using the $H(z)$ for the
chosen $z$, but we do not increment the $z$ as we move down a sight line,
and we do not change the H(z).  These frozen spectra are pixel to pixel
identical if we pass through the box from the top to the bottom or in the
reverse direction.

Each spectrum starts on a face of the box and spans the length
of the simulation box to the opposite face, along the z-direction.
We made $N^2$ spectra that filled one side of the box. As for the CDM,
have explored the other two orthogonal directions and found no
difference of interest.

We encountered two problems with the spectrum generator. We initially
truncated the integrations of the H~I number density to obtain the opacity
 at $\pm 500$~\kms\ which lead to sharp cutoffs
in the edges of the lines when high density cells first enter or
leave an integration.
In J05 we truncated at $\pm 200$ cells, which was approximately 250~\kms\
in the highest resolution simulation (B2) and 2000~\kms\ for the A series.
The second problem was
an error in the spectrum generator which made it fail
when the optical depth exceeded $10^7$, giving a near vertical spike in the
spectrum. The first problem
made the power spectra in all simulations turn up to higher than
correct power on scales near the Nyquist frequency. The second problem effected 15
spectra from the A simulations alone, since the others did not have
such high optical depths. We do not expect the last problem to have
affected the results in J05 because
we most likely did not sample high density regions with our random lines of sight.
However, we believe that the limited integration range could have contaminated
the spectra we reported in J05 and may explain some of the odd behaviour of
the flux power that J05 discuss.

\section{Evolving Spectra}
\label{secevol}

In addition to the spectra that we have discussed so far, all
frozen in time, we also made
spectra that include an approximation to the evolution expected
in the IGM as light travels to us.
The matter power in the IGM increases with
decreasing $z$ while the flux power decreases because the mean amount of
absorption drops rapidly \citep{weymann98, janknecht06a, kirkman07a}.
These changes will produce some power on their own, since they make the signals
(matter density, flux, flux power) non-stationary, because they
change systematically with $z$.

The evolving spectra are intended to mimic the cosmological evolution of the
IGM as seen in QSO spectra. The parameters of the simulation are read out and
stored for
some $z$, and we then make the evolutions using scaling laws applied as the
rays propagate through the data dump.
We leave the radiation intensity constant, we scale the matter
density as $(1+z)^3$ and
the velocities as $(1+z)$. The ionization then decreases as $z$ increases.
Since we increment the redshift as we move along a sight line,
we obtain different spectra if we reverse our direction of
passage through the box.

Over the length of a single box, the evolution is unlikely to be significant.
For example, crossing the A box, the redshift change is 0.05 and the mean
flux changes by 0.004, with is about half of the measurement errors
with current spectra. Moreover, we see a ten times smaller change in the mean
flux when we average over a range of redshifts that is symmetric about the
central redshift: the evolution of the mean flux with z is nearly linear
over small intervals. However, evolution is useful if we intend to traverse
many data dumps at different redshifts, since with evolution we can more
readily make  smoothly evolving spectra.

\begin{figure}
\includegraphics[width=84mm]{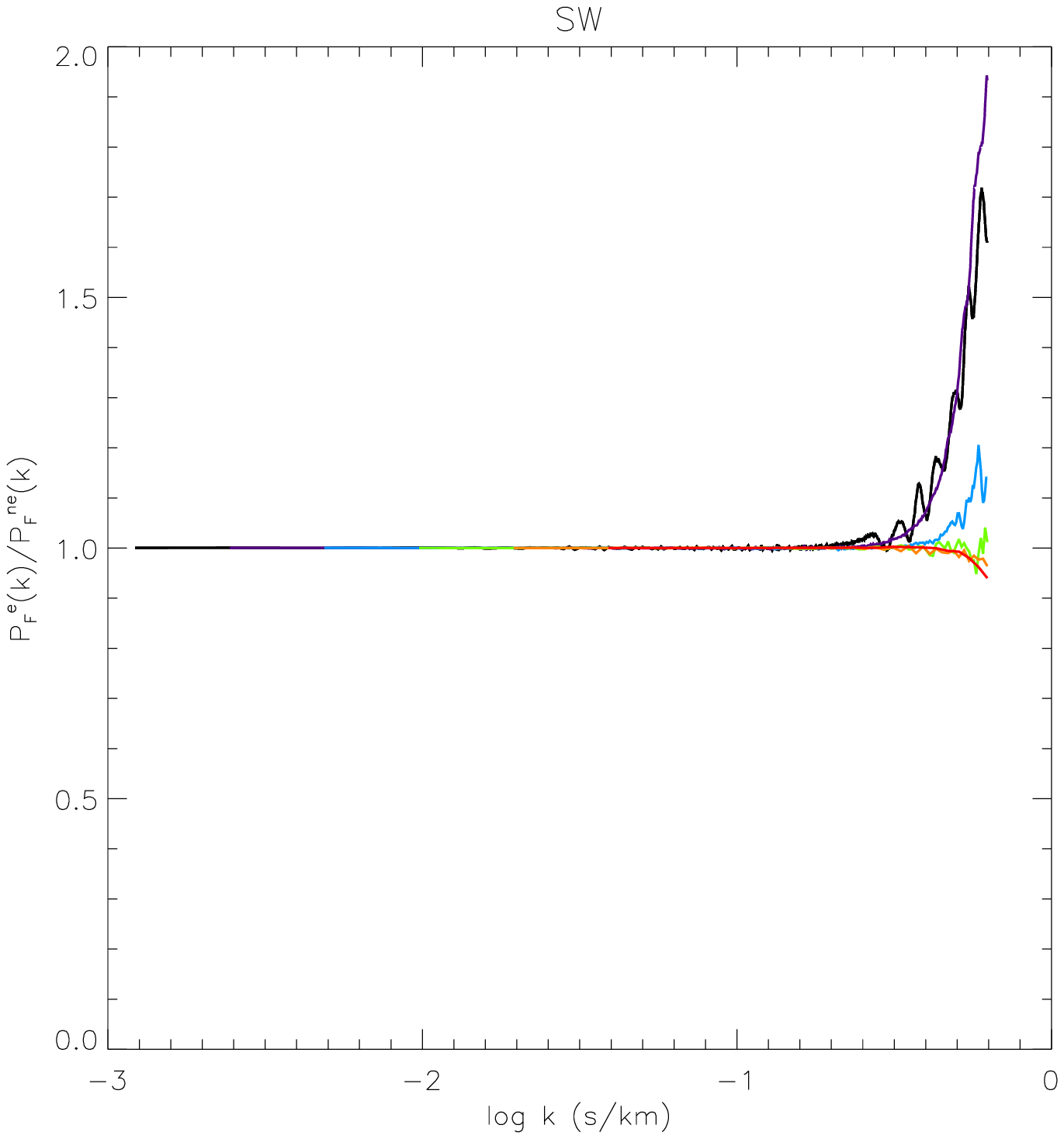}
\caption{
The effect of cosmological evolution along the sight lines on the power of
the flux. We plot the ratio of the power of the spectra with evolution to
the same without evolution.
}
\label{figevnw}
\end{figure}

In Fig. \ref{figevnw}
we show the power of the flux from evolving spectra divided by the same for the
stationary spectra that we have used up to now.
The values for both the evolving and non-evolving power spectra were given
in Table \ref{fluxpowerA},
Evolution has a dramatic effect on power spectra at high frequencies
obtained using the FFT algorithm.
The mean flux is different at either end of an evolving spectrum,
and hence the spectrum itself will not
join, but instead has a flux jump. When the box edge is in the wings of a
line, this jump can be 5\% in flux.
These jumps can increase the power in a single spectrum at frequencies within
a factor of a few of the Nyquist frequency by orders of magnitude, and
increase the mean power from all sight lines by ten times.

\begin{figure}
\includegraphics[width=84mm]{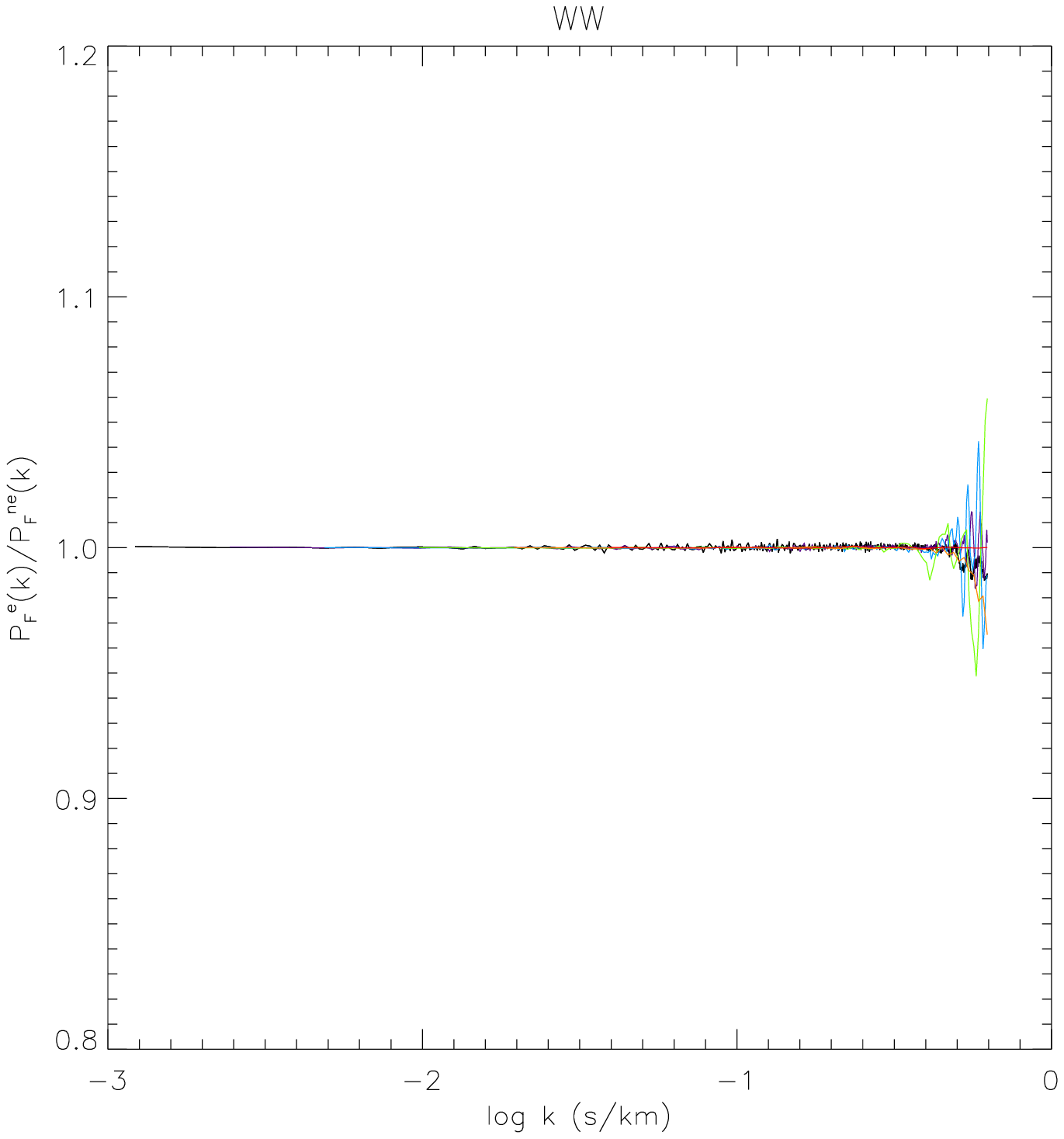}
\caption{
As Fig. \ref{figevnw} but after applying a Welch type window to both the
evolving and non-evolving flux spectra.
}
\label{figenvyw}
\end{figure}

There are several ways to recover the expected power from spectra with 
evolution. The evolution has
an extremely mild effect on the flux and the conversion from Mpc into velocity.
Both change slowly with velocity, in ways that have almost no effect on the
small scale power. The effects that we show in Fig. \ref{figevnw}
are artifacts of the use of the FFT on data with a discontinuity.
To avoid the artifact,
we could fit and remove the long term trends in the spectra before using the
FFT, we can use a non-FFT algorithm, or we can window the
data reducing the signal to zero at either end of each spectrum.
In Fig. \ref{figenvyw}
we see that applying a Welch type window before using
the FFT removes nearly all the artifacts.

\section{Extended Sight Lines in Random Directions}
\label{random}

We made all the spectra that we discussed so far parallel to the edges
of a box, with the length of the box edge.

We now discuss spectra that we made that can be of arbitrary length, by
passing through the box multiple times in random directions.
We can begin these spectra at random places in the simulation box, and
send them in random directions. They loop through the simulation cube
following the periodic boundary conditions, so that all the fields that
specify the simulation vary smoothly and continuously along the sight line.
A sight line that exits a face at 30 degrees to the normal will
enter the opposite face travelling in the original direction, and in general
the spectra do not duplicate.
Since the direction is random, we interpolate the pixel values using a
spline fit.

We have two different ways of adding evolution along these sight lines.
For short sight lines, say $\delta z = 0.1$, we use the passive
cosmological evolution that we described in \S \ref{secevol}.
For longer sight lines we can patch together the data dumps from the
simulation for different redshifts. When combined with the passive evolution,
this allows us to make spectra that are as long as the \lyaf\ in a
QSO spectrum. These spectra do not capture the full variation expected of
the QSO sight line, because we have only a single volume evolving in time.
However, they are often useful. Since they have the same length as real spectra,
they respond in the same way to division by the mean flux in each spectrum.
They also contain the entire line profile from a DLA with a high \nhi , a 
line that can make the flux zero across the width of a small box.

\section{Lack of Realistic Variation in the Simulations}

When we compare to data, we are conscious
that the simulation boxes have less variation for many connected reasons.
We can see this difference by eye since the simulated spectra are more
uniform and lack the large variations and strong clumping that we see
in real spectra with the same total absorption from the low density \lyaf\ alone. 
This comparison is difficult
because the real spectra include strong \lya\ lines and metals that we need to mask
and ignore. 

At the physical level, our simulations all have identical ionizing radiation, both
within the volume of each box and from box to box. They have exactly the
mean density of the universe, and they begin with the
mean power of the universe. Because of the
periodic boundary conditions, they contain only modes that fit inside the box.
All the simulations shown here have the same random number seed for the
initial phases of their power.

\begin{figure}
   \includegraphics[width=84mm]{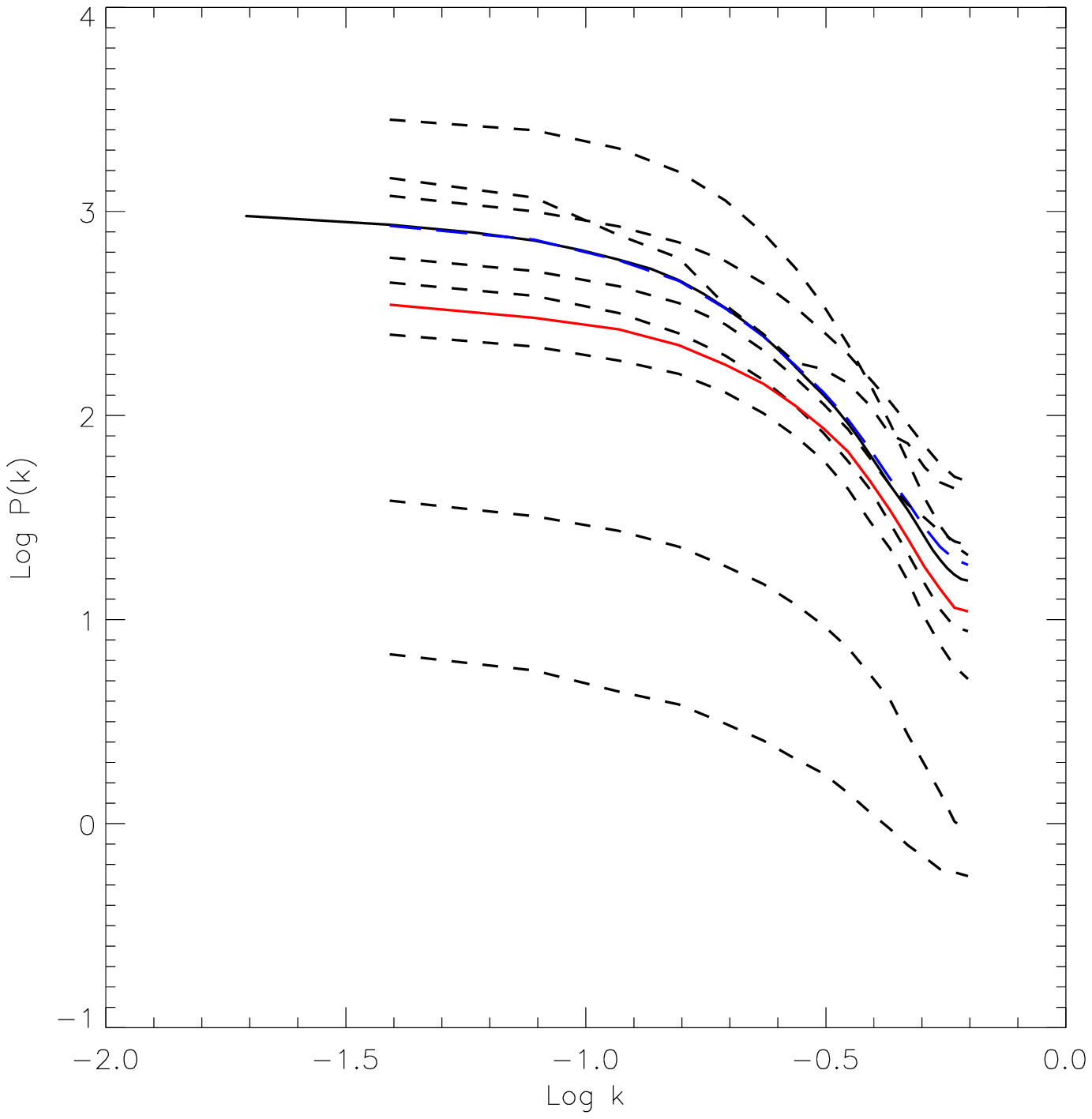}
   \caption{\label{subpower} The 1D power spectra of \dc -1 for sight lines
that fill each of the eight sub-cubes that exactly fill the A6 simulation.
Each sub-cube has the same volume of A7, but unlike A7, the sub-cubes are not
periodic, and they do not have the mean power or the mean density of the
universe. The solid (blue) line extending farthest to the left is the power
for A6. The eight dashed lines show the power for the sub-cubes. The short
solid (red) line is the power for the A7 box.
   }
\end{figure}

In Figure \ref{subpower}
we see the power of \dc\ for each of the eight sub-cubes that exactly
fill simulation A6.
Each sub-cube has the same volume as A7, and the same number and length of
sight lines, but unlike A7, these sub-cubes are not periodic,
they do not have the mean density of the universe, and they did not begin
with the mean power in the universe.
The amplitudes of the power spectra in the sub-cubes differ
by many orders of magnitude. This is much more variation that seen in
Figure \ref{subpowerrandom}
where we selected sight lines  at random from all eight sub-cubes. The
variation in density amongst the sub-cubes is large and it has has a
large effect on the power.

\begin{figure}
   \includegraphics[width=84mm]{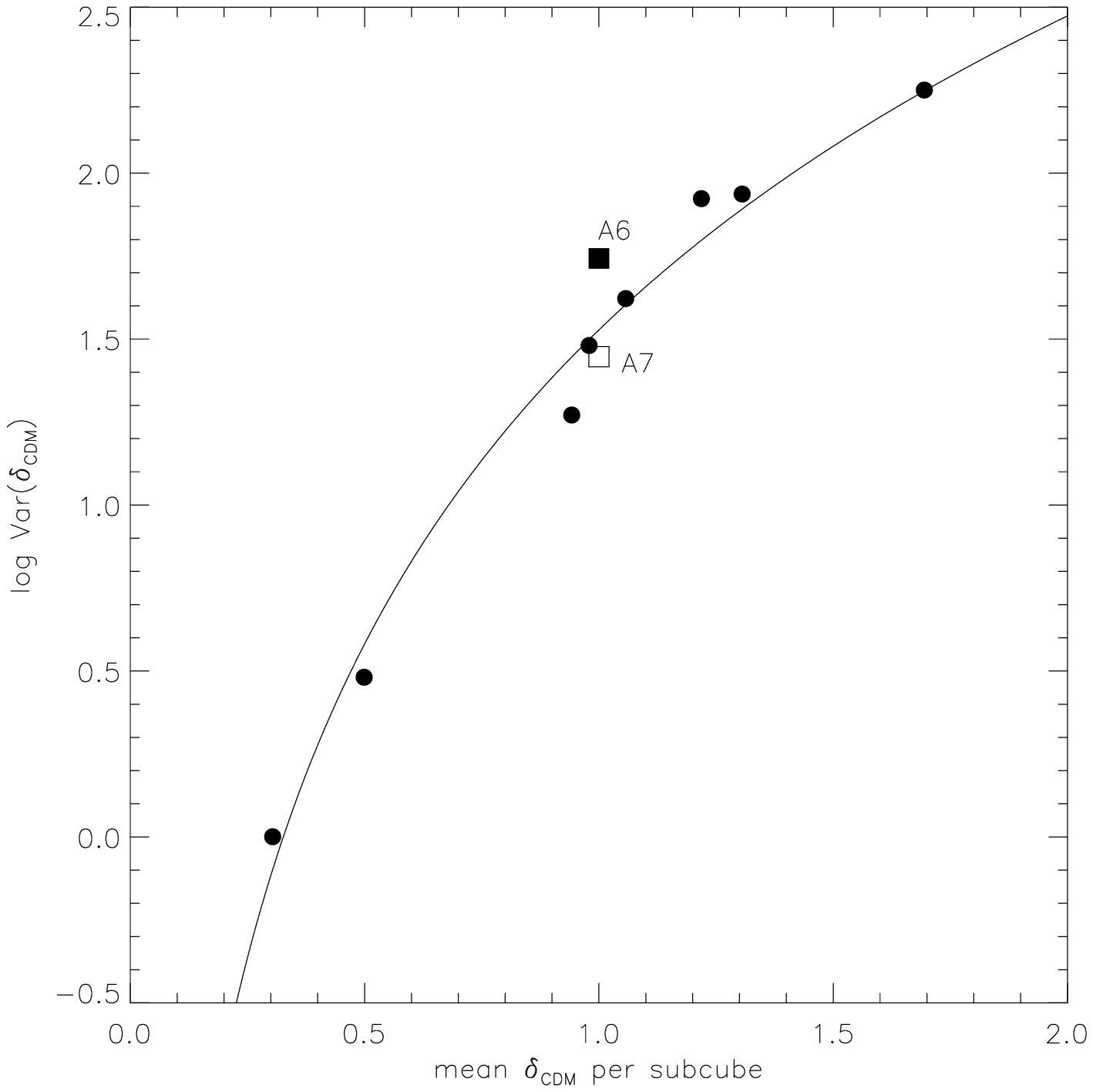}
   \caption{\label{subvar} The variance of \dc\ in sub-cubes of A6 as a
function of their mean \dc . We show the log of the mean of the
dark matter variance values vertically, and
the mean \dc\ horizontally. Each circle applies to one of the eight sub-cubes
that together fill A6 exactly. Each sub-cube has the same volume of A7. A7
is shown by the open square symbol and A6 by the filled square symbol, both at \dc = 1.0.
   }
\end{figure}

In Figure \ref{subvar}
we show the strong
correlation between the mean density in a sub-cube and the variance of
the \dc . There is a very large variation in the mean density amongst the
sub-cubes because they are only 2.4 Mpc on a side.

\clearpage

\end{document}